# Quantifying the interplay between fine structure and geometry of an individual molecule on a surface

**Authors:** Manuel Steinbrecher[1,†], Werner M. J. van Weerdenburg[1,†], Etienne F. Walraven[1,†], Niels P. E. van Mullekom[1], Jan W. Gerritsen[1], Fabian D. Natterer[2], Danis I. Badrtdinov[3,1], Alexander N. Rudenko[4,3,1], Vladimir V. Mazurenko[3], Mikhail I. Katsnelson[1,3], Ad van der Avoird[1], Gerrit C. Groenenboom[1], Alexander A. Khajetoorians[1,*]

**Affiliations:**

[1] Institute for Molecules and Materials, Radboud University, 6525 AJ Nijmegen, The Netherlands

[2] Department of Physics, University of Zürich, Winterthurerstrasse 190, 8057 Zürich, Switzerland

[3] Theoretical Physics and Applied Mathematics Department, Ural Federal University, 620002 Ekaterinburg, Russia

[4] School of Physics and Technology, Wuhan University, Wuhan 430072, China

*Correspondence to: a.khajetoorians@science.ru.nl

†The authors contributed equally to this work.

**Abstract:**

The pathway toward the tailored synthesis of materials starts with precise characterization of the conformational properties and dynamics of individual molecules. Electron spin resonance based scanning tunneling microscopy can potentially address molecular structure with unprecedented resolution. Here, we determine the fine structure and geometry of an individual TiH molecule, utilizing a combination of a newly developed mK ESR-STM in a vector magnetic field and *ab initio* approaches. We demonstrate a strikingly large anisotropy of the *g*-tensor unusual for a spin doublet ground state, resulting from a non-trivial orbital angular momentum stemming from the molecular ground state. We quantify the relationship between the resultant fine structure, hindered rotational modes, and orbital excitations. Our model system provides new avenues to determine the structure and dynamics of individual molecules.

# I. INTRODUCTION

Precisely determining the fine structure, dynamics, and geometry of an individual molecule, with sub-molecular resolution, is a grand challenge in numerous fields of nanoscience. Scanning probe microscopy (SPM) has emerged as a surface imaging approach capable of intramolecular resolution of individual molecules [1,2], quantifying conformational modifications like the static Jahn-Teller distortion [3], or light-assisted conformational changes [4]. Complementary to imaging, SPM-based inelastic excitation spectroscopy (ISTS) has been successfully applied to infer the various intramolecular vibrational [5], rotational [6,7] or hindered rotational modes [8]. However, these methods lack the precision to quantify the interplay between structure and molecular geometry like methods such as electron spin resonance (ESR) [9,10]. These methods are also not well suited for studying low-energy dynamics, such as the quantum zero-point motion of hydrogen and other light elements that are quenched by strong tip-sample interactions. Moreover, the resolution of traditional SPM, particularly scanning tunneling microscopy (STM), is limited by both convolution [1,11,12] and current preamplifier related bandwidth issues that preclude insight into the structure and rotational dynamics of individual molecules.

Hybrid methods have recently emerged, combining the spatial resolution of STM with temporal resolution [13,14] driven by continuous wave excitation [15]. THz-based STM [16,17] has been used to excite and quantify the vibrational motion of an individual phthalocyanine molecule with picosecond precision [18]. Likewise, electron paramagnetic/spin resonance has been established [15,19,20], based on a combination of microwave excitation of the STM junction, with the detection of spin-polarized current [21] of individual atoms. This technique, referred to as ESR-STM, has been used to quantify magnetic interactions, hyperfine couplings, and the coherent dynamics of individual magnetic impurities with unprecedented resolution [22-24]. However, in the spirit of traditional EPR/ESR, ESR-STM has yet to be applied to infer the molecular structure and the related low-energy modes of an individual molecule.

Here, we quantify the fine structure of an individual titanium-hydride molecule (TiH) on the surface of magnesium oxide (MgO), and use this to determine the molecular geometry and low-energy excitations of TiH with picometer precision. Utilizing a newly developed ESR-STM to access previously unmeasured low-frequency bands in multi-directional magnetic fields at milliKelvin temperature, we observe a giant anisotropy in the *g*-tensor concomitant with a spin-½ (doublet) ground state. Along a field direction parallel to the surface, the *g*-factor nearly has the free electron value, as expected for an ideal doublet. However, the *g*-factor is strongly renormalized in a field direction perpendicular to the surface, which has thus far not been measured. In the light of the inability of conventional density functional theory (DFT), as well as the mean-field DFT+*U* approach, to describe these experimental results, we adapted an approach based on quantum chemistry and exact quantum dynamics, to properly account for the correlations in this molecular system. Using this approach, we include the Coulomb interactions generated by the ions of the surface, and illustrate how the spin quartet electronic ground state of the isolated TiH molecule transforms into a doublet state as it approaches the MgO surface. We reveal that the origin of the strongly anisotropic *g*-tensor stems from a sizable orbital angular momentum of the electronic ground state, which DFT-based methods fail to predict, when the molecule is near the surface. Moreover, we quantify the *g*-tensor in embedded cluster calculations, which yield good agreement with the experiments, and enable determination of the structure and low-energy excitations of the molecule.

## II. RESULTS

### A. Electronic ground state calculation

TiH is a molecular radical relevant in molecular astrophysics and astrochemistry due to its abundance in space [25]. It has been predicted to host an electronic $^4\Phi$ ground state [25], with three parallel unpaired electrons and orbital occupation [Ar]$4\sigma^2(4\sigma^*)^1 3\mathrm{d}^1_{x^2-y^2/xy} 3\mathrm{d}^1_{xz/yz}$, where $4\sigma$ denotes the bonding molecular orbital formed between Ti and H and $4\sigma^*$ its anti-bonding counterpart. Nevertheless, there is no experimental data identifying its electronic structure in the gas phase. Starting from *ab initio* quantum chemistry, we consider the TiH molecule in the gas phase with $C_{\infty v}$ symmetry, in order to fully account for the orbital angular momentum. In the absence of the surface, TiH indeed resides in the $^4\Phi$ state, with a projected orbital angular momentum $\Lambda = 3$ and spin angular momentum ($S$) in a quartet configuration, $S = 3/2$. In the gas phase, the $^4\Phi$ state is favored (green) over the excited $^2\Delta$ state (orange), with $\Lambda = 2$ and a doublet $S = 1/2$, with an energy separation of 343 meV (Fig. 1(a)). The $^2\Delta$ state contains a single unpaired electron with orbital occupation of [Ar]$4\sigma^2(4\sigma^*)^2 3\mathrm{d}^1_{x^2-y^2/xy}$.

Magnetic atoms and molecules on surfaces are most often treated within density functional theory (DFT), where the effects of electron correlations are commonly considered in mean field approaches, such as DFT+$U$ and DMFT [26,27]. By contrast, quantum chemistry (QC) approaches can more precisely capture the electron correlations, but are not often used in combination with surfaces due to the computational complexity introduced by the distance-dependent coupling with the band structure of the surface. Due to the ionic and insulating nature of MgO (band gap is ≈ 6 eV) [28,29], we consider the surface with a point charge model where all charges are highly localized. This is a first approximation to the surface, and later we discuss higher-level theory where the surface is more properly treated in an embedded cluster approach. We also utilized DFT+$U$ (see supplemental), but this approach leads to an inaccurate prediction of the splitting of the doublet state and to a trivial value of the orbital angular momentum. In the QC approach, we accounted for the Mg and O atoms directly below TiH and mimic the rest of the surface by a finite lattice of point charges, $\pm 2e$, having the four-fold rotational symmetry of MgO(100), and used the lattice parameters obtained from the relaxed DFT+$U$ calculations (supplementary section S8 [30]).

The adsorption of TiH onto MgO strongly modifies the electronic structure of the molecule, and limits its angular motion due to the ionic environment. Starting from gas-phase calculations, we computed the state energies of the $^4\Phi$ and $^2\Delta$ electronic states of the TiH molecule as a function of distance ($d$) normal to the MgO surface (Fig. 1(a)). We considered TiH adsorption on top of oxygen (top site), in order to directly compare to the experimental data. As the molecule approaches the surface, there is a crossover in the favored ground state from the $^4\Phi$ state to the $^2\Delta$ state, below $d = 2.7$ Å. At the relaxed height from DFT+$U$ (dashed line), at $d \approx 2.50$ Å [22,31], the $^2\Delta$ state is therefore the ground state. We later confirm the favorability of the $^2\Delta$ state, at the relaxed height from theory, with embedded cluster calculations. The change in the ground state can be attributed to a higher electron density below the TiH in the $^4\Phi$ state than in the $^2\Delta$ state, resulting in a larger repulsion from the underlying MgO unit with the $^4\Phi$ state compared to the $^2\Delta$ state. This also forces the H atom to reside above the Ti atom, while the latter is closer to the O site. For $d$ = 2.42 Å, the degeneracy between the two preferred orbitals $^2\Delta_{x^2-y^2}$ (orange) and $^2\Delta_{xy}$ (cyan) is broken. This leads to a partially quenched orbital moment at short distances to the surface, which will be discussed later together with the precise values of the splitting and $g$-tensor. The striking difference between QC and DFT+$U$ [22,31], is that the molecule retains a sizeable orbital angular momentum compared to the negligible values resulting from DFT+$U$ (supplementary Table S2). Additionally, DFT+U overestimates the splitting of the ground state by roughly a factor of five. Although the $^2\Delta$ ground state maintains a non-trivial orbital angular momentum, it hosts a spin doublet and should not be susceptible to residual magnetic anisotropy in line with previous experimental observations [24].

### B. ESR-STM of an individual TiH molecule

In order to probe the orbital angular momentum of the TiH molecule and the possible effect of a crystal field, we adopt ESR-STM [15] down to mK temperatures [32,33] in magnetic field orientations parallel and perpendicular to the surface to extract the $g$-tensor, as schematically depicted in Fig. 1(b). By operating at lower temperature, we access new frequency bands corresponding to absolute energies roughly an order of magnitude lower than previously studied [15]. Our use of magnetically stable bulk Cr probes [19,34] with additionally picked up Fe atoms ensures spin polarization at zero field and enables ESR-STM at both magnetic field polarities. After cold deposition of Ti, TiH molecules appeared on both top and bridge sites of a two monolayer (ML) thick patch of MgO grown on a Ag(100) surface (see appendix A). We additionally co-deposited Fe atoms on the surface for tip preparation and calibration, and both species can be identified by a combination of their apparent height (Fig. 1(c)) and spectroscopic fingerprints (supplementary section S1). For the ESR measurements, we operated in two complementary modes, namely external magnetic field ($B$) sweep ($B$-sweep) mode or frequency ($f$) sweep ($f$-sweep) mode. In the first mode ($B$-sweep), we measured at selected values $f$, while the external field $B_{\text{ext}}$ was swept, and in the second mode ($f$-sweep), $B_{\text{ext}}$ was kept constant while $f$ was swept (see Appendix A).

We observed resonant excitations in both previously probed frequency bands as well as new band regimes (0.3 – 21 GHz). We note that the out of plane excitations have not been previously studied in detail. The resonance peak shifts at rates that depend on the orientations of $B_{\text{ext}}$ (Fig. 2). Measurements in $B$-sweep mode for $B_{\text{ext}}$ in the ⊥- (red) and ∥-directions (blue) to the surface are shown in Fig. 2(b), measured at constant RF amplitude ($V_{\text{RF}}$ = 7.9 mV) and at selected frequencies ranging from 1.165 to 10.92 GHz. Resonance peaks are each fitted with a Lorentzian (not shown), enabling precise identification of the peak location and width.

Strikingly, these linear trends have distinct slopes depending on the orientation of $B_{\text{ext}}$. The description of these different slopes resides in the anisotropy of the $g$-tensor, which is schematically depicted in Fig. 2(a). The allowed $\Delta\Omega = \pm1$ (with $\Omega = \Lambda + \Sigma$, see definitions in section E) ESR transitions at a given frequency occur for different amplitudes of $B_{\text{ext}}$, depending on the magnetic field orientation. Likewise, the linear behavior was observed in both orientations for both polarities of $B_{ext}$, enabling the determination of the offset magnetic fields due to the magnetic probe. We performed $f$-sweep mode measurements on the same molecule in the band $f = [7.9 – 8.5$ GHz]. Similar to the $B$-sweep mode, the peak positions extracted from Lorentzian fitting (solid lines) revealed an identical linear trend with slopes depending on the orientation of $B_{ext}$. Additional raw data sets of such experiments are presented in supplementary section S3, Fig. S7, S8 and S9.

### C. Anisotropic *g*-tensor

In order to ascertain the $g$-tensor, as well as set an upper bound of a potential zero-field splitting, we performed repeated measurements in both modes, in bands from 382 MHz to 22 GHz. Fig. 3(a) shows the extracted resonance peak positions of 21 experimental data sets with both measurement modes. All data points were recorded with the same measurement parameters for distinct micro-tips, different TiH molecules, as well as for various frequencies or $B$-field ranges and orientations. Both the $B$- and $f$-dependencies remain linear down to frequencies of 382 MHz and up to ≈ 21 GHz (inset). We note that we have not observed the hyperfine splitting on the oxygen binding site as reported in ref. [23]. We did observe non-linear trends for some, but not all, data sets at low frequencies (Figs. 3(a) and S10) that we attributed to a significant stray field of the STM tip $\vec{B}_{\text{tip}}$ with a component oriented orthogonal to the applied field $\vec{B}_{\text{ext}}$ (supplementary section S5). To extract values of the slopes, all data sets were fitted individually with linear functions, excluding the non-linear data (supplementary section S4). A $g$-factor was then extracted from each data set assuming a fixed magnetic moment of 1 $\mu_B$. All individually extracted $g$-factors are plotted in Fig. 3(b). The error bars represent the standard deviation from the cumulative error of the slope from linear regression. The weighted means were calculated from all data points and reveal $g_{\parallel} = 1.67 \pm 0.16$ and $g_{\perp} = 0.61 \pm 0.09$. Here, the error corresponds to twice the weighted standard deviation. The weighted average of these values represents the extracted $g$-factor for each distinct direction (∥ and ⊥). In this comprehensive analysis, we also included 10 additional data sets taken with different stabilization parameters and $V_{RF}$, which are shown in Fig. S10. The error bars, which are often smaller than the symbol size, emphasize the high precision of the ESR-STM method [22] where $g$ was determined with an error as small as $\Delta g = 0.0011$. Therefore, we can conclude that the scatter in the values of $g$ stems from a physical mechanism and not from the precision of the measurement (see supplementary section S9.5).

We note that there are a number of mechanisms that lead to a variation in the reported $g$-tensor. In addition to systematic variations resulting from the measurements, we observe a variation in the $g$-tensor based on the given tip as well as for a given molecule. Not all molecules were measured in both measurement modes, nor were all molecules measured in both field directions with the same tip, leading to an apparent difference in the $g$-factor value in the $y$-direction for the two different modes. We observe no variation in the precision between the two measurement modes, in the $z$-direction, where more statistics were measured.

The strong $g$-tensor anisotropy can result from a variety of phenomena. We can rule out the presence of a Jahn-Teller distortion, which would be accompanied by a crystal field splitting that is absent down to 382 MHz. This corresponds to an upper bound for a possible zero-field splitting of only 1.58 μeV, an energy precision inaccessible by other methods like ISTS. Complementary ISTS experiments without applied $V_{RF}$ and a non-spin-polarized tip confirmed the measured $g$-tensor anisotropy seen in ESR ($g_{\parallel,ISTS} = 1.84 \pm 0.01$; $g_{\perp,ISTS} = 0.50 \pm 0.01$) (Fig. S14). As we also illustrate later, TiH experiences a strong potential barrier with nearly cylindrical symmetry, which rules out a multi-well potential and a dynamic Jahn-Teller description. Due to the lower temperature of our present setup, compared to previous studies [22,24], the spin-½ ground state may enter a Kondo screening regime below a critical temperature. However, the magnetization of a spin-½ Kondo impurity exhibits a non-linearity at energy scales below/near $T_K$ [35], and we observe strictly linear behavior up to 21 GHz. These trends, including the $g$-tensor anisotropy, persist at elevated temperature up to 1.1 K (see supplementary section S6). Likewise, we observe no signature of a Kondo resonance in STS (Fig. S3). Therefore, we conclude that the $g$-tensor anisotropy concomitant with the lack of any non-linear trend or zero-field splitting is a direct result of the $^2\Delta$ molecular ground state, demonstrating the sizable orbital angular momentum of the molecule.

### D. Linewidth analysis

In addition to the anisotropic $g$-tensor of the molecule, we also observe a finite linewidth in a given resonance with an intrinsic linewidth comparable to that measured up to 50 times higher in temperature [19,22,24,36,37]. In Fig. 3(c), we illustrate the power dependence of one TiH molecule measured in $f$-sweep mode for different currents. We measured and subsequently fitted the extracted widths (top graph) and the intensities (bottom graph) of the resonance peaks for $V_{RF}$ ranging from 8 to 28 mV, similar to the reports in refs. [22,36]. We observed an asymptotic trend towards a resonance linewidth of ≈ 6 MHz for $I_t = 1$ pA, which compares to the ≈ 3.5 MHz reported for the same settings at higher temperature [22]. The resonance linewidth is broadened, e.g., by scattering with electrons, variations in magnetic field or $V_{RF}$, and variations of the Rabi frequency caused by mechanical motion of the tip relative to the sample [24,38]. In particular, it has been shown that the linewidth can be strongly increased with increasing applied power [38,39]. Our experimental findings indicate that the likely broadening mechanisms are temperature independent in the measured temperature range. This rules out other temperature dependent broadening mechanisms such as substrate electron scattering and spin-orbit coupling. It also suggests that the hyperfine coupling from the hydrogen nuclear moment or another degree of freedom may play a role in determining the saturated linewidth for the molecule.

### E. Theoretical model of the structure and excitations

Having established that the electronic state of the TiH molecule on the surface has $^2\Delta$ symmetry, we present the model used to calculate the $g$-tensor. We first discuss the free molecule, then introduce the effect of the crystal surface assuming the TiH molecule is vertical on the surface, and finally consider the effect of angular motion of the molecule on the $g$-factors. We present results for a simple point-charge model of the surface, as well as embedded-cluster calculations on the complete-active-space multiconfigurational self-consistent-field (CASSCF) level and the internally contracted multireference configuration interaction (MRCI) level, which we did with the Molpro quantum chemistry code [40].

For the free molecule, the spin $S = 1/2$ Hund's case (a) wave functions are $|\Lambda, \Sigma\rangle$, where $\Lambda = \pm 2$ and $\Sigma = \pm 1/2$ are the orbital and spin angular momentum projection quantum numbers. The spin-orbit coupling for these wave functions is given by $A_{\mathrm{SO}}\Lambda\Sigma$, and since the spin-orbit coupling constant $A_{\mathrm{SO}}$ is positive, $\Lambda$ and $\Sigma$ have opposite signs in the lower fine-structure state and $\Omega = \Lambda + \Sigma = \pm 3/2$, while the upper state has $\Omega = \pm 5/2$. In the field-free case, the states $|\Lambda, \Sigma\rangle$ and $|-\Lambda, -\Sigma\rangle$ are degenerate. Since spin-orbit coupling as well as the potential that describes the interaction with the surface commute with the time-reversal operator, it is convenient to use a time-reversal symmetry adapted basis

$$\Psi_{\pm}(\Lambda, \Sigma) = \frac{1}{\sqrt{2}}\{|\Lambda, \Sigma\rangle \pm |-\Lambda, -\Sigma\rangle\}. \quad (1)$$

The two-dimensional basis $\{\Psi_{+}(2, -1/2), \Psi_{-}(2, -1/2)\}$ describes the doubly degenerate lower fine-structure state. The degeneracy is lifted by the interaction with the magnetic field, which is described by the Zeeman Hamiltonian, in atomic units,

$$\hat{H}_{\mathrm{Zeeman}} = \mu_B\big(\hat{\boldsymbol{L}} + g_e\hat{\boldsymbol{S}}\big) \cdot \boldsymbol{B} \quad (2)$$

where $\hat{\boldsymbol{L}}$ and $\hat{\boldsymbol{S}}$ are the orbital and spin angular momentum vector operators, $g_e \approx 2.0023$ is the electron spin *g*-factor, $\mu_B$ is the Bohr magneton ($\mu_B = 1/2$ in atomic units), and $\boldsymbol{B}$ is the magnetic field. A *g*-factor is related to the derivative of the energy splitting with respect to the strength of the magnetic field, $B \equiv |\boldsymbol{B}|$, and must be divided by $\mu_B$. When the field is perpendicular to the surface, the degeneracy is lifted by the *z*-components of the angular momentum operators, which couple the two basis functions and we find for $\Lambda = 2$ and $\Sigma = -1/2$,

$$g_{\perp} = 2\big|\big\langle\Psi_{-}(\Lambda, \Sigma)\big|\hat{L}_z + g_e\hat{S}_z\big|\Psi_{+}(\Lambda, \Sigma)\big\rangle\big| \quad (3)$$

$$= 2(\Lambda + g_e\Sigma) = 4 - g_e \approx 2 \quad (4)$$

When the field is parallel to the surface, the $\hat{S}_x$ operator in the Zeeman Hamiltonian couples the lower and upper fine-structure states,

$$\hat{S}_x\Psi_{\pm}(\Lambda, \Sigma) = \pm\frac{1}{2}\Psi_{\pm}(\Lambda, -\Sigma) \quad (5)$$

But this coupling is second order and because it is the same for both time reversal symmetries the field does not lift the degeneracy of the lower state, and so $g_{\parallel} = 0$. However, there will be first order coupling when we consider the fine-structure states mixed by the crystal field.

The effect of the crystal potential $\hat{V}_S$ is to break the ($C_{\infty,v}$) cylinder symmetry of the TiH molecule, and lift the degeneracy of the $\Delta_{x^2-y^2}$ and $\Delta_{xy}$ components of the $^2\Delta$ state. With the TiH molecule vertically on top of an O-ion the system has four-fold symmetry, but we will use the Abelian symmetry group $C_{2v}$, for which the $\Delta_{x^2-y^2}$ and $\Delta_{xy}$ states are of $A_1$ and $A_2$ symmetry, respectively. The energies of the states are given by

$$V_1 = \big\langle\Delta_{x^2-y^2}\big|\hat{V}_S\big|\Delta_{x^2-y^2}\big\rangle \quad (6)$$

$$V_2 = \langle \Delta_{xy} | \hat{V}_S | \Delta_{xy} \rangle \tag{7}$$

and the off-diagonal element is zero by symmetry. The $\hat{L}_z$ orbital angular momentum eigenstates are related to the real functions through

$$|\Lambda = \pm 2\rangle = \frac{1}{\sqrt{2}}(\Delta_{x^2-y^2} \pm i\Delta_{xy}) \tag{8}$$

so we find that the crystal field quenches the orbital angular momentum by coupling the $\Lambda = 2$ and $\Lambda = -2$ states,

$$\langle \Lambda = \pm 2 | \hat{V}_S | \Lambda = \mp 2 \rangle = (V_1 - V_2)/2 \equiv V_c/2 \tag{9}$$

Just like the spin-orbit coupling, the crystal field cannot couple states with different time-reversal symmetry

$$\langle \Psi_+(\Lambda_1, \Sigma_1) | \hat{V}_S | \Psi_-(\Lambda_2, \Sigma_2) \rangle = 0 \tag{10}$$

but it will couple the fine-structure states with $\Omega = 3/2$ and $\Omega = 5/2$,

$$\langle \Psi_\pm(2, -1/2) | \hat{V}_S | \Psi_\pm(2, 1/2) \rangle = \pm V_c/2 \tag{11}$$

Thus, in the presence of both spin-orbit coupling and the crystal field, the wave functions are found variationally by diagonalizing a $2 \times 2$ Hamiltonian matrix

$$\boldsymbol{H}_\pm = \begin{pmatrix} -A_{SO} & \pm V_c/2 \\ \pm V_c/2 & A_{SO} \end{pmatrix} \tag{12}$$

The ground state is still doubly degenerate and the *g*-factors are found in closed form by computing how the magnetic field lifts this degeneracy, as for the free molecule. The eigenvectors of the Hamiltonian matrix only depend on the ratio of the diagonal and off-diagonal elements

$$r \equiv \left| \frac{V_c}{2A_{SO}} \right| \tag{13}$$

and so the *g*-factors are functions of the ratio *r*. In the SI (section S9.3), we show that

$$g_\parallel = g_e \frac{r}{\sqrt{1+r^2}}, \tag{14}$$

$$g_\perp = \left| g_e - \frac{4}{\sqrt{1+r^2}} \right|. \tag{15}$$

Both the crystal field coupling $V_c$ and the spin-orbit coupling depend on the height of the TiH above the surface. While the spin-orbit coupling varies only slightly from the gas phase to the equilibrium height, the crystal-field splitting, obviously, starts at zero in the gas phase, but increases exponentially near the surface [41]. As a result, not only the coupling which arises from the breaking of the cylinder symmetry must be known, but also the forces that determine the height of the molecule above the surface must be computed.

## Electrostatic approximation

We first consider a point charge model for the crystal-field splitting. In Fig. 4(b) we show how $g_{\parallel}$ and $g_{\perp}$ depend on the height above the surface. For $g_{\perp}$ we left out the absolute value from Eq. (15) so it is easier to distinguish the strong coupling regime where $r > 1$ and the weak coupling where $r < 1$. In the figure, the dashed line indicates the height above the surface as calculated at the DFT+$U$ level, and we see that the $g$-factors at that distance are in good agreement with experiment.

We computed the electrostatic crystal field splitting in an MRCI calculation where we included a $21 \times 21 \times 2$ grid of point charges to represent the crystal. In these calculations, we also included the O-ion right below the TiH and the Mg-ion in the second layer of the crystal below the central O-ion, to account for the Pauli-repulsion. These results are well converged with respect to the number of point charges. The adsorption height at this level of theory is 2.42 Å, close to the DFT+$U$ value. Furthermore, we found that if we set the point charges to (+1.7,-1.7) rather than the formal (+2,-2), the crystal field is reduced, but at the same time the adsorption height becomes smaller, and the effect on the $g$-tensor is small (see supplementary information section S9.5.6).

In our model, both $g_{\parallel}$and $g_{\perp}$ are determined by a single parameter, the interaction ratio $r$ [Eq. (13)]. In the appendix Fig. A4, we show the curve of possible combinations of $g_{\parallel}$ and $g_{\perp}$ within this model. This figure also shows the experimental values for the $g$-tensor and the error bars, which are near the curve of possible model results.

**Second order effects**

It is straightforward to extend the model to include other low-lying states such as the $^2\Pi$ and $^4\Phi$ states. These states are coupled to the $^2\Delta$ state through spin-orbit coupling, giving rise to second order spin-orbit effects. This introduces the energy separation and the spin-orbit coupling between the states as extra parameters, so Eqs. (14) and (15) no longer hold and so the results are no longer restricted to the curve in Fig. A4. However, since these low-lying states are not well described by single excitations of the $^2\Delta$ state, the couplings are most likely smaller than the coupling within the states. Some crude estimates suggest that the second order spin-orbit effects are very small, so we do not consider them further. In principle, the crystal field can couple the $^2\Delta$ state to neighboring doublet states, but because of the four-fold symmetry, coupling between $\Lambda = 2$ and $\Lambda = 1, 3$ only occurs in high order, so we also ignore these second order effects.

**Embedded cluster calculations**

To investigate the effect of going beyond the point-charge model for the crystal field splitting and the adsorption height, we performed a series of embedded cluster calculations, with four different clusters, up to $Mg_9O_9$, a $3 \times 3 \times 2$ cluster of ions. In these calculations, point charges were added again to extend the clusters to $21 \times 21 \times 2$. Convergence with respect to the one-electron basis set and with respect to the treatment of electron correlation was studied. Details of these calculations and their results are given in the appendix. The conclusion is that increasing the size of the cluster, using a larger one electron basis, and treating the electron correlation at a higher level, all tend to get the result closer to the experimental value. Our best embedded cluster results give $g_{\parallel} = 1.85$ and $g_{\perp} = 0.48$. This is surprisingly close to experiment and to the much simpler electrostatic calculations. We note that we verified that the $^2\Delta$ state is the ground state at the equilibrium adsorption height in the embedded cluster calculations.

**Spin-orbit coupling**

The spin-orbit coupling constant of the $^2\Delta$ state of TiH in the gas phase is about 119 $cm^{-1}$ when calculated at the full-valence CASSCF level in a aug-cc-pVTZ basis, using the full Breit-Pauli Hamiltonian as implemented in Molpro. In embedded cluster calculations at the same level of theory, with the TiH 5 Å above the surface we get almost the same result. At the equilibrium height (≈ 2.6 Å) for the $Mg_9O_9$ cluster (extended with point charges) we find the slightly smaller value of 116 $cm^{-1}$ at the CASSCF level of theory. When we improve the description of the electron correlation by computing the spin-orbit coupling at the MRCI level, using the full-valence active space of the CASSCF calculation as reference space, we find that the gas-phase value becomes quite a bit smaller: 110.17 $cm^{-1}$, compared to 119 $cm^{-1}$ at the CASSCF level. In the embedded cluster calculations, we also find that the spin-orbit coupling is smaller at the MRCI level, but only by about 2 $cm^{-1}$. The most likely explanation for this is that because of the larger numbers of electrons and orbitals in the embedded cluster calculations, we cannot include all the orbitals located on TiH that we included in the gas-phase calculation. Hence, our best estimate for the spin-orbit coupling constant at the equilibrium adsorption height is 107 $cm^{-1}$, about 3 $cm^{-1}$ below the gas-phase MRCI value. The results in the figures and tables were obtained using the gas-phase value. If we use our best estimate for the spin-orbit coupling from our most accurate embedded cluster calculation we find $g_{\parallel} = 1.87$ and $g_{\perp} = 0.56$, even closer to the experimental values.

**Dynamic model**

With TiH in the $^2\Delta$ state, the Ti-atom is slightly positive and the H-atom slightly negative. Since the TiH is on top of the negative O-ion, the equilibrium position is vertical in the electrostatic approximation. Still, there is always zero-point energy resulting in angular motion. This motion lowers the symmetry, which makes *ab initio* calculations harder. More importantly, since the two electronic states are nearly degenerate, angular motion will give rise to strong, possibly singular, nonadiabatic coupling. In the supplementary information section S9, we present an electrostatic model to construct diabatic electronic states that allow us to compute the coupled electronic-nuclear motion. The results of the static calculations at the vertical geometry suggest that the electrostatics captures much of the physics of the system. In Fig. 4(c), we plot the angular potential and the inclination angle dependent probability densities. The main conclusion is that the zero-point motion only has a small effect on the $g$-tensor.

As the potential is nearly cylindrically symmetric, the hindered rotations are only weakly affected by the four-fold symmetry of the underlying MgO (see Fig. S20). We therefore considered the probability density $\rho$ as a function of the polar angle $\theta$ for azimuthal angle $\varphi = 0°$, where the latter is defined to be towards one of the surrounding Mg atoms (or one of its 4-fold symmetric equivalents). The perturbed $^2\Delta$-state is defined by two electronic states correlating with the two components of the $^2\Delta$-state that are separated by an energy of about 70 meV. This suggests that the step observed in STS (Fig. S3, [19,20,22]) around ±90 mV may represent orbital excitation. As these excitations show variations in the presence of a magnetic tip but weak changes in magnetic field, we attribute these excitations to orbital excitations linked through spin-orbit coupling.

The low-energy hindered rotational modes of TiH are well described by a 2D quantum oscillator (Fig. 4(c)). The lowest rotational level is about 35 meV above the bottom of the potential well. This is rather large compared to the rotational constant of TiH of 674 μeV. Considering the steep potential energy barrier and its near cylindrical symmetry, we can rule out a multi-well potential and any tunneling of hydrogen related to the dynamic Jahn-Teller effect. The classical turning point of the lowest level at ≈ 15 degrees from the azimuth reflects the magnitude of the zero-point motion and the delocalization of the hydrogen wavefunction. We note that the $g$-tensor is insensitive to small perturbations of the internal Ti-H bond length ($\Delta g \approx 0.005$ for $\Delta R = 2$ pm, see Fig. 4(b) inset). The potential resembles an anharmonic quantum oscillator, which is also signified by the uneven energy spacing of the low-lying states. As the excitation of the quantum oscillation and the orbital excitation will lead to a spatial variance of hydrogen, this most likely will lead to fluctuations of the spin polarization measured directly above the molecule with the STM probe. These points may need to be considered in measurements of the coherent properties of TiH on MgO based on pulsed-ESR [24] which hitherto was reduced to a two-state system.

## III. CONCLUSIONS

In conclusion, utilizing a newly developed mK ESR-STM together with quantum modeling, based on quantum chemistry and density functional theory, we quantified the interplay between the fine structure, geometry and hindered rotations of an individual TiH molecule with unprecedented precision. Our measurements are exemplified by the striking observation of a giant anisotropy in the $g$-tensor concurrent with a doublet ground state. Adopting quantum chemistry calculations to account for electron correlation within the molecule and the effect of the surface, we demonstrated that the electronic ground state of TiH is modified near the surface, and that this electronic ground state hosts a sizeable orbital angular momentum. TiH at the surface of MgO provides a clear example of a system, which cannot qualitatively or quantitatively be described by mean-field approaches and where correlation effects play a crucial role. Our calculations show that DFT strongly overestimates the splitting of the $d$-states compared to the quantum chemistry calculations. This finding is relevant in understanding the ESR mechanism for the TiH molecule [31]. From detailed embedded cluster calculations, we reproduced the observed anisotropic $g$-tensor, which is highly sensitive to the height of the molecule above the surface. With this model, we also quantified the hindered rotational modes of the molecule and the orbital excitations of the molecule on the surface, which exhibit signatures similar to previous experimental observations. The combination of experiment and theory here provides an extremely powerful method to map out the fine structure of the molecule and relates it to the molecular geometry, going beyond what conventional scanning probe microscopy methods can provide. Moreover, the development of QC calculations for these classes of experiments provides a more accurate way of handling the electronic behavior and correlations of small molecules on surfaces as well as their resultant structural dynamics. In future experiments, it will be interesting to probe how the $g$-tensor is modified by varying the adsorption site (e.g., bridge site) as well as by varying the insulator. Likewise, there are many new questions raised concerning the ESR mechanism, utilizing a stable magnetic probe, such as the role of a non-collinear magnetization and spin pumping in the measured signal. In parallel, the observation of low frequency bands with mK ESR-STM signal demonstrates the exquisite energy resolution of mK based STM. This development opens up the possibilities to explore spin coherence in novel quantum states of matter, as well as the response of superconducting materials to RF fields [42,43].

## ACKNOWLEDGEMENTS

We acknowledge funding from NWO, and the VIDI project: "Manipulating the interplay between superconductivity and chiral magnetism at the single-atom level" with project number 680-47-534. This project has received funding from the European Research Council (ERC) under the European Union's Horizon 2020 research and innovation programme (SPINAPSE: grant agreement No 818399). F.D.N. thanks the Swiss National Science Foundation for financial support under grant PP00P2_176866. The work of D.I.B., A.N.R. and V.V.M. was supported by Act 211 Government of the Russian Federation, contract 02.A03.21.0006.

## APPENDIX A: MATERIALS & METHODS

### A. Experimental setup

Experiments were performed in a home-built UHV-STM system , which was upgraded for ESR measurements [33]. If not stated otherwise, the base temperature for all experiments was 30 mK $\leq T_{\text{base}} \leq$ 55 mK. The system houses a vector magnetic field with maximal out-of-plane value of 9 T and a maximal in-plane value of 4 T. The field was swept while the STM tip was in tunneling contact. The in-plane magnetic field direction is oriented 23.8° with respect to the oxygen rows of the MgO surface. Electrochemically etched Cr bulk tips with a diameter of 0.5 mm were used. Tips were in-situ cleaned by field emission prior to experiments. Additional Fe atoms were picked up to enhance spin-contrast.

The DC bias voltage $V_{\text{DC}}$ was applied to the tip and the sample was virtually grounded, unlike in previous publications, e.g. ref. [32], where $V_{\text{DC}}$ was applied to the sample. STS ($\mathrm{d}I/\mathrm{d}V$) was recorded via a lock-in technique with the feedback loop opened after applying stabilization parameters $V_{\text{DC}}$ and tunneling current $I_t$. A modulation voltage $V_{\text{mod}}$ (RMS) was added to $V_{\text{DC}}$ with modulation frequency $f_{\text{mod}}$ = 809 Hz. For the ESR experiments, a radio frequency (RF) voltage was generated with an analog microwave signal generator (Keysight N5183B) and added to $V_{\text{DC}}$ with a bias-tee at frequencies $f_{\text{RF}}$ ranging from MHz to GHz. $\tilde{P}_{\text{RF}}$ and $\tilde{V}_{RF}$ denote the output power or voltage at the generator and $V_{RF}$ represents the RF voltage at the junction. $V_{RF}$ is given as the zero-to-peak value throughout this manuscript. To measure a current-signal compatible with the bandwidth of the preamplifier (≈ 1 kHz), we used a well-established chopping scheme (15) at $f_{chop}$ = 877 Hz. The difference of the spin-polarized current $\Delta I_{ESR}$ is measured with a lock-in amplifier to optimize the signal-to-noise ratio.

### B. Sample preparation

Samples were prepared *in-situ* with a base pressure of $p \approx 10^{-10}$ mbar. Ag(100) was cleaned by repeated cycles of $Ar^+$ sputtering ($p_{Ar} \approx 2\cdot10^{-5}$ mbar, $V_{HV}$ = 1.5 kV) and annealing ($T \approx$ 570°C). MgO was grown on Ag(100) by depositing Mg from an effusion cell at $T_{sample} \approx$ 380°C for five minutes in an oxygen atmosphere of $p_{O2} \approx 3\cdot10^{-7}$ mbar. Fe and Ti atoms were co-deposited onto the cold surface (< 80 K) after MgO preparation.

## APPENDIX B: EMBEDDED CLUSTER CALCULATIONS

We compute the *g*-factors for TiH interacting with embedded clusters of four different sizes. The smallest cluster, MgO, consists of the O-ion to which the TiH is attached and the Mg-ion right below it. The second cluster, $Mg_5O$, has four additional Mg-ions, which are the four *xy*-plane nearest neighbors of the central O-ion. The third cluster, $Mg_5O_5$, also includes O-ions below these four additional Mg-ions. The largest cluster, $Mg_9O_9$, consists of 3x3x2 ions, i.e., it has four additional O-ions on the surface, with Mg-ions below them. In all cases the central O-ion is moved up by 0.48 Å, and the Mg-ion just below is moved down by 0.2 Å compared to the other ions in the second layer. All these clusters were extended with (2+, 2-) point charges to create a 21x21x2 cluster. The TiH is perpendicular to the surface in these calculations and the TiH bond distance is taken to be 1.773 Å. The distance of the center-of-mass of TiH to the top layer of the crystal ($d$) was varied from 2.38 to 5.29 Å (on a grid with steps of 0.1 $a_0$ between 4.5 and 5.5 $a_0$ and additional points at d = 5.75, 6, 6.5, 7, 8, 9, and 10 $a_0$).

All calculations were done with the Molpro 2015 quantum chemistry program [40]. Molecular orbitals were calculated at the complete-active-space self-consistent field (CASSCF) level, with either the cc-pVDZ or cc-pVTZ correlation-consistent polarized-valence one-electron basis sets of double-zeta or triple-zeta quality. The calculations were done in $C_{2v}$ point group symmetry. In all calculations the active space consists of seven molecular orbitals and contains five electrons, corresponding to the valence orbitals and electrons of TiH. In $C_{2v}$ symmetry this gives four orbitals of $a_1$ symmetry, and an additional orbital in each of the remaining irreps ($b_1$, $b_2$ and $a_2$), which we denote as [4,1,1,1]. The two electronic states correlating with the two components of the $^2\Delta$ state have symmetries $a_1$ and $a_2$. We note that if the LATTICE keyword of Molpro 2015 is used to add point charges, it is not possible to use symmetry. However, with the MATROP matrix operation facility of Molpro it is possible to add Coulomb operators to the one-electron Hamiltonian, while still using symmetry.

To generate the initial orbital guess we first calculated the molecular orbitals of the crystal and those of TiH separately. The TiH orbitals were calculated using the LQUANT,2 option of the CASSCF program to effectively use $C_{\infty,v}$ symmetry and force $\Lambda = 2$, i.e., select the $^2\Delta$-state. The Molpro MERGE option was used to generate the orbital guess of the cluster by combining molecular orbitals of TiH with molecular orbitals of the cluster. The initial guess of the molecular orbitals of the crystal was obtained by merging orbitals of the $O^{2-}$ and $Mg^{2+}$ sub-lattices. This somewhat elaborate procedure guarantees convergence to the proper states and avoids artificial symmetry breaking and other convergence issues.

To investigate the effect of dynamic correlation we performed internally-contracted multi-reference configuration-interaction calculations with single and double excitations (MRCI). The active space of the CASSCF calculations was used as reference space. A total of thirteen electrons were correlated, with single and double excitations from four core orbitals, two $a_1$, one $b_1$, and one of $b_2$ symmetry. The effect of higher excitations was estimated with the Pople size-consistency correction (MRCI+Q).

The computation of the *g*-factors requires matrix elements of the orbital angular momentum operator $\hat{L}_z$ between the two electronic states correlating with the $^2\Delta$-state and also the spin-orbit interaction between these states.

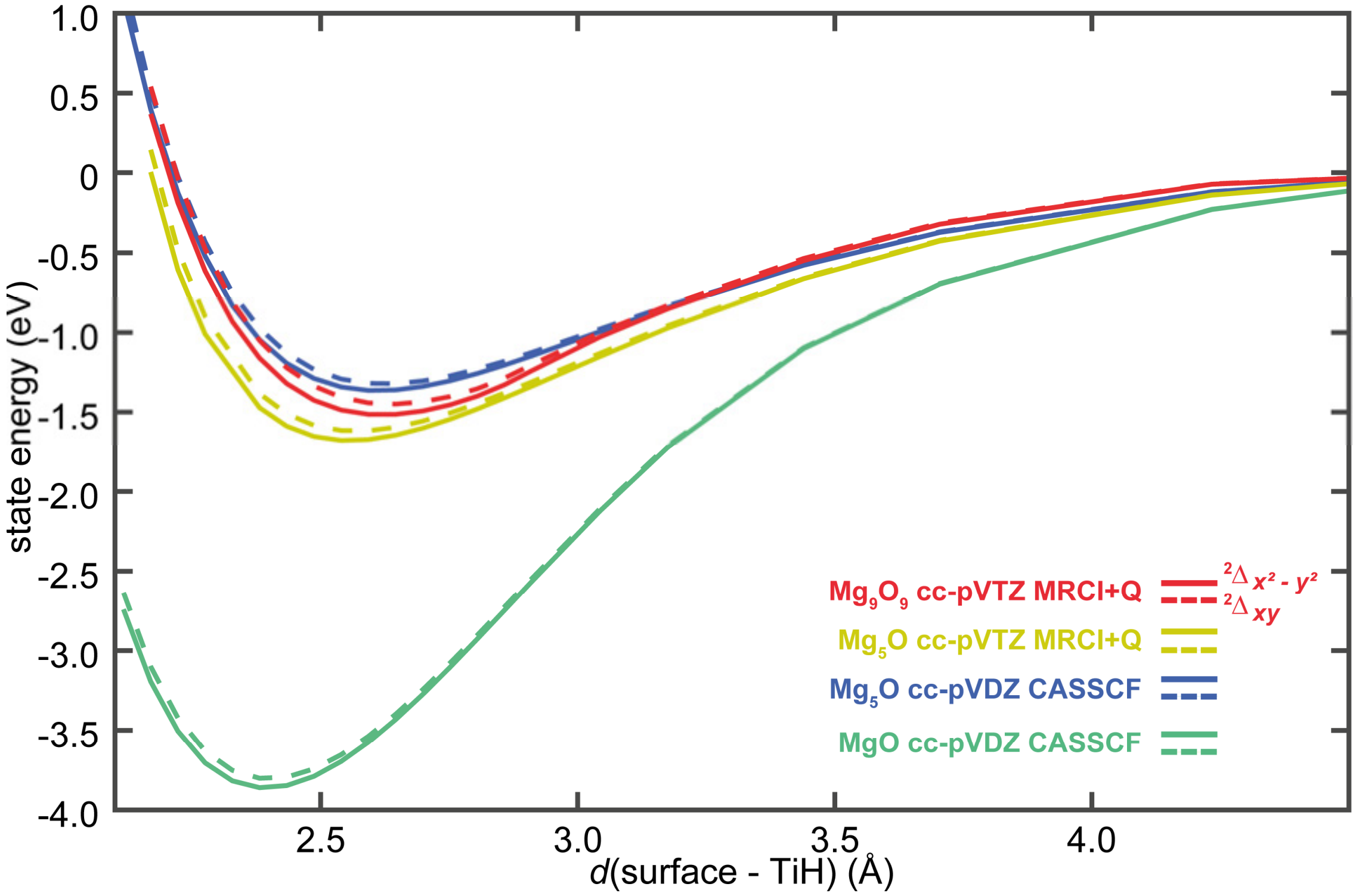

FIG. A1. Potential energy curves of the $^2\Delta$ state as a function of the height above the surface in various embedded cluster calculations.

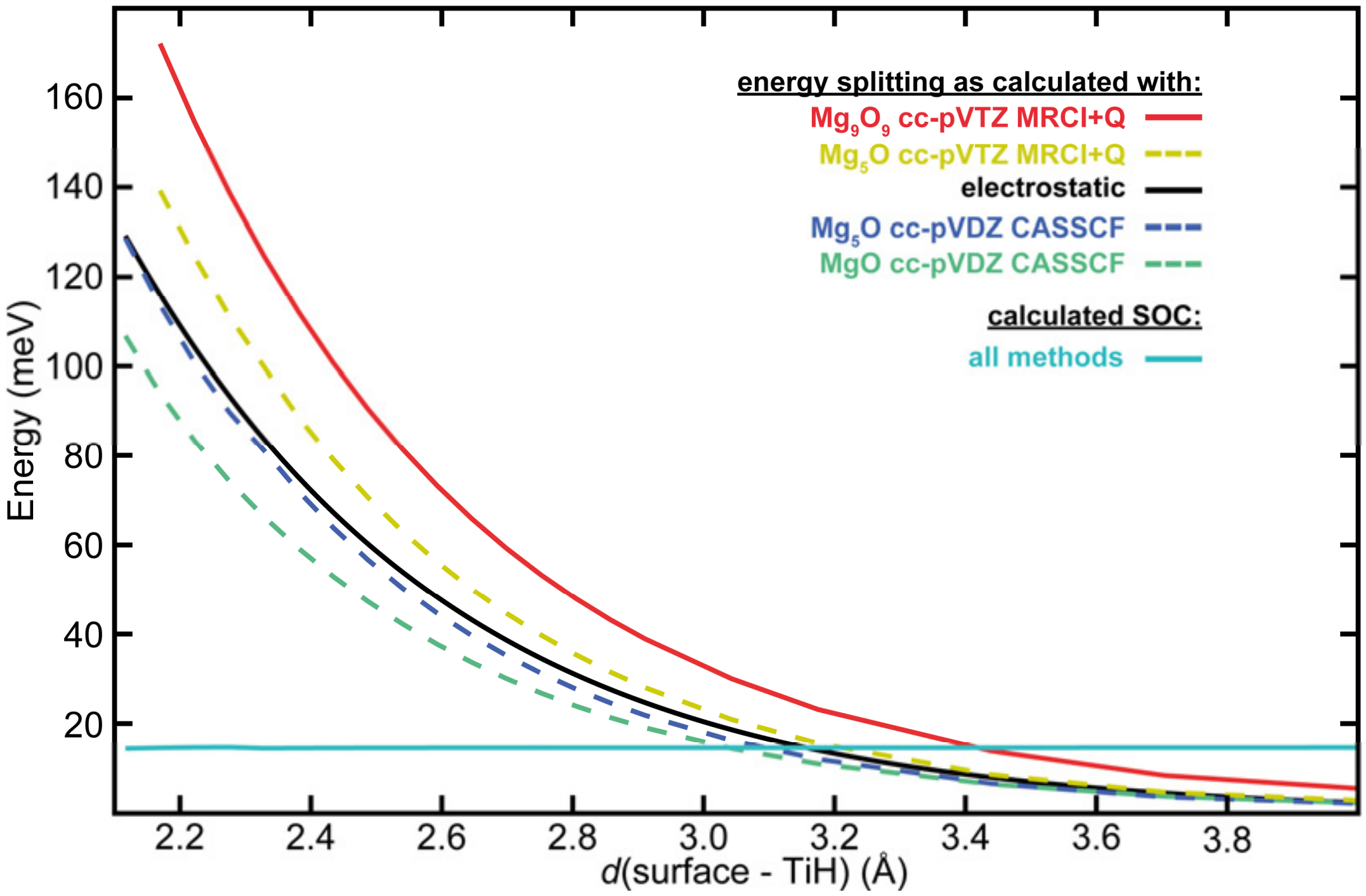

FIG. A2 Energy splitting and spin-orbit coupling constant of the $^2\Delta$ state as a function of the height above the surface in various embedded cluster calculations.

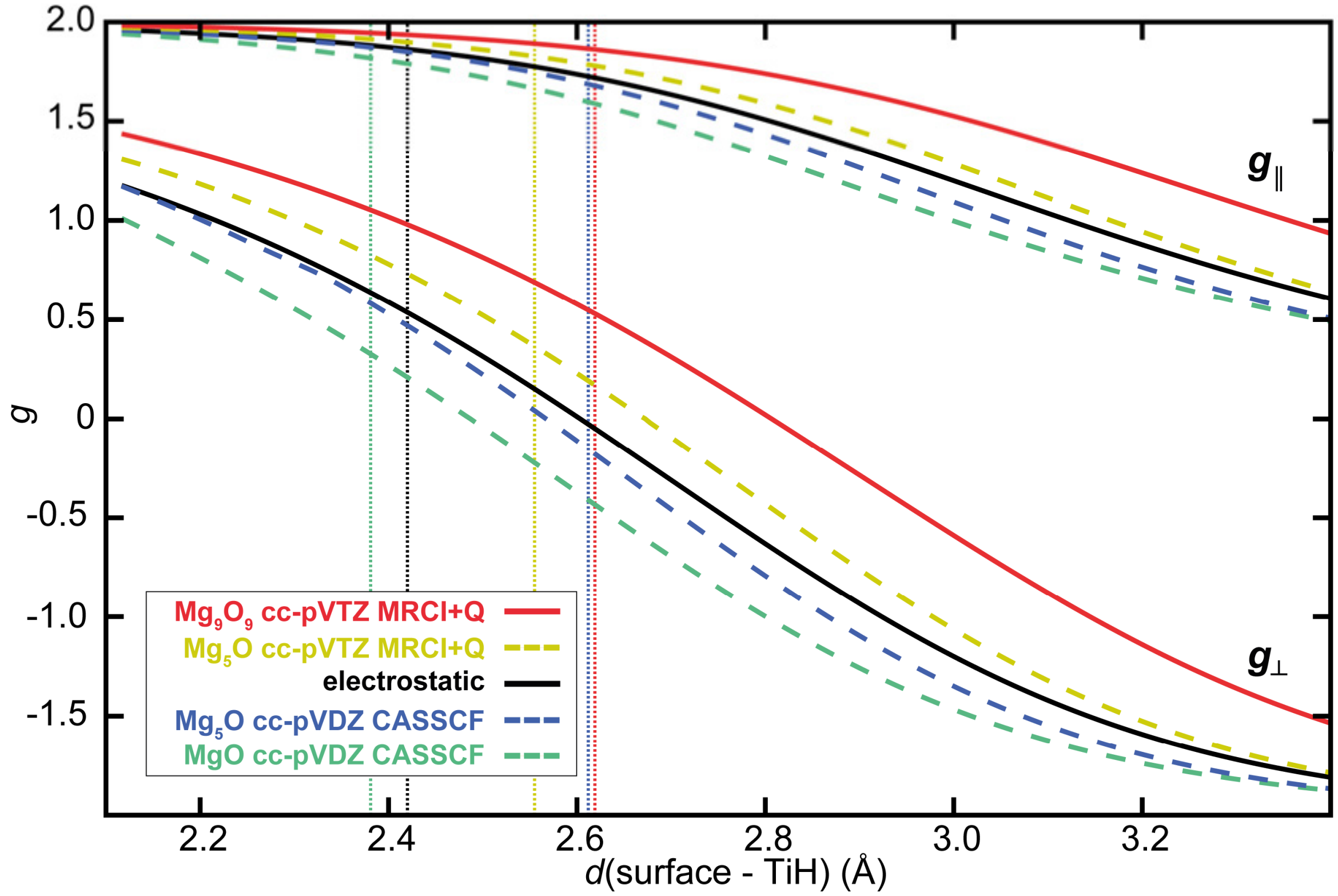

FIG. A3. $g$-factors of the $^2\Delta$ state as a function of the height above the surface in the electrostatic model and using various embedded cluster calculations. Dotted vertical lines denote the optimal height above the surface in that method.

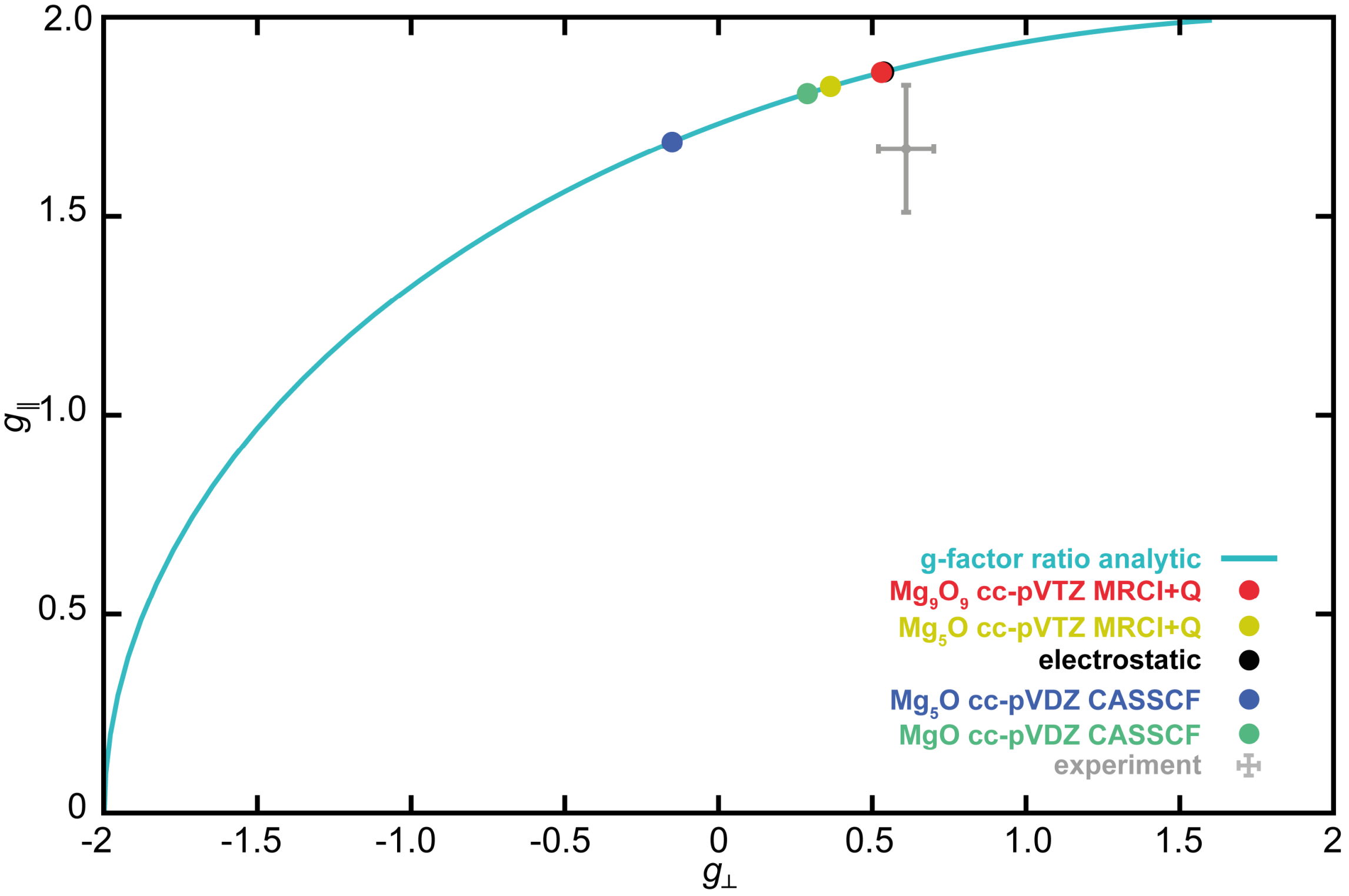

FIG. A4. Relation between $g_\parallel$ and $g_\perp$ for the $^2\Delta$ state. The curve represents the one-dimensional space in the theoretical model, parametrized by the ratio between the energy splitting and spin-orbit coupling. Results from the electrostatic model and embedded cluster calculations are represented by dots, together with experimental results given in grey with error bars. We note that the black and red dots nearly overlap.

| Cluster | | cc-pVDZ | | cc-pVTZ | |
|---|---|---|---|---|---|
| | | CASSCF | CASSCF | MRCI | MRCI+Q |
| MgO | $d$ (Å) | 2.39 | 2.40 | 2.37 | 2.36 |
| | $D_e$ (eV) | -3.99 | -3.16 | -3.72 | -3.84 |
| | $V_c$ (cm$^{-1}$) | 464.41 | 470.85 | 565.47 | 602.48 |
| | $g_\parallel$ | 1.81 | 1.81 | 1.87 | 1.88 |
| | $g_\perp$ | 0.29 | 0.31 | 0.55 | 0.63 |
| $Mg_5O$ | $d$ (Å) | 2.61 | 2.60 | 2.56 | 2.56 |
| | $D_e$ (eV) | -1.43 | -1.36 | -1.69 | -1.75 |
| | $V_c$ (cm$^{-1}$) | 344.59 | 374.57 | 462.25 | 490.40 |
| | $g_\parallel$ | 1.69 | 1.73 | 1.81 | 1.83 |
| | $g_\perp$ | -0.15 | 0.02 | 0.28 | 0.36 |
| $Mg_5O_5$ | $d$ (Å) | 2.60 | | | |
| | $D_e$ (eV) | -1.66 | | | |
| | $V_c$ (cm$^{-1}$) | 355.78 | | | |
| | $g_\parallel$ | 1.70 | | | |
| | $g_\perp$ | -0.10 | | | |
| $Mg_9O_9$ | $d$ (Å) | 2.59 | 2.58 | 2.61 | 2.62 |
| | $D_e$ (eV) | -1.68 | -1.60 | -1.00 | -0.93 |
| | $V_c$ (cm$^{-1}$) | 485.24 | 521.18 | 543.95 | 555.96 |
| | $g_\parallel$ | 1.82 | 1.85 | 1.86 | 1.86 |
| | $g_\perp$ | 0.35 | 0.45 | 0.50 | 0.53 |

TAB. A1. Convergence of embedded cluster calculations of $g$-factors of TiH on MgO. Here, $d$ denotes the height of the molecule above the surface, $D_e$ the dissociation energy and $V_c$ the energy splitting of the $^2\Delta$ states. $A_{SO} = 110.05$ cm$^{-1}$ was used here. The experimental values are $g_\parallel = 1.67 \pm 0.16$ and $g_\perp = 0.61 \pm 0.09$.

## SUPPLEMENTARY MATERIALS

Supplementary Text
Figures S1 to S22
Tables S1 to S5
References

## REFERENCES & NOTES

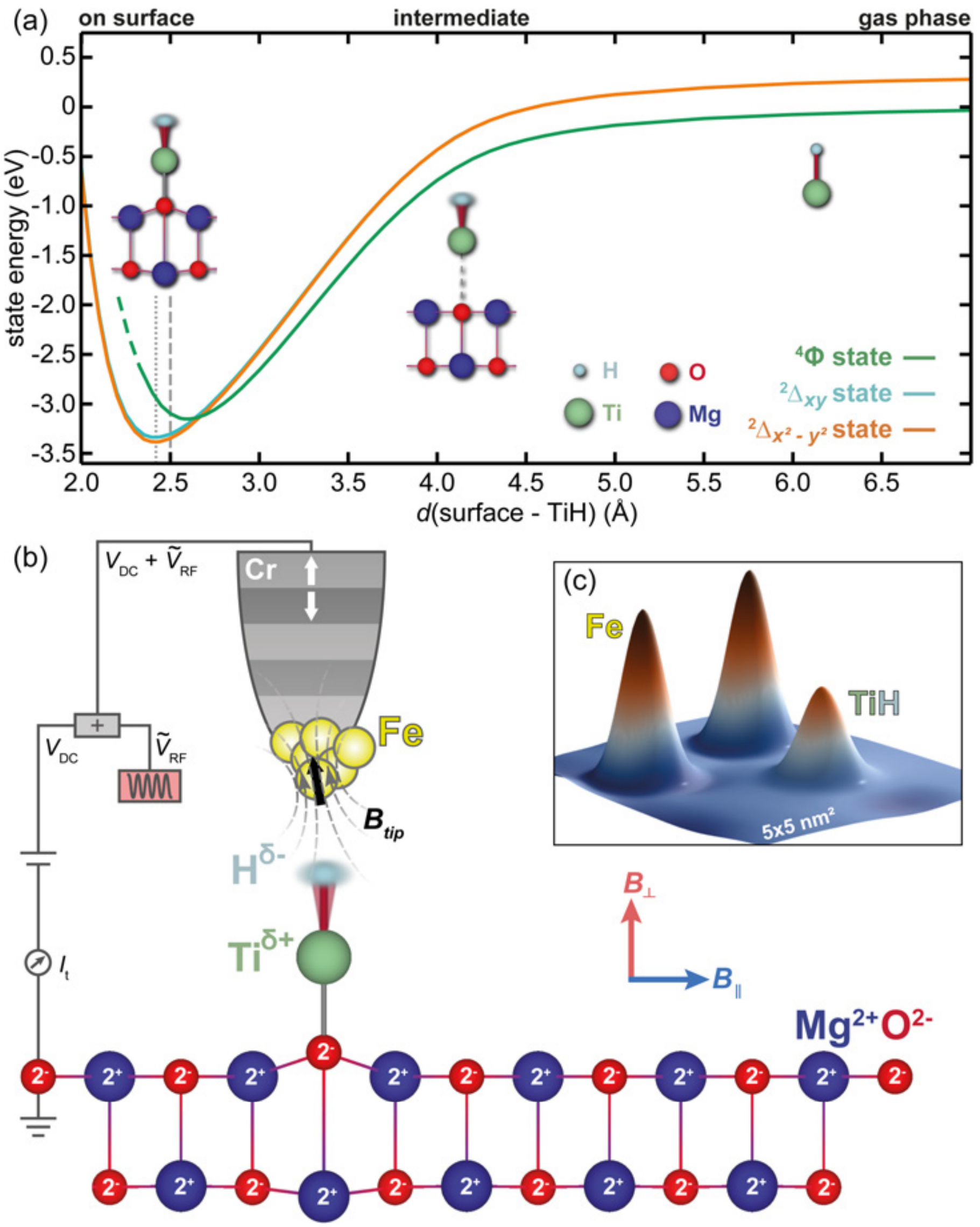

FIG. 1. Structure of the TiH molecule and influence of the surface. (a) Evolution of the $^2\Delta$ and $^4\Phi$ states of the TiH molecule as a function of distance from the surface of MgO, obtained from *ab initio* QC calculations. The electronic ground state of the system changes from the $^4\Phi$ to the $^2\Delta$ state for the adsorbed TiH. (b) Illustration of the TiH molecule adsorbed on the oxygen site of an ionic $Mg^{2+}O^{2-}$ surface. The $\tilde{V}_{RF}$ is applied on the magnetic Cr tip and is added to $V_{DC}$ as schematically sketched. (c) 3D representation of a constant-current STM image of a TiH molecule adsorbed next to two Fe atoms. Different apparent heights for Fe (151 ± 8 pm) and TiH (103 ± 8 pm) clearly distinguish both atom types. ($V_{DC}$ = 30 mV, $I_t$ = 10 pA).

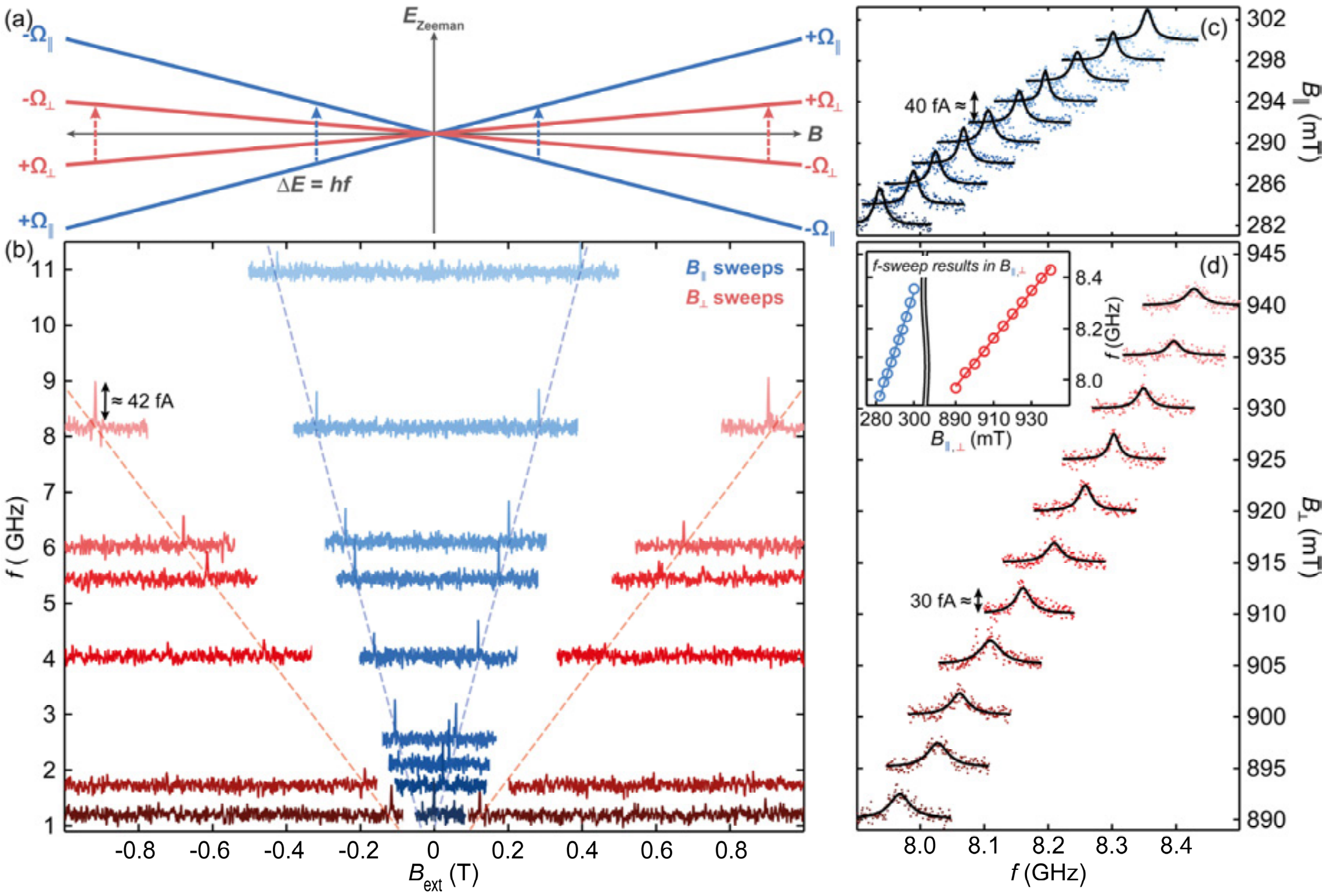

FIG. 2. mK ESR-STM of a TiH molecule in variable field orientations. (a) Sketch of the Zeeman-diagram for the level-splitting of the doublet state in different field orientations $B_{\parallel}$ (blue) and $B_{\perp}$ (red). Dashed arrows of the same lengths indicate the allowed $\Delta\Omega = \pm 1$ transitions for a specific $f$ (with $\Omega = \Lambda + \Sigma$). (b) $B$-sweep mode ESR measurements with the same micro-tip for two TiH molecules with the magnetic field swept in $\pm B_{\parallel}$ (blue) or $\pm B_{\perp}$ (red) direction. Peak positions are extracted from Lorentzian fits and subsequently fitted with a linear model (dashed lines). For the same selected frequencies, they appear at very different magnetic fields for the two directions, revealing an anisotropic $g$-tensor with $g_{\parallel} = 1.80 \pm 0.02$ ([25.2 ± 0.2] GHz/T) and $g_{\perp} = 0.63 \pm 0.01$ ([8.8 ± 0.1] GHz/T). ($V_{DC}$ = 50 mV, $I_t$ = 2 pA, $f_{chop}$ = 877 Hz, $V_{RF}$ = 7.9 mV). (c,d) $f$-sweep mode ESR measurements in $B_{\parallel}$ (c) and $B_{\perp}$ (d) direction with the same micro-tip and on the same TiH molecule as in the $B_{\parallel}$-sweep in (b). Solid lines represent Lorentzian fits to the experimental data. Linear fits to the extracted peak positions (inset in (d)) reveal $g_{\parallel} = 1.62 \pm 0.06$ ([22.7 ± 0.9] GHz/T) and $g_{\perp} = 0.66 \pm 0.02$ ([9.3 ± 0.3] GHz/T). ($V_{DC}$ = 50 mV, $I_t$ = 2 pA, $f_{chop}$ 877 Hz, $V_{RF}$ = 8.0 mV).

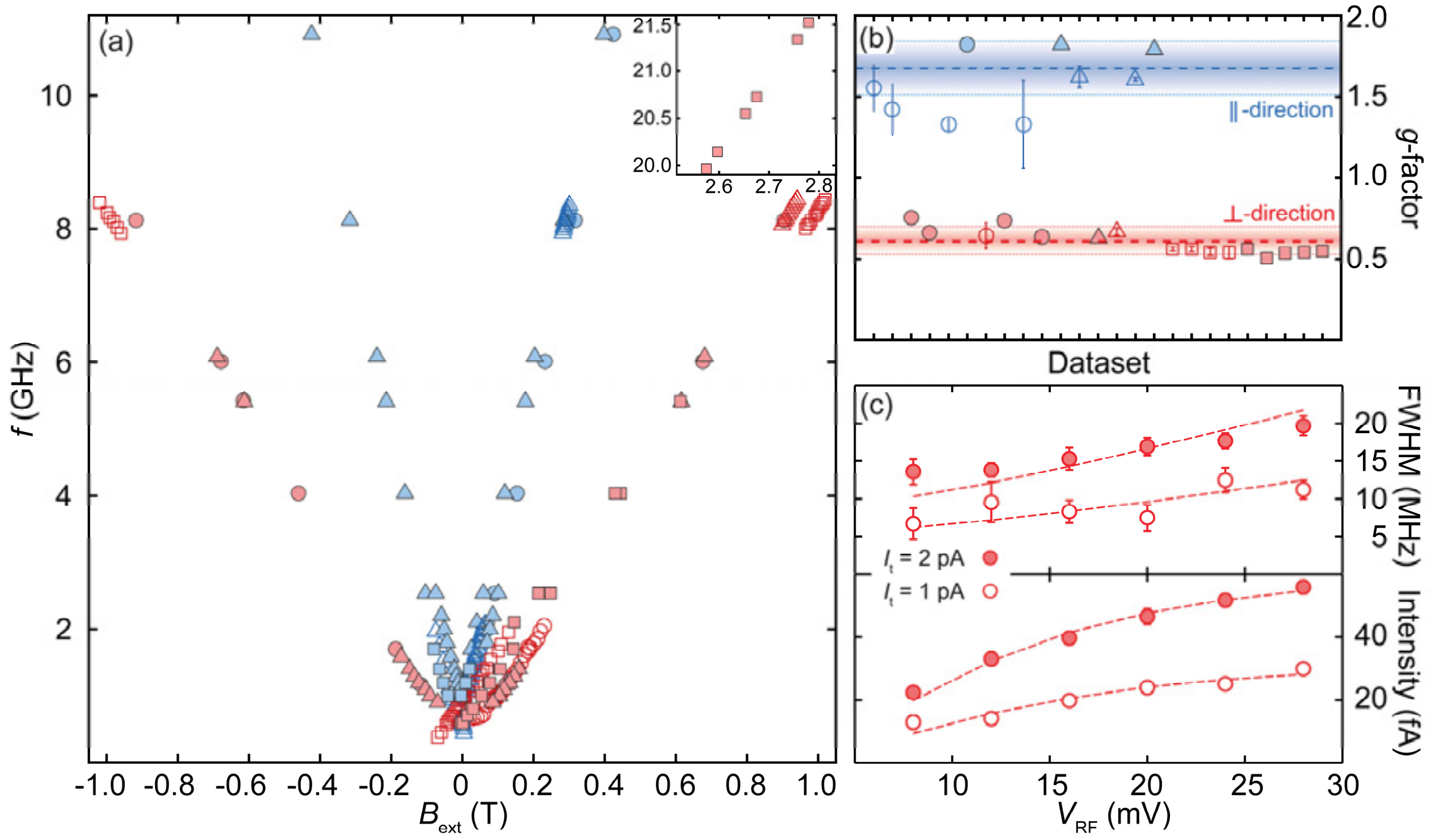

FIG. 3. Giant $g$-tensor anisotropy of TiH. (a) Extracted ESR peak positions from 21 data sets on different molecules (indicated by different symbols) measured in two $B$-field directions ⊥ (red) or ǁ (blue). Filled or open symbols correspond to $B$- or $f$-sweep mode, respectively. Measurement parameters throughout all experiments were the same ($V_{DC}$ = 50 mV, $I_t$ = 2 pA, $V_{RF}$ = 8 mV), except for the inset ($V_{DC}$ = 50 mV, $I_t$ = 10 pA, $\tilde{V}_{RF}$ = 4.466 V (uncalibrated $V_{RF}$)). (b) Experimental $g$-factors obtained from linear fits to the data in A as well as additionally included data sets with varied experimental parameters and tips (see Fig. S7 for a plot of all data). From 30 data sets in total we obtain $g_{\parallel}$ = 1.67 ± 0.16 ([23.4 ± 2.29] GHz/T) and $g_{\perp}$ = 0.61 ± 0.09 ([8.49 ± 1.19] GHz/T). (c) Plots of the FWHM (top) and ESR peak intensity (bottom) from $V_{RF}$ power-dependent measurements of a TiH molecule extracted from fitting Fano lineshapes to the experimental data. Dashed lines indicate simultaneous fits within the 1 pA and the 2 pA data set, respectively and reveal an asymptotic trend for the FWHM with $V_{RF}$.

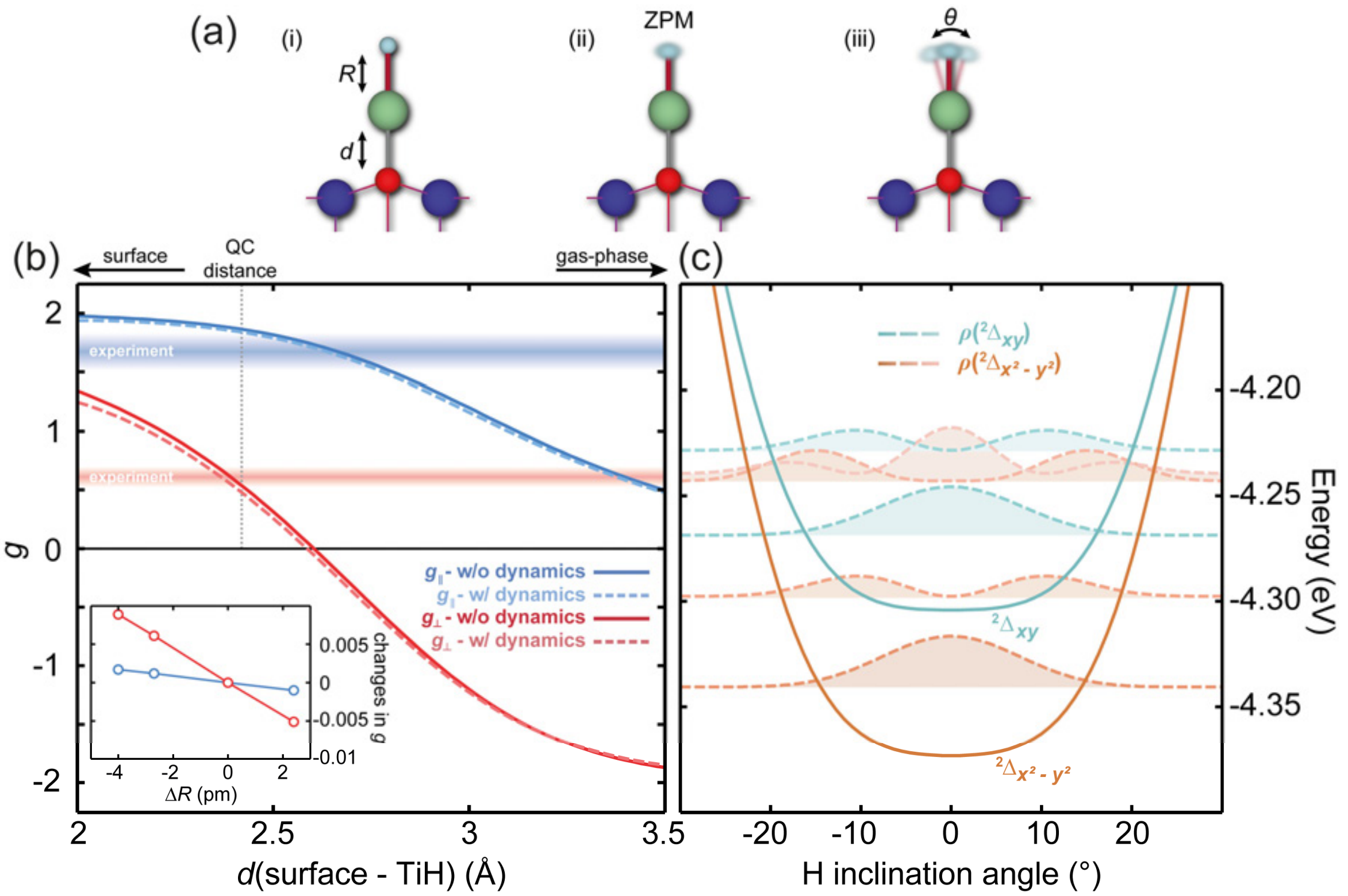

FIG. 4. Modeling the *g*-tensor, the molecular geometry and the hindered rotations. (a) Illustrations indicating the structural degrees of freedom of the TiH molecule: (i) bonding distance between Ti and O (*d*) or Ti and H (*R*), (ii) zero-point motion of H and (iii) excited rotational mode. (b) Calculated in-plane (blue) and out-of-plane (red) *g*-factors of the TiH without (solid lines) or with (dashed lines) rotational dynamics. The *g*-tensor is highly sensitive to the adsorption height of the molecule. At $d = 2.42$ Å, the minimum found in QC calculations (Fig. 1A), the experimentally observed anisotropic *g*-tensor can very well be reproduced. The inset shows the small variations in *g* for changes of *R*. (c) Calculated potential wells (solid lines) for both $^2\Delta$ orbital states, and densities $\rho$ (dashed lines) of the wave functions for the corresponding equidistant energy levels as a function of inclination angle $\theta$. Both are reminiscent of an anharmonic 2D quantum oscillator, where an amplitude of $\approx 15°$ can be deduced for the zero-point motion of the hydrogen.

# SUPPLEMENTARY MATERIALS

for

## Quantifying the interplay between fine structure and geometry of an individual molecule on a surface

Manuel Steinbrecher[1,†], Werner M. J. van Weerdenburg[1,†], Etienne F. Walraven[1,†], Niels P. E. van Mullekom[1], Jan W. Gerritsen[1], Fabian D. Natterer[2], Danis I. Badrtdinov[3,1], Alexander N. Rudenko[4,3,1], Vladimir V. Mazurenko[3], Mikhail I. Katsnelson[1,3], Ad van der Avoird[1], Gerrit C. Groenenboom[1], Alexander A. Khajetoorians[1,*]

[1] Institute for Molecules and Materials, Radboud University, 6525 AJ Nijmegen, The Netherlands

[2] Department of Physics, University of Zürich, Winterthurerstrasse 190, 8057 Zürich, Switzerland

[3] Theoretical Physics and Applied Mathematics Department, Ural Federal University, 620002 Ekaterinburg, Russia

[4] School of Physics and Technology, Wuhan University, Wuhan 430072, China

*Correspondence to: a.khajetoorians@science.ru.nl

†The authors contributed equally to this work.

**This PDF file includes:**

Supplementary Text with sections S1 to S9
Figs. S1 to S22
Tables S1 to S5
References

**TABLE OF CONTENTS**

## SUPPLEMENTARY TEXT

### S1 – STM/STS sample characterization

We characterized the properties of Fe and TiH, as well as the MgO film with STM/STS. A typical constant-current image is shown in Fig. S1(a). MgO islands on top of Ag(100) can be identified by a lower apparent height compared to the bare Ag surface. Fe atoms and TiH molecules adsorbed on the oxygen sites were distinguished by their different apparent heights. A histogram of numerous measurements for both species is shown in Fig. S1(b) resulting in average heights of 103 ± 8 pm for TiH and 151 ± 8 pm for Fe, in agreement with previously reports [37,44].

We identified the adsorption sites by imaging the oxygen atoms of the MgO lattice and creating a reference grid (Fig. S2). Utilizing this grid (white lines), we ascertained that both TiH and Fe adsorbates that exhibit the aforementioned apparent heights resided on top of an oxygen atom. Additionally, TiH can be identified by two particular spectroscopic fingerprints, namely the orbital excitation observed at ≈ ±90 mV (Fig. S3), as well as the spin excitation (see section S6). The position of the orbital excitation varied by several mV dependent on the tip, stabilization parameters or investigated molecule and was not observed for bridge-site TiH molecules.

We identified the thickness of the MgO film by using point-contact measurements, as previously reported [44]. Fig. S4 shows the results measured on two different Fe atoms adsorbed on two and three monolayers (ML) of MgO. Point contact was defined as the *z*-piezo position with the highest measured current and was set to be 0 pm in the plot. For 2 ML films, we measured $I_t$ = 11.1 nA at point-contact, corresponding to a conductance of $G$ = 0.143 $G_0$. This is in agreement with previous work [44]. For Fe adsorbed on 3 ML films, we measured $I_t$ = 1.78 nA. This corresponded to $G$ = 0.023 $G_0$, manifesting the expected strong reduction of conductance for thicker layers of MgO. All experiments reported in the main manuscript were performed on 2 ML films.

### S2 – RF transmission measurement and compensation

Prior to all *f*-sweep mode measurements, we corrected for the frequency-dependent variations in transmission due to the transfer function [45]. Fig. S5 illustrates an example of the frequency-dependent transmission in a wide range of 1 – 22 GHz as measured on the non-linearity at $V_{DC}$ = -78 meV of a TiH spectrum with constant $\tilde{P}_{RF}$. To compensate for this frequency-dependent transmission and achieve a constant $V_{RF}$, $\tilde{P}_{RF}$ was adjusted accordingly via the generator. Fig. S6 shows an example of compensation in a frequency range of 7.9 – 8.5 GHz. The blue/red curve shows $V_{RF}$ for a constant/adjusted $\tilde{P}_{RF}$ before/after compensation.

### S3 – Additional ESR raw data

In this section, we present ESR raw data sets supplementary to Fig. 2 of the main manuscript and that were later used in the analysis in Fig. 3 and supplementary section S4. The data set presented in Fig. S7 was taken with the same micro tip and on the same molecule as the data shown in the $B_{\parallel}$-sweep as well as both *f*-sweeps of Fig. 2. Further data sets on another TiH molecule are presented in Fig. S8 and S9 for the *B*-sweep and *f*-sweep mode, respectively.

### S4 – ESR of TiH molecules for different stabilization parameters

All the 31 data sets used for the *g*-tensor analysis in Fig. 3(b) are plotted in Fig. S10. We found that different tip-sample distances did not significantly affect the extracted *g*- factors, although this did result in different horizontal and vertical offsets due to variable influence of the tip stray field. The corresponding measurement parameters for all data sets are indicated in the figure legend, with $V_{DC}$ = 50 mV for all sets. To analyze the $g$-tensor for these data sets, we excluded non-linear data points in the low frequency region around zero-field. For this, we first fitted the data with a hyperbolic function to mimic the stray field effect that was subsequently modeled thoroughly in section S4:

$$f(x) = \frac{b}{a}\sqrt{(x-c)^2 + a^2} + d,$$

where $c$ is the center of the hyperbola, $d$ a vertical offset and $b/a$ denotes the slope $k_A$ of the corresponding linear asymptotes. Via the minima and maxima of the third derivative, we found $x_{min,max} = c \pm \frac{1}{2} a$. Using a scaling factor $S$ we moved from these positions along the $x$-axis to $x_{1,2} = c \pm \frac{1}{2} aS$. We related the algebraic expression for the hyperbola's slope $k_H$ at $x_{1,2}$ as a function of $S$ with respect to $k_A$:

$$k_H = \frac{S}{\sqrt{4+S^2}} \cdot k_A = F \cdot k_A.$$

In our analysis, we considered the slope $k_H$ to be within 95 % of $k_A$ ($F$ = 0.95) and neglect experimental data not fulfilling this criterion.

## S5 – Modeling the tip stray field

Probing low frequency bands requires the application of only small magnetic field values, while maintaining a magnetically stable tip. It has been shown previously, that Cr bulk tips retain a stable magnetization from zero to several Tesla [46]. Furthermore, despite the antiferromagnetic layering of the Cr, such magnetic tips also exhibit magnetic stray fields on the order of 50-200 mT, which is a common observation in the literature [21,47-50]. As the value of the externally applied magnetic field $B_{ext}$ is reduced, it can become comparable to this stray field $B_{tip}$, making it a non-negligible component of the total field $B_{tot}$. We constructed a classical model that includes both $B_{ext}$ and $B_{tip}$, in order to calculate the expected Zeeman splitting for a classical TiH spin. We note that variations of the quantization axis in the low-field limit is not captured by this treatment. Furthermore, temperature-related effects were neglected. The externally applied magnetic field, $\vec{B}_{ext}$, is oriented in the $yz$-plane:

$$\vec{B}_{ext} = \begin{pmatrix} 0 \\ B_{y,ext} \\ B_{z,ext} \end{pmatrix}.$$

Here, we define $B_{y,ext} = B_{\parallel}$ and $B_{z,ext} = B_{\perp}$. During the experiment, one of the components ($B_{y,ext}$ or $B_{z,ext}$) was varied while the other component is held at a constant value . We consider an arbitrary $\vec{B}_{tip}$ toward a total applied magnetic field $\vec{B}_{tot}$:

$$\vec{B}_{tip} = \begin{pmatrix} B_{x,tip} \\ B_{y,tip} \\ B_{z,tip} \end{pmatrix},$$

$$\vec{B}_{tot} = \vec{B}_{ext} + \vec{B}_{tip} = \begin{pmatrix} B_{x,tip} \\ B_{y,ext} + B_{y,tip} \\ B_{z,ext} + B_{z,tip} \end{pmatrix}.$$

We consider the spin of the TiH molecule as $|\vec{S}| = 1/2$ with its orientation parallel to the total magnetic field, $\vec{B}_{\text{tot}} \parallel \vec{S}$, hence:

$$\vec{S} = \begin{pmatrix} S_x \\ S_y \\ S_z \end{pmatrix} = \begin{pmatrix} \frac{B_{x,\text{tot}}}{|\vec{B}_{\text{tot}}|} * |\vec{S}| \\ \frac{B_{y,\text{tot}}}{|\vec{B}_{\text{tot}}|} * |\vec{S}| \\ \frac{B_{z,\text{tot}}}{|\vec{B}_{\text{tot}}|} * |\vec{S}| \end{pmatrix}.$$

Lastly, we define a *g*-tensor that allows for anisotropy in axial symmetry ($g_x = g_y \neq g_z$) by

$$\overline{\overline{g}} = \begin{pmatrix} g_x & 0 & 0 \\ 0 & g_y & 0 \\ 0 & 0 & g_z \end{pmatrix} = \begin{pmatrix} g_\parallel & 0 & 0 \\ 0 & g_\parallel & 0 \\ 0 & 0 & g_\perp \end{pmatrix}.$$

With these components, the Hamiltonian for the Zeeman interaction can be described as:

$$\mathcal{H}_{\text{Zeeman}} = \mu_{\text{B}} \vec{B}_{\text{tot}} \cdot (\overline{\overline{g}} \cdot \vec{S})$$

$$= \mu_{\text{B}} |\vec{S}| \, \frac{g_\parallel \left(B_{x,\text{tot}}{}^2 + B_{y,\text{tot}}{}^2\right) + g_\perp \left(B_{z,\text{tot}}{}^2\right)}{|\vec{B}_{\text{tot}}|} \tag{1}$$

$$= \mu_{\text{B}} |\vec{S}| \, \frac{g_\parallel \left( \left(B_{x,\text{tip}}\right)^2 + \left(B_{y,\text{ext}} + B_{y,\text{tip}}\right)^2 \right) + g_\perp \left( \left(B_{z,\text{ext}} + B_{z,\text{tip}}\right)^2 \right)}{\sqrt{\left(B_{x,\text{tip}}\right)^2 + \left(B_{y,\text{ext}} + B_{y,\text{tip}}\right)^2 + \left(B_{z,\text{ext}} + B_{z,\text{tip}}\right)^2}}. \tag{2}$$

Fig. S11(a) depicts how $\vec{B}_{\text{tot}}$ varies as $\vec{B}_{\text{ext}}$ is swept in one direction, and how the spin $\vec{S}$ is reoriented as a result. Note, that only the $\vec{B}_{\text{tip}}$ component parallel to $\vec{B}_{\text{ext}}$ can be compensated, while the remaining part provides a constant offset field. This latter point is what leads to non-linearity in the data in the low field limit. In Fig. S11(b), two example data sets with strong plateau-like deviations are shown. The measurement conditions were identical, except for a difference in $\vec{B}_{\text{tip}}$ for the two different measurements. The observed difference in this non-linear behavior for the same atom measured with distinct micro tips rules out that the non-linear behavior results from intrinsic effects from the atom, such as the crystal field. This was further substantiated as we observed measurements with a linear trend down to smaller frequencies for other tips. Modeling the data sets with equation (2) (solid lines in Fig. S11(b), fit performed within the global fit of Fig. S12), we found tip magnetic fields (see figure legend) with an expected magnitude, as discussed within this section.

In Fig. S12, we fitted all data sets of Fig. S10 simultaneously with equation (2) by using $g_\parallel$ and $g_\perp$ as global fitting parameters, and allowing for a unique $\vec{B}_{\text{tip}}$ for every data set. Two additional data sets with strong plateau-like behavior of the peak shifts were included compared to section S3. We found $g_\parallel = 1.896$ and $g_\perp = 0.638$. Note, that the perfect agreement with a S=1/2 system and keep in mind that the fitting procedure does not capture variations in *g*-factors.

The average magnitude $|\vec{B}_{\text{tip}}|$ for $I_t$ = 2 pA is ≈ 44 mT, in line with expectations for stray fields of Cr bulk tips of around 50 mT [21,47]. As expected, by reducing the tip-sample distance, i.e. for a larger value of $I_t$ the extracted $\vec{B}_{\text{tip}}$ values increased (≈ 123 mT for $I_t$ = 5 pA, ≈ 229 mT for $I_t$ = 10 pA). The average angle of $\vec{B}_{\text{tip}}$ with respect to the *z*-axis is ≈ 54° (with decreasing angle for larger $I_t$).

Furthermore, via simple arguments one can rule out that the observed non-linearities are a result of an unsaturated tip magnetization. Firstly, we observe these non-linearities for different tips, even though they were all prepared with Fe clusters at the apex. Secondly, for a magnetic moment of $3\mu_B$ (assuming Fe with a spin of 2 at the apex), a paramagnetic tip at 40 mK should saturate below 25 mT. Since we observe the non-linearities typically at even larger fields and the stray fields we find are on the order of what is reported in literature (see above), the stray field assumption is the most likely explanation in the experimental observations.

## S6 – ESR of a TiH molecule at elevated temperature

As shown in Fig. S3 and below in section S7, we see no signature of a Kondo-like resonance in STS. To further rule out Kondo-related effects, we performed ESR experiments at 1.1 K in *f*-sweep mode in Fig. S13. This allows a comparison of the measured *g*-tensor at a temperature where previous experiments were performed. We observed the renormalized *g*-factors of $g_{\parallel}$ = 1.57 ± 0.04 and $g_{\perp}$ = 0.55 ± 0.01, comparable to what was observed at 30-50 mK. This further rules out that the anisotropy in the *g*-tensor results from Kondo screening at mK temperatures, since the magnetization should strongly change as a function of temperature near $T_K$.

## S7 – ISTS of TiH molecules and Fe atoms

We investigated ISTS of TiH and Fe with inelastic scanning tunneling spectroscopy (ISTS), to complement the observation of the *g*-tensor anisotropy. Magnetic field dependent ISTS measurements, with a non-magnetic tip, in a very small energy window of ±1.5 meV on a TiH molecule are plotted in Fig. S14(a,b). There is no zero-field splitting (ZFS) observed in the case of zero magnetic field, further corroborating that the TiH molecule on the surface of MgO resides in a spin doublet ground state as previously reported [22].

We observed an inelastic spin excitation in variable magnetic field, for both $B_{\parallel}$ (Fig. S14(a)) and $B_{\perp}$ (Fig. S14(b)). For both orientations, the inelastic step shifts toward higher energy in increasing field, as expected for a spin doublet. The magnitude of the shifts differed for the two applied magnetic field directions, with the change in magnetic field being smaller for $B_{\perp}$. The ISTS measurements were done on four different TiH molecules and the extracted averaged step position is plotted in Fig. S14(c). We obtained $g_{\parallel}$ = 1.84 ± 0.01 and $g_{\perp}$ = 0.50 ± 0.01, confirming the anisotropic *g*-tensor found in our ESR-STM measurements. We note that in STS, there was no $V_{RF}$ applied. This illustrates the anisotropic *g*-tensor results from an intrinsic property of the TiH molecule, and not from the ESR method.

In order to calibrate the applied magnetic field, we also performed magnetic field-dependent ISTS measurements on individual Fe atoms. Two Fe atoms were investigated as shown in the inset of Fig. S14(c). A spin-excitation step at zero field was observed at ≈ 14.4 meV, in agreement with literature [15,51]. As previously reported [51], the inelastic step energy increases linearly with $B_{\perp}$. We extracted an effective $g$-factor of $g^*_{\perp} = 2.48 \pm 0.18$. When the ISTS measurements were done with a spin-polarized tip, the inelastic spin excitation steps of Fe showed spin pumping features [51] that were used to confirm that the tip was spin polarized. Additionally, the emerging asymmetry of the spin excitation steps around $E_F$ of the TiH molecules (Fig. S3) was also used for this purpose [22].

## S8 – DFT+*U* calculations

We calculated the electronic and crystal structure of the TiH molecule on the MgO surface using density-functional theory (DFT) methods. We considered a 3×3 unit cell of bilayer MgO with periodic boundary conditions and 18 Å vacuum space along the $z$ direction. The TiH molecule was placed on top of the central oxygen atom, as depicted in Fig. S15(a). The DFT calculations were performed within generalized gradient approximation (GGA-PBE) of exchange-correlation functionals [52] as implemented in the projector augmented wave based Vienna ab initio simulation package (VASP) [53]. In these calculations, we set the energy cutoff to 500 eV and the energy convergence criteria to $10^{-6}$ eV. For the Brillouin zone integration, a 8x8x1 Γ-centered Monkhorst-Pack mesh was used. Electronic correlations were taken into account in a mean field way using the DFT+$U$ method in rotationally invariant form [54]. In this approach, the intra-atomic exchange interaction $J_H$ was set to 0.9 eV, while $U$ was varied within the calculations. All atoms of the constructed unit cell were allowed to relax until all the residual force components of each atom were less than $5 \cdot 10^{-3}$ eV/Å. In agreement with ref. [31], the resulting unit cell oxygen atom was distorted upwards to the TiH molecule, while the bottom Mg atom relaxed downwards due to weakening of the Mg-O bond. For DFT, without the $U$ correction, we found the optimized distances: $d$(Ti-H) = 1.79 Å, $d$(Ti-O) = 1.93 Å and $d$(O-MgO) = 0.46 Å. Small changes in bond distances were observed with DFT+$U$, e.g. for $U$ = 4 eV: $d$(Ti-H) = 1.80 Å, $d$(Ti-O) = 1.99 Å and $d$(O-MgO) = 0.49 Å.

The MgO substrate with $C_{4v}$ point group symmetry causes a crystal field splitting of the Ti($d$) states, which can be seen in the DFT ($U$ = 0) band structure and densities of states in Fig. S15(b,c). Since the titanium $d_{z^2}$ orbital is oriented towards the hydrogen atom, it leads to a strong hybridization of the Ti($d_{z^2}$) and H($s$) states, transferring a valence electron from titanium to H($s$), filling the $s$ shell. From the band structure point of view, Ti($d_{z^2}$) has a finite bandwidth due to the hybridization, while the rest of the titanium states appear as flat energy levels. Another valence electron occupies the orbital of $d_{x^2-y^2}$ symmetry, which is oriented in the direction of positively charged magnesium atoms, giving an unpaired spin $S = ½$ in the system. The obtained energy spectra were in good agreement with earlier calculations [22,31].

The orbital contribution to energy bands near the Fermi energy is represented in Table S1. Due to the geometry of orbitals, the hydrogen $s$ state is hybridized with the titanium $d_{z^2}$ and bottom oxygen $p_z$ states. At the same time, doubly degenerated energy bands are mainly composed of titanium $d_{xz}$ and $d_{yz}$ states and bottom oxygen states with different symmetry combinations. Finally, bands close to the Fermi energy are composed of purely atomic titanium $d_{xy}$ and $d_{x^2-y^2}$ states.

The projected densities of states for different values of $U$ are represented in Fig. S16. The spin up channel of $d_{x^2-y^2}$ states is fully occupied, while the spin down one is unoccupied, yielding S = ½. The variation of the $U$ parameter leads to a stronger splitting of spin up and down channels of the $d_{x^2-y^2}$ states and shifts the titanium $d$ states to higher energies. The hydrogen $s$ state was not influenced by this variation.

Within these DFT calculations, we computed the orbital moment in the presence of spin-orbit coupling. However, the resulting orbital moment on the titanium atom was found to be very small and only amounts to ≈ 0.05 $\mu_B$ and nearly independent of the $U$ parameter (Table S2). The calculations also predict that the orbital moment is oppositely oriented to the spin moment. Thus, the DFT+$U$ calculations predict that the orbital moment is quenched in this system, which contradicts the experimental and QC observations.

Finally, we calculated the potential energy landscape of the TiH molecule by rotating the hydrogen atom around the Ti at a fixed Ti-H distance. We found that this landscape depended strongly on the choice of $U$, with multiple solutions. This reveals a strong dependence on the on-site Coulomb interaction parameter. For large enough $U$, the energy minimum corresponds to the situation, where hydrogen sits on the top of the titanium atom. A smaller $U$ changes the favorable position of H, which is then tilted off the $z$-axis towards the MgO surface. The $C_{4v}$ symmetry of the substrate then allows four energy minimum positions, forming a multi-well potential as shown in Fig. S17. However, this requires a significant lowering of the $U$ parameter (down to $U$ = 3 eV) than is commonly used for titanium systems ($U \geq$ 5eV).

## S9 – Quantum chemistry modelling

To compute the $g$-factors of TiH on the MgO surface in the presence of a magnetic field, we use two models: a static model, in which we assume the center of mass of the TiH molecule as well as the orientation of the internuclear axis to be fixed with respect to the crystal surface and the magnetic field, and a dynamic model, in which we still keep the center of mass of the molecule fixed, but include the (hindered) rotation of the internuclear axis quantum mechanically.

The static model is described in section S9.1. It requires an interaction potential between the surface and the molecule, which is described in section S9.1.3. The dynamic model is described in section S9.2. The models are approximate, but have no empirical parameters. Instead, the parameters are obtained from *ab initio* quantum chemistry calculations described in section S9.3 (see also Fig. 1 of the main paper). The results are given in section S9.4 and in Fig. 4 of the main paper.

### S9.1 The static model

The $^2\Delta$ electronic state of TiH can be described by Hund's case (a) wave functions $|\Lambda S \Sigma\rangle$ [55], where the electron spin quantum number $S = 1/2$, the projection of the orbital angular momentum onto the internuclear axis (the $z$-axis) is $\Lambda\hbar = \pm 2\hbar$, and the projection of the electron spin angular momentum is $\Sigma\hbar = \pm 1/2\ \hbar$. In the static model, the electronic wave function is a linear combination of these four functions:

$$\Psi = \sum_{\Lambda=\pm 2} \sum_{\Sigma=\pm 1/2} |\Lambda S \Sigma\rangle\, c_{\Lambda\Sigma}. \tag{1}$$

**S9.1.1 Zeeman Hamiltonian**

The interaction with the magnetic field is described by the Zeeman Hamiltonian

$$\hat{H}_{\mathrm{Z}} = -\hat{\boldsymbol{\mu}} \cdot \boldsymbol{B} = \frac{\mu_B}{\hbar}(\hat{\boldsymbol{L}} + g_e \hat{\boldsymbol{S}}) \cdot \boldsymbol{B}, \tag{2}$$

where $g_e \approx 2.0023$ is the electron spin $g$-factor, $\mu_B = \frac{e\hbar}{2m_e}$ is the Bohr magneton, $\hat{\boldsymbol{\mu}}$ the magnetic moment of the molecule, $\boldsymbol{B}$ the magnetic field, and $e$ and $m_e$ are the electron charge and mass, respectively. In atomic units we have $e = m_e = \hbar = 1$ and $\mu_B = 1/2$. Furthermore, a magnetic field of 1 tesla has a strength of $4.254 \cdot 10^{-6}$ in atomic units. Only the $z$-component $\hat{L}_z$ of the orbital angular momentum operator has nonzero matrix elements and it is diagonal in the case (a) basis:

$$\langle \Lambda' S \Sigma' |\, \hat{L}_z\, | \Lambda S \Sigma \rangle = \delta_{\Lambda'\Lambda} \delta_{\Sigma'\Sigma}\, \Lambda\hbar. \tag{3}$$

The $z$-component of the spin operator has diagonal elements

$$\langle \Lambda' S \Sigma' |\, \hat{S}_z\, | \Lambda S \Sigma \rangle = \delta_{\Lambda'\Lambda} \delta_{\Sigma'\Sigma}\, \Sigma\hbar. \tag{4}$$

The spin-operator, however, also has off-diagonal elements. They are most easily expressed for the ladder operators $\hat{S}_\pm = \hat{S}_x \pm i\hat{S}_y$ by

$$\langle \Lambda' S \Sigma' |\, \hat{S}_\pm\, | \Lambda S \Sigma \rangle = \delta_{\Lambda'\Lambda} \delta_{\Sigma',\Sigma\pm 1} \sqrt{S(S+1) - \Sigma(\Sigma \pm 1)}\hbar = \delta_{\Lambda'\Lambda} \delta_{\Sigma',\Sigma\pm 1}\hbar, \tag{5}$$

for $S = ½$. To compute matrix elements of the Zeeman Hamiltonian, we express the scalar product using spherical components of the vectors

$$\hat{H}_Z = -\hat{\boldsymbol{\mu}} \cdot \hat{\boldsymbol{B}} = -\sum_{m=-1}^{1} (-1)^m \hat{\mu}_m B_{-m}. \tag{6}$$

The spherical components $\hat{T}_{-1}$, $\hat{T}_0$ and $\hat{T}_1$ of a vector operator (such as $\hat{\boldsymbol{\mu}}$) are related to the Cartesian components $\hat{T}_x$, $\hat{T}_y$, and $\hat{T}_z$ through

$$\hat{T}_0 = \hat{T}_z$$
$$\hat{T}_{\pm 1} = \mp \frac{\hat{T}_x \pm i\hat{T}_y}{\sqrt{2}}. \tag{7}$$

The same transformation can be used for vectors, such as $\boldsymbol{B}$. The spherical components of the magnetic moment operator are given by

$$\hat{\mu}_0 = -\frac{\mu_B}{\hbar}(\hat{L}_z + g_e \hat{S}_z)$$
$$\hat{\mu}_{\pm 1} = \mp \frac{\mu_B g_e}{\hbar\sqrt{2}} \hat{S}_\pm\,. \tag{8}$$

We consider two additional terms in the Hamiltonian: the spin-orbit coupling and the electrostatic interaction between the $\Lambda = 2$ and $\Lambda = -2$ components due to the crystal field.

**S9.1.2 Spin-orbit Hamiltonian**

The spin-orbit Hamiltonian is given by

$$\hat{H}_{\mathrm{SO}} = A \hat{\boldsymbol{L}} \cdot \hat{\boldsymbol{S}}. \tag{9}$$

In our *ab initio* calculations described below, we find a spin-orbit coupling constant $A = 110.05$ cm$^{-1}$. In the case (a) basis, this operator has only diagonal elements

$$\langle \Lambda' S \Sigma' | \hat{H}_{\rm SO} | \Lambda S \Sigma \rangle = \delta_{\Lambda' \Lambda} \delta_{\Sigma' \Sigma} A \Lambda \Sigma \ . \tag{10}$$

Before we include the crystal field and present the full calculation, it is instructive to consider four special cases:

(i) If we only consider the $S = 1/2$ electron spin, i.e., for a $^2\Sigma$ state, the Zeeman Hamiltonian has eigenvalues $\pm \mu_B g_e B/2$, which are independent of the angle between the internuclear axis and the direction of the magnetic field. The $g$-factor is reported as the energy of the upper level minus the energy of the lower level, divided by the Bohr magneton and the magnetic field $B$, i.e., it is $g_e (\approx 2.0023)$.

(ii) If we only consider the orbital angular momentum, i.e., for a $^1\Delta$ state, the $g$-factor, which is non-negative by definition, is $2|\Lambda| = 4$ when the magnetic field is parallel to the internuclear axis, and it is zero when they are perpendicular.

(iii) When the magnetic field is parallel to the internuclear axis, and we ignore the crystal field, the case (a) wave functions are eigenfunctions of the Hamiltonian consisting of the Zeeman interaction and the spin-orbit coupling, and the four eigenvalues are

$$\langle \Lambda S \Sigma | \hat{H}_{\rm SO} + \hat{H}_{\rm Z} | \Lambda S \Sigma \rangle = A \Lambda \Sigma + \mu_B (\Lambda + g_e \Sigma) B. \tag{11}$$

We assume the observed transition involves the lowest two levels. Since the Zeeman interaction is much weaker than the spin-orbit coupling, the lowest two energy levels have $A\Lambda\Sigma < 0$, and since $A > 0$ we also have $\Lambda\Sigma < 0$. This means that the magnetic moments due to orbital angular momentum and spin angular momentum have opposite signs. Since $2|\Lambda| > g_e$, the $g$-factor is $2|\Lambda| - g_e \approx 1.9977$.
Thus, for a $^2\Delta$ state with a positive spin-orbit coupling constant, with the magnetic field parallel to the internuclear axis and no crystal field interaction, we expect a $g$-factor of about $2$, just as for a $^2\Sigma$ state.

(iv) The last limiting case we consider is where the spin-orbit coupling is negligible and the crystal field is strong. When the crystal field is cylindrically symmetric around the internuclear axis it will have no effect on the mixing of the case (a) functions and the $g$-factor, but when it has a Fourier component $\cos(4\phi)$ or $\sin(4\phi)$, where $\phi$ is the azimuthal angle, it will mix the $\Lambda = 2$ and $\Lambda = -2$ functions. When the mixing is 50/50, the orbital angular momentum will be completely quenched, i.e., the matrix elements of $\hat{L}_z$ (and also the other components), will be zero. In this case, the $g$-factor will again be equal to $g_e$, due to the electron spin.

Thus, perhaps somewhat surprisingly, we expect a $g$-factor of about two for the $^2\Delta$ state when *either* the orbital angular momentum is quenched by the crystal field, *or* when instead the spin-orbit interaction dominates, and the magnetic field is parallel to the internuclear axis. However, in intermediate coupling cases and when the magnetic field is not parallel to the axis, which we consider below, we expect $g$-factors between 0 and 2.

### S9.1.3 Crystal Field

We adopt a phase convention for the $|\Lambda\rangle$ states with $\Lambda = \pm 2$ that matches the symmetry of one-electron functions with $e^{i\Lambda\varphi}$ dependence on the azimuthal angle $\varphi$ of the electron. Thus, we define real wave functions by

$$\Delta_{x^2-y^2} = \frac{|\Lambda = 2\rangle + |\Lambda = -2\rangle}{\sqrt{2}} \tag{12}$$

$$\Delta_{xy} = \frac{|\Lambda = 2\rangle - |\Lambda = -2\rangle}{i\sqrt{2}} \tag{13}$$

i.e.,

$$|\Lambda = \pm 2\rangle = \frac{\Delta_{x^2-y^2} \pm i\Delta_{xy}}{\sqrt{2}}. \tag{14}$$

Thus, the matrix elements of the surface potential operator $\hat{V}_\mathrm{s}$ in the complex basis are related to the real matrix elements as

$$\langle \Lambda = \pm 2| \hat{V}_\mathrm{s} |\Lambda = \pm 2\rangle = \frac{1}{2}\left[\langle \Delta_{x^2-y^2}| \hat{V}_\mathrm{s} |\Delta_{x^2-y^2}\rangle + \langle \Delta_{xy}| \hat{V}_\mathrm{s} |\Delta_{xy}\rangle\right] \tag{15}$$

$$\langle \Lambda = \pm 2| \hat{V}_\mathrm{s} |\Lambda = \mp 2\rangle = \frac{1}{2}\left[\langle \Delta_{x^2-y^2}| \hat{V}_\mathrm{s} |\Delta_{x^2-y^2}\rangle - \langle \Delta_{xy}| \hat{V}_\mathrm{s} |\Delta_{xy}\rangle\right] \mp i\,\langle \Delta_{x^2-y^2}| \hat{V}_\mathrm{s} |\Delta_{xy}\rangle\,. \tag{16}$$

A direct calculation of the lowest two electronic states of the TiH molecule near the surface would give the eigenvalues of the $\hat{V}_s$ operator and the corresponding *adiabatic* states, which would be a linear combination of $|\Delta_{x^2-y^2}\rangle$ and $|\Delta_{xy}\rangle$. Since these two electronic states are degenerate in the free TiH molecule and their splitting by the crystal-field remains relatively small, one may expect the Born-Oppenheimer or adiabatic approximation to break down. Therefore, some kind of *diabatization* procedure is required to obtain the $2 \times 2$ interaction matrix. We determine the real adiabatic states $|\Delta_{x^2-y^2}\rangle$ and $|\Delta_{xy}\rangle$ in an *ab initio* calculation that does not include the surface, and we include the Coulomb interaction in the surface potential $\hat{V}_\mathrm{s}$.

Initially, we used a multipole expansion of the electric field of the crystal at the position of the TiH molecule. Since for the $^2\Delta$ state $\Lambda = 2$, the first moment that couples $\Lambda = 2$ and $\Lambda = -2$ is of rank $l = 2|\Lambda| = 4$ because of angular momentum constraints. This moment couples through the transition hexadecapole multipole moment between the $\Delta_{x^2-y^2}$ and $\Delta_{xy}$ states. The ranks of other terms that contribute are multiples of four because of the four-fold symmetry. However, we found that the multipole starts to diverge before it reaches a stable value. This is not too surprising, since the electron clouds of the crystal and the molecule must overlap at the equilibrium height: the Coulomb interaction is attractive, and the Pauli exchange repulsion between the electron clouds result in a stable equilibrium. Thus, instead we calculated the electrostatic interaction between the crystal, with the ions modeled by point charges, and the molecule exactly. With $|1\rangle$ denoting the $\Delta_{x^2-y^2}$ state and $|2\rangle$ denoting the $\Delta_{xy}$ state, we compute the diagonal elements of the operator $\hat{V}_\mathrm{s}$

$$V_{i,i} = \langle i| \hat{V}_\mathrm{s} |i\rangle = \sum_k q_k \left[\iiint \frac{-\rho_i(\boldsymbol{r})}{|\boldsymbol{r} - \boldsymbol{R}_k|}\,\mathrm{d}\boldsymbol{r} + \sum_A \frac{Z_A}{|\boldsymbol{R}_A - \boldsymbol{R}_k|}\right]. \tag{17}$$

for $i = 1$ and $i = 2$. The sum over $k$ is a sum over the ions in the crystal with charges $q_k$ and Cartesian positions $_k$, the sum over $A$ is over the two nuclei of the TiH molecule, with Cartesian positions $_A$ and nuclear charges $Z_A e$ (i.e., $Z_\mathrm{H} = 1$ and $Z_\mathrm{Ti} = 22$). The electron densities $\rho_i(\boldsymbol{r})$ for the two states are computed *ab initio* as described in section S9.4.

The off-diagonal element of $\hat{V}_\mathrm{s}$ in the real basis is given by

$$V_{1,2} = V_{2,1} = \sum_k q_k \iiint \frac{-\rho_{1,2}(\boldsymbol{r})}{|\boldsymbol{r} - \boldsymbol{R}_k|}\,\mathrm{d}\boldsymbol{r}, \tag{18}$$

where again we have a sum over the ions in the crystal, but there is no contribution from the molecular nuclei because of the orthogonality of the molecular states. The *transition density* $\rho_{1,2}(\boldsymbol{r})$ is the product of the wave functions of the two states.
Since both expressions consist of sums over the ions, and all terms are linear in the ionic charges $q_k$, we can write the potential matrix elements as

$$V_{i,j} = \sum_k q_k V_{i,j}^{(+)}(\boldsymbol{R}_k), \tag{19}$$

where $V_{i,j}^{(+)}(\boldsymbol{R})$ is the interaction potential for the molecule and a single ion with charge $+e$ at position $\boldsymbol{R}$.

The functional form of the potentials $V_{i,j}^{(+)}(\boldsymbol{R})$ is well known, since they correspond to the potentials of a complex consisting of an atom (or ion) and a diatomic molecule in a spatially degenerate state [56]. These potentials are most easily described in a frame with the $z$-axis along the molecular axis and the origin at the center of mass of the molecule. With $\boldsymbol{R}$ as the Cartesian coordinates of the ion in this frame and $R \equiv |\boldsymbol{R}|$, we express the unit vector $\hat{\boldsymbol{R}}$ along the vector $\boldsymbol{R}$ in spherical polar coordinates $(\beta, \alpha)$, i.e.,

$$\hat{\boldsymbol{R}} = \boldsymbol{R}_z(\alpha)\boldsymbol{R}_y(\beta)\boldsymbol{e}_z, \tag{20}$$

where $\boldsymbol{e}_z$ is the unit vector along the $z$-axis, and $\boldsymbol{R}_z(\alpha)$ and $\boldsymbol{R}_y(\beta)$ are $3 \times 3$ matrices representing rotations around the $z$- and $y$-axes, respectively. Note that the potentials $V_{i,j}^{(+)}(R, \beta, \alpha)$ depend not only on the distance $R$ and Jacobi "bending" coordinate $\beta$, but also on the azimuthal angle $\alpha$. For a molecule in a $\Sigma$ state, the electron density would be cylindrically symmetric, and the potential would be independent of $\alpha$. Since the sum of the electron densities of the $\Delta_{x^2-y^2}$ and $\Delta_{xy}$ also has cylinder symmetry, the sum of $V_{1,1}^{(+)}(R, \beta, \alpha)$ and $V_{2,2}^{(+)}(R, \beta, \alpha)$ is also independent of $\alpha$. For the one-electron $\Delta_{x^2-y^2}$ state, the azimuthal angle ($\varphi$)-dependence of the wave function is $\cos(2\varphi)$ and for $\Delta_{xy}$ it is $\sin(2\varphi)$, so the transition density $\rho_{1,2}$ depends on

$$\cos(2\varphi)\sin(2\varphi) = \frac{1}{2}\sin(4\varphi). \tag{21}$$

As a result, the off-diagonal element depends on $\alpha$ as

$$V_{1,2}^{(+)}(R, \beta, \alpha) = V^{(+)}(R, \beta)\sin(4\alpha). \tag{22}$$

The eigenvalues of the $2 \times 2$ interaction matrix $\boldsymbol{V}^{(+)}$, with elements $V_{i,j}^{(+)}(R, \beta, \alpha)$ are

$$\epsilon_\pm = \frac{V_{1,1}^{(+)} + V_{2,2}^{(+)}}{2} \pm \sqrt{\left|V_{2,2}^{(+)} - V_{1,1}^{(+)}\right|^2 + |V_{1,2}^{(+)}|^2}. \tag{23}$$

and since they clearly must be independent of $\alpha$, the difference potential, $V_{2,2}^{(+)} - V_{1,1}^{(+)}$, should be

$$V_{1,2}^{(+)}(R, \beta, \alpha) = V^{(+)}(R, \beta)\cos(4\alpha). \tag{24}$$

Since the $\alpha$-dependence is known, and the off-diagonal element is related to the difference of the diagonal elements, we can restrict the *ab initio* calculations to $\alpha = 0$, and determine $V^{(+)}(R, \beta)$ from the difference potential. Since the sum potential is also independent of $\alpha$, we only need to compute the $\Delta_{x^2-y^2}$ and $\Delta_{xy}$ states with the molecule along the $z$-axis and the point charge in the $xz$-plane (i.e., for $\alpha = 0$).

Note that this derivation does not give the relative sign of the difference potential and the off-diagonal potential. Since the off-diagonal potential depends on a phase convention for the wave-functions, this is unimportant in the static model. In the *diabatic* dynamic model, discussed in the next section, the phase convention for the electronic wave function must be consistent for different orientations of the molecule.

The derivation above seems to depend on the explicit angular dependence of a one-electron $\Delta$ wave function. However, a slightly more formal derivation [56] only uses the rotational symmetry of the complex states

$$\hat{R}_z(\alpha)\,|\Lambda\rangle = e^{-\frac{i}{\hbar}\alpha\hat{L}_z}\,|\Lambda\rangle = e^{-i\alpha\Lambda}\,|\Lambda\rangle\,, \tag{25}$$

which also holds for arbitrary $N$-electron $\Delta$ states. The symmetry-based derivation also clearly shows why no higher Fourier components such as $\cos(4n\alpha)$ or $\sin(4n\alpha)$ for $n = 2, 3, \ldots$ appear in the potential and it solves the phase convention issue. Specifically, it shows that the matrix elements of the interaction potential for a single point charge in the complex basis can be expanded as

$$\langle\Lambda' S\Sigma'|\,\hat{V}^{(+)}\,|\Lambda S\Sigma\rangle = \delta_{\Sigma'\Sigma}\sum_{l=|\Lambda-\Lambda'|}^{\infty} C_{l,\Lambda-\Lambda'}(\beta,\alpha)c_{l,\Lambda-\Lambda'}(R), \tag{26}$$

where $C_{l,\Lambda-\Lambda'}$ are Racah normalized spherical harmonics, which depend on $\alpha$ through $e^{i(\Lambda-\Lambda')\alpha}$. We can relate the potentials in the complex basis to the matrix element for the real components of the $^2\Delta$ function with Eqs. (12)-(13), and we can easily re-derive the $\alpha$ dependence of the diagonal and off-diagonal matrix elements in the real basis.
Before we can take the sum of Eq. (19) to obtain the potential for the entire crystal, we transform Eq. (26) to a "crystal-fixed" coordinate system, in which the $z$-axis is the normal of the MgO crystal surface, and the origin is on the oxygen ion of the unperturbed crystal underlying the TiH molecule. The $x$-axis points from the origin to one of the nearest $Mg^{2+}$ ions. We denote the spherical polar coordinates of ion $k$ in this frame by $(R_k, \theta_k, \phi_k)$, while the unit vector along the TiH axis has polar angles $(\theta, \phi)$. The Racah spherical harmonics in Eq. (26) can be expressed in these new coordinates by [56]

$$C_{l,\Lambda-\Lambda'}(\beta,\alpha) = \sum_{m=-l}^{l} C_{l,m}(\theta_k,\phi_k)D^{(l)}_{m,\Lambda-\Lambda'}(\phi,\theta,0), \tag{27}$$

where $D^{(l)}_{m,\Lambda-\Lambda'}$ is a Wigner D-matrix element. Thus, the matrix elements of the interaction potential for the crystal and the molecule in the case (a) basis are

$$\langle\Lambda' S\Sigma'|\,\hat{V}_{\rm s}\,|\Lambda S\Sigma\rangle = \delta_{\Sigma',\Sigma}\sum_{l=|\Lambda'-\Lambda|}^{\infty}\sum_{m=-l}^{l} D^{(l)}_{m,\Lambda'-\Lambda}(\phi,\theta,0) \times \sum_k q_k C_{l,m}(\theta_k,\phi_k)c_{l,\Lambda-\Lambda'}(R_k). \tag{28}$$

### S9.1.4 Computation of the *g*-factor in the static model

The Hamiltonian consists of the Zeeman Hamiltonian [Eq. (2)], the spin-orbit coupling [Eq. (9)], and the crystal field $\hat{V}_{\rm s}$,

$$\hat{H}_{\rm static} = \hat{H}_{\rm Z} + \hat{H}_{\rm SO} + \hat{V}_{\rm s}. \tag{29}$$

The wave function is expanded in a basis of four Hund's case (a) functions [Eq. (1)]. The matrix elements of the Hamiltonian are computed with Eqs. (3), (4), (10), (17) and (18). We compute the energies as the eigenvalues of the Hamiltonian matrix for a magnetic field of $B$ = 1 T, and compute the *g*-factor as

$$g_{\perp/\parallel} = \frac{E_1 - E_0}{\mu_B B_{\perp/\parallel}}\,. \tag{30}$$

For a derivation of the analytic expressions of both *g*-factors as provided in the main text, we refer to section S9.3.

## S9.2 Dynamical model

In the dynamical model, the polar angles $\theta$ and $\phi$ of the TiH internuclear axis are no longer fixed, but we still keep the center of mass of the molecule at a fixed position above the surface. We expand the wave function in a free rotor Hund's case (a) dynamical basis

$$|\Lambda S\Sigma JM\Omega\rangle = \sqrt{\frac{2J+1}{4\pi}} D^{J,*}_{M,\Omega}(\phi,\theta,0)\hat{R}(\phi,\theta,0)|\Lambda S\Sigma\rangle_\perp . \tag{31}$$

The electronic wave function $|\Lambda S\Sigma\rangle_\perp$ is the case (a) wave function as in the static model, with the TiH internuclear axis perpendicular to the surface. It is rotated by the rotation operator $\hat{R}(\phi,\theta,0)$ so that the axis has spherical polar angles $\theta$ and $\phi$. The rotation operator is defined by

$$\hat{R}(\phi,\theta,0) = \hat{R}_z(\phi)\hat{R}_y(\theta) = e^{-\frac{i}{\hbar}\phi\hat{J}_Z} e^{-\frac{i}{\hbar}\theta\hat{J}_Y} , \tag{32}$$

where the $\hat{J}_Z$ and $\hat{J}_Y$ are the components of the total angular momentum operator in the crystal frame. This somewhat formal notation simplifies the computation of matrix elements below. The rotation of the TiH axis is described by the complex conjugate of normalized Wigner-$D$ functions, where $\Omega \equiv \Lambda + \Sigma$, and $M$ is the quantum number corresponding to the projection of the total angular momentum $\hat{J}$ onto the normal of the surface.

The Hamiltonian consists of $\hat{H}_{\text{static}}$ with two additional terms, the rotational kinetic energy $\hat{T}_{\text{R}}$ and an extra term in the potential $\hat{V}_{\text{r}}$,

$$\hat{H} = \hat{T}_{\text{R}} + \hat{H}_{\text{SO}} + \hat{H}_{\text{Z}} + \hat{V}_{\text{s}} + \hat{V}_{\text{r}} . \tag{33}$$

### S9.2.1 Rotational kinetic energy

The rotational kinetic energy is given by

$$\hat{T}_{\text{R}} = B_{\text{TiH}}\hat{\boldsymbol{R}}^2 = B_{\text{TiH}}(\hat{R}_x^2 + \hat{R}_y^2) , \tag{34}$$

where $\hat{\boldsymbol{R}} = \hat{\boldsymbol{J}} - \hat{\boldsymbol{L}} - \hat{\boldsymbol{S}}$ is the nuclear angular momentum, which only has contributions perpendicular to the molecular axis. The rotational constant of TiH is $B_{\text{TiH}} = 1/(2\mu r^2)$, with reduced mass $\mu$ and bond length $r$. Using the *ab initio* bond distance from section S9.4, we find a rotational constant of $B_{\text{TiH}} = 5.4344$ cm$^{-1}$. The rotational term can be rewritten to

$$\begin{aligned}\hat{T}_{\text{R}} = B_{\text{TiH}}(&\hat{\boldsymbol{J}}^2 - \hat{J}_z^2 + \hat{\boldsymbol{S}}^2 - \hat{S}_z^2 + \hat{\boldsymbol{L}}^2 \\ &- \hat{L}_z^2 + \hat{L}_+\hat{S}_- + \hat{L}_-\hat{S}_+ - \hat{J}_+\hat{L}_- - \hat{J}_-\hat{L}_+ - \hat{J}_+\hat{S}_- - \hat{J}_-\hat{S}_+) .\end{aligned} \tag{35}$$

Some of these terms only result in an overall shift in energy and may be neglected (assuming a doublet state, this also holds for $\hat{S}_z$). Furthermore, the ladder operators $\hat{L}_\pm$ only couple other electronic states, which we assume to be far apart in energy. The last two terms give rise to a Coriolis effect and couple nearby $\Omega$, which we also assume to be separated more in energy than states with different $J$. We can then greatly simplify Eq. (35) to

$$\hat{T}_{\text{R}} \approx B_{\text{TiH}}(\hat{\boldsymbol{J}}^2 - \hat{J}_z^2) . \tag{36}$$

The matrix elements of this operator are diagonal in the dynamic basis and are given by

$$\langle \Lambda' S\Sigma' J' M'\Omega'|\hat{T}_{\text{R}}|\Lambda S\Sigma JM\Omega\rangle = \delta_{\Lambda'\Lambda}\delta_{\Sigma'\Sigma}\delta_{J'J}\delta_{M'M}\delta_{\Omega'\Omega}\, B_{\text{TiH}}\left[J(J+1) - \Omega^2\right] . \tag{37}$$

Note, that the operator $\hat{J}_z$ is a body-fixed operator and therefore yields a factor $\Omega$ instead of $M$

.

**S9.2.2 Repulsive potential**

A crystal surface modeled by point charges completely lacks the Pauli repulsion between the electrons of the molecule and the electrons of the ions. This leads to unphysically strong interactions between, e.g., the partly charged hydrogen atom and the negative oxygen ions at large values of $\theta$. To mimic the Pauli repulsion we include a repulsive potential, which increases the energy for spatial orientations in the downward direction without modifying the potential for small $\theta$. This potential therefore does not affect the energy at small $\theta$ and is mainly used to avoid nonphysical local minima at larger inclination angles. For computational convenience, we expand the repulsive potential in Legendre polynomials, which are related to Racah normalized spherical harmonics,

$$V_{\mathrm{r}}(\theta) = \sum_{\ell=0}^{\ell_{\max}} c_\ell P_\ell(\cos\theta) = \sum_{\ell=0}^{\ell_{\max}} c_\ell C_{\ell,0}(\theta, 0)\,. \tag{38}$$

The coefficients $c_\ell$ are chosen such that the potential is zero for $\theta = 0$ and has some maximal value $V_{\max}$ for $\theta = \pi$, and such that its derivatives up to some order $n_{\max}$ for $\theta = 0$ and $m_{\max}$ for $\theta = \pi$ are zero as well. These conditions give rise to a system of $2 + n_{\max} + m_{\max}$ equations for which evaluation of the Legendre polynomials and its derivatives for $\cos\theta = -1$ and $\cos\theta = 1$ are needed. These can be calculated by the explicit representation of Legendre polynomials:

$$P_l(z) = \frac{1}{2^l} \sum_{k=0}^{\lfloor \frac{l}{2} \rfloor} (-1)^k \binom{l}{k} \binom{2l-2k}{l} z^{l-2k}\,. \tag{39}$$

The repulsive potential is taken the same for both electronic states, so the matrix elements are given by

$$\langle \Lambda' S \Sigma' | \hat{V}_{\mathrm{r}} | \Lambda S \Sigma \rangle = \delta_{\Lambda'\Lambda} \delta_{\Sigma'\Sigma} \sum_{\ell=0}^{\ell_{\max}} c_\ell C_{\ell,0}(\theta, \phi)\,. \tag{40}$$

This repulsive potential is only of importance for the dynamical model. From the static model we determine the value of $V_{\max}$ needed to remove the unphysical minima for large $\theta$. In our calculations we have used $V_{\max} = 0.06$ atomic units, $n_{\max} = 2$ and $m_{\max} = 4$.

**S9.2.3 Matrix elements of the spin-orbit Hamiltonian**

Since the spin orbit Hamiltonian only acts on the electronic wave functions, and the normalized Wigner rotation matrix elements constitute an orthonormal basis, the matrix elements are given by

$$\langle \Lambda' S \Sigma' J' M' \Omega' | \hat{H}_{\mathrm{SO}} | \Lambda S \Sigma J M \Omega \rangle = \delta_{\Lambda'\Lambda} \delta_{\Sigma'\Sigma} \delta_{J'J} \delta_{M'M} \delta_{\Omega'\Omega}\, A \Lambda \Sigma\,. \tag{41}$$

**S9.2.4 Matrix elements of the Zeeman Hamiltonian**

The Zeeman Hamiltonian is the same as for the static model [Eq. (2)]. Using the spherical representation of the dot product we find for the matrix elements

$$\langle \Lambda' S \Sigma' J' M' \Omega' | \hat{H}_{\mathrm{Z}} | \Lambda S \Sigma J M \Omega \rangle = \frac{\sqrt{(2J'+1)(2J+1)}}{4\pi} \sum_{m=-1}^{1} (-1)^{m+1} B_{-m}$$
$$\times \iint D^{J'}_{M'\Omega'}(\phi, \theta, 0)\ {}_\perp\langle \Lambda' S \Sigma' | \hat{R}^\dagger \hat{\mu}_m \hat{R} | \Lambda S \Sigma \rangle_\perp D^{J,*}_{M,\Omega}(\phi, \theta, 0)\, \mathrm{d}\phi\, \mathrm{d}\cos\theta\,. \tag{42}$$

The spherical components of the magnetic moment transform as rank one tensor operators

$$\hat{R}^\dagger \hat{\mu}_m \hat{R} = \sum_{m'=-1}^{1} \hat{\mu}_{m'} D^{(1)}_{m'm}(\hat{R}^\dagger) = \sum_{m'=-1}^{1} \hat{\mu}_{m'} D^{1,*}_{mm'}(\phi, \theta, 0)\,. \tag{43}$$

Rotating the magnetic moments is done using Eq. (43). For the integral we introduce the third Euler angle by multiplying the equation with

$$1 = \frac{1}{2\pi}\int_0^{2\pi} e^{i\chi(-\Omega'+m'+\Omega)}\,\mathrm{d}\chi\ . \tag{44}$$

This relation holds, since the matrix element of the magnetic moment $\hat{\mu}_{m'}$ gives $\Omega' = m' + \Omega$. The integral then evaluates to

$$\frac{1}{2\pi}\iiint D^{J'}_{M',\Omega'}(\phi,\theta,\chi) D^{1,*}_{m,m'}(\phi,\theta,\chi) D^{J,*}_{M,\Omega}(\phi,\theta,\chi)\,\mathrm{d}\phi\,\mathrm{d}\cos\theta\,\mathrm{d}\chi$$
$$= 4\pi(-1)^{M'-\Omega'}\begin{pmatrix} J' & 1 & J \\ -M' & m & M \end{pmatrix}\begin{pmatrix} J' & 1 & J \\ -\Omega' & m' & \Omega \end{pmatrix}\ , \tag{45}$$

where we have introduced the 3-$j$ symbols resulting from the coupling of angular momenta. Moreover, these coefficients restrict the values for $m$ and $m'$ to $m = M' - M$ and $m' = \Omega' - \Omega$, both attaining integer values of -1 to 1. Due to these constraints, the summations over $m$ and $m'$ will only yield a value for one specific term. Evaluating the matrix elements for $\hat{\mu}_{m'}$ in the same way as for the static case then results in the following matrix elements for the Zeeman Hamiltonian:

$$\langle \Lambda' S \Sigma' J' M' \Omega' | \hat{H}_\mathrm{Z} | \Lambda S \Sigma J M \Omega \rangle = \delta_{\Lambda',\Lambda}[J',J]^{\frac{1}{2}}(-1)^{M'-\Omega'}$$
$$\times \begin{pmatrix} J' & 1 & J \\ -M' & M'-M & M \end{pmatrix}\begin{pmatrix} J' & 1 & J \\ -\Omega' & \Omega'-\Omega & \Omega \end{pmatrix}$$
$$\times \left[\frac{1}{\sqrt{2}}(B_x + iB_y)\delta_{M',M-1} + B_z\delta_{M',M} - \frac{1}{\sqrt{2}}(B_x - iB_y)\delta_{M',M+1}\right]$$
$$\times \mu_B\left(\frac{g_e}{\sqrt{2}}C_-(S,\Sigma)\,\delta_{\Sigma',\Sigma-1} + (\Lambda + g_e\Sigma)\,\delta_{\Sigma',\Sigma} - \frac{g_e}{\sqrt{2}}C_+(S,\Sigma)\,\delta_{\Sigma',\Sigma+1}\right). \tag{46}$$

where $[J',J] \equiv (2J'+1)(2J+1)$ and

$$C_\pm(S,\Sigma) = \sqrt{S(S+1) - \Sigma(\Sigma \pm 1)} = 1 \quad (\text{for } S = 1/2). \tag{47}$$

**S9.2.5 Matrix elements of the surface potential**

Using the matrix elements for the Coulomb interactions from Eq. (28), we find for the dynamic matrix elements that

$$\langle \Lambda' S \Sigma' J' M' \Omega' | \hat{V}_\mathrm{s} | \Lambda S \Sigma J M \Omega \rangle = \delta_{\Sigma',\Sigma}\frac{[J',J]}{4\pi}\sum_k q_k \sum_{\ell,m} c_{\ell,\Lambda-\Lambda'}(r_k) C_{\ell,m}(\theta_k,\phi_k)$$
$$\times \iint D^{J'}_{M',\Omega'}(\phi,\theta,0) D^{\ell}_{m,\Lambda-\Lambda'}(\phi,\theta,0) D^{J,*}_{M,\Omega}(\phi,\theta,0)\,\mathrm{d}\phi\,\mathrm{d}\cos\theta\ . \tag{48}$$

The integral is evaluated in the same way as before. We now find the restrictions $m = M - M'$ and $\Lambda' - \Lambda = \Omega' - \Omega$. The final matrix elements are then given by

$$\langle \Lambda' S \Sigma' J' M' \Omega' | \hat{V}_\mathrm{s} | \Lambda S \Sigma J M \Omega \rangle = \delta_{\Sigma',\Sigma}(-1)^{M'-\Omega'}[J,J']$$
$$\times \sum_\ell \begin{pmatrix} J' & \ell & J \\ -M' & M'-M & M \end{pmatrix}\begin{pmatrix} J' & \ell & J \\ -\Omega' & \Lambda'-\Lambda & \Omega \end{pmatrix}$$
$$\times \sum_k q_k c_{\ell,\Lambda-\Lambda'} C_{\ell,M-M'}(\theta_k,\phi_k)\,. \tag{49}$$

**S9.2.6 Matrix elements of the repulsive potential**

The repulsive potential is given by Eq. (38). It is the same for both electronic states. The matrix elements are

$$\langle \Lambda' S \Sigma' J' M' \Omega' | \hat{V}_{\mathrm{r}} | \Lambda S \Sigma J M \Omega \rangle = \delta_{\Lambda'\Lambda} \delta_{\Sigma'\Sigma} \frac{[J', J]}{4\pi} \times \sum_{\ell=0}^{\ell_{\max}} c_\ell \iint D^{J'}_{M'\Omega'}(\phi, \theta, 0) C_{\ell,0}(\theta, \phi) D^{J,*}_{M\Omega}(\phi, \theta, 0) \, \mathrm{d}\phi \, \mathrm{d}\cos\theta \ . \quad (50)$$

Evaluating the integral as before then yields the matrix elements

$$\langle \Lambda' S \Sigma' J' M' \Omega' | \hat{V}_{\mathrm{r}} | \Lambda S \Sigma J M \Omega \rangle = \delta_{\Lambda',\Lambda} \delta_{\Sigma',\Sigma} \delta_{M',M} \delta_{\Omega',\Omega} (-1)^{M-\Omega} [J', J] \times \sum_{\ell} c_\ell \begin{pmatrix} J' & \ell & J \\ -M & 0 & M \end{pmatrix} \begin{pmatrix} J' & \ell & J \\ -\Omega & 0 & \Omega \end{pmatrix} . \quad (51)$$

The coefficients $c_\ell$ are determined by the procedure as given in the static case, such that only a finite number of coefficients is non-zero.

**S9.2.7 Computation of the *g*-factor in the dynamic model**

Using Eqs. (37), (41), (46), (49) and (51), the matrix elements for the total Hamiltonian (33) can now be calculated. In all calculations, a basis is used containing elements with $J = 1.5$ up to 20.5. Diagonalizing this Hamiltonian then yields the eigenenergies and the eigenfunctions as linear combinations of the complex basis $|\Lambda S \Sigma J M \Omega\rangle$. The *g*-factor is then again determined by Eq. (30).

**S9.2.8 Adding an external electric field**

In the dynamic model we can also incorporate an external electric field in any direction and induce a Stark effect. This interaction is given by

$$H_{\mathrm{S}} = -\boldsymbol{E} \cdot \hat{\boldsymbol{\mu}} \, , \quad (52)$$

where $\boldsymbol{E}$ is the electric field and $\hat{\boldsymbol{\mu}}$ is the electric dipole moment. The derivation of the matrix elements is analogous to the one for the Zeeman Hamiltonian with the magnetic dipole moment, where we note that only $\langle \Lambda' | \hat{\mu}_z | \Lambda \rangle$ remains and is equal for both states. We then find the matrix elements

$$\langle \Lambda' S \Sigma' J' M' \Omega' | \hat{H}_{\mathrm{S}} | \Lambda S \Sigma J M \Omega \rangle = \delta_{\Lambda',\Lambda} \delta_{\Sigma',\Sigma} \delta_{\Omega',\Omega} (-1)^{M'-\Omega} [J, J'] \langle \Lambda | \hat{\mu}_z | \Lambda \rangle \times \begin{pmatrix} J' & 1 & J \\ -M' & M'-M & M \end{pmatrix} \begin{pmatrix} J' & 1 & J \\ -\Omega & 0 & \Omega \end{pmatrix} \times \left( -\frac{1}{\sqrt{2}} (E_x + iE_y) \delta_{M',M-1} - E_z \delta_{M',M} + \frac{1}{\sqrt{2}} (E_x - iE_y) \delta_{M',M+1} \right) . \quad (53)$$

**S9.3 Derivation of analytic expression for the *g*-tensor**

Here, we provide a derivation of the analytic expressions for both *g*-factors as given in the main text. For both time-reversal symmetries, the lower eigenvalue of the Hamiltonian matrix is

$$E_0 = -\sqrt{A_{SO}^2 + \left(\frac{1}{2} V_c\right)^2} \quad (54)$$

and the upper eigenvalue is $E_1 = -E_0$. Rewrite the Hamiltonian matrices (in the bases $\{\Psi_{\pm}(2, -1/2), \Psi_{\pm}(2, 1/2)\}$, assuming $A_{SO} \geq 0$),

$$\boldsymbol{H}_{\pm} = E_1 \begin{pmatrix} -\cos\alpha & \pm\sin\alpha \\ \pm\sin\alpha & \cos\alpha \end{pmatrix}, \tag{55}$$

where

$$\tan\alpha = \frac{V_c}{2A_{SO}}. \tag{56}$$

The eigenvector with the lower eigenvalue ($E_0$) is

$$\boldsymbol{u}^{\pm} = \begin{pmatrix} -\cos(\alpha/2) \\ \pm\sin(\alpha/2) \end{pmatrix}, \tag{57}$$

so that the relative sign of the components of the eigenvector depends on the time-reversal symmetry. It is obviously true for $\alpha = 0$. When $\alpha$ goes from zero to $2\pi$, the eigenvector changes sign. This is the famous geometric phase you get when you go around a conical intersection.

To determine $g_{\parallel}$ consider the effect of $\hat{S}_x$: it couples to the fine-structure states, but the sign of the coupling depends on the time-reversal symmetry:

$$\langle \Psi_{\pm}(2, 1/2) | \hat{S}_x | \Psi_{\pm}(2, -1/2) \rangle = \pm\frac{1}{2}. \tag{58}$$

Thus, the lower eigenvalue $E_0$ splits due to the Zeeman interaction as

$$E_0^{(\pm)} = -\sqrt{A_{\mathrm{SO}}^2 + \left(\frac{1}{2}V_c \pm \frac{1}{2}\mu_{\mathrm{B}} g_e B\right)^2}. \tag{59}$$

To first order in $B$ this gives, using $\sqrt{1+\epsilon} = 1 + \epsilon/2 + O(\epsilon^2)$ for small $\epsilon$,

$$E_0^{(\pm)} = -\sqrt{E_1}\left(1 \pm \frac{\mu_{\mathrm{B}} g_e V_c B}{4E_1}\right) \tag{60}$$

The $g$-factor is the derivative of the energy difference with respect to $B$, divided by $\mu_B$, and so the $g_{\parallel}$-factor is

$$g_{\parallel} = \frac{g_e V_c}{2\sqrt{E_1}} = g_e \frac{r}{\sqrt{1 + r^2}}, \tag{61}$$

with $\mu_B = 1/2$ in atomic units and where

$$r = \frac{V_c}{2A_{\mathrm{SO}}}. \tag{62}$$

When the field is perpendicular to the surface, the degeneracy is lifted by the $z$-component of the Zeeman Hamiltonian which breaks time-reversal symmetry, see Eq. (3) of the main text. The contribution of the fine-structure states are determined by the eigenvectors $\boldsymbol{u}^{\pm}$, so we find

$$g_{\perp} = \langle u_1^- \Psi_-(2, -1/2) + u_2^- \Psi_-(2, 1/2) | \hat{L}_z + g_e \hat{S}_z | u_1^+ \Psi_+(2, -1/2) + u_2^+ \Psi_+(2, 1/2) \rangle \tag{63}$$

$$= |u_1|^2(4 - g_e) - |u_2|^2(4 + g_e) \tag{64}$$

$$= 4(|u_1|^2 - |u_2|^2) - g_e. \tag{65}$$

The eigenvector dependent part can be written as

$$|u_1|^2 - |u_2|^2 = \cos(\alpha/2)^2 - \sin(\alpha/2)^2 = \cos\alpha = \frac{A_{\text{SO}}}{E_1} = \frac{1}{\sqrt{1+r^2}}, \tag{66}$$

where in the before last step we used the expression for $H_{1,1}$ of Eq. (55), so we find

$$g_\perp = \frac{4}{\sqrt{1+r^2}} - g_e \,. \tag{67}$$

**S9.4 *Ab initio* calculations**

For the *ab initio* electronic structure calculations we used the Molpro [40] program package. Using this software, we determined the electronic ground state of TiH in the adsorbed phase. Moreover, for use in the model we computed molecular properties of TiH in the gas phase, in particular: the spin-orbit coupling constant and the molecular electrostatic potential for calculations of the Coulomb interaction with the surface.

Two electronic states of the TiH molecule are of special interest: ${}^4\Phi$ and ${}^2\Delta$. In the gas phase they are the lowest quartet and doublet spin states, respectively. The former is also the overall ground state of free TiH. Decreasing the distance between the TiH and the surface changes the order of electronic states and, as shown in Fig. 1 of the main text, the ${}^2\Delta$ state becomes the ground state at smaller heights above the MgO surface.

**S9.4.1 Gas phase**

For the gas-phase calculations the actual symmetry of the diatomic TiH molecule is $C_{\infty v}$, but Molpro adopts $C_{2v}$ symmetry. This group has irreducible representations (irreps) $A_1, B_1, B_2$, and $A_2$, and the two components of the ${}^2\Delta$ state belong to the $A_1$ and $A_2$ irreps. We denote these by ${}^2\Delta_{x^2-y^2}$ and ${}^2\Delta_{xy}$, respectively, since they represent the character of the corresponding occupied *d*-orbitals. In the gas phase, these are degenerate, but they will split upon introduction of the surface with its four-fold symmetry. The degenerate ${}^4\Phi$ states belong to the $B_1$ and $B_2$ irreps, but are not split in energy in this system.

All calculations have been performed using an aug-cc-pVTZ basis set [57]. A first orbital guess is generated by Hartree-Fock (HF) calculations, which are followed by a complete-active-space self-consistent field (CASSCF) [58,59] calculation with irrep occupation 9,3,3,1 ($A_1, B_1, B_2, A_2$) and closed orbitals 5,2,2,0 for both electronic states. To obtain the correct states, however, a prior CASSCF calculation was needed with a larger number of active orbitals in the $A_1$ irrep, i.e., occupied orbitals 13,3,3,1 and closed orbitals 6,2,2,0. Following these CASSCF calculations, a multireference configuration interaction (MRCI) [60-62] calculation was made in each case to obtain the final energy. The wavefunction definition remained the same as in the CASSCF calculations.

With the final wave function thus obtained we calculated the spin-orbit coupling constant, as well as the molecular electronic potential using the one-electron operator ‘pot’.

**S9.4.2 Adsorbed phase**

For the MgO lattice we used the parameters resulting from DFT calculations. The in-plane lattice constant was set to 2.04 Å and the out-of-plane constant to 2.20 Å. Moreover, the oxygen atom directly below the TiH molecule is vertically displaced by 0.48 Å relative to the rest of the surface. For all calculations we used a finite lattice consisting of two layers and stretching 21 atoms along both in-plane directions, after which the total *ab initio* energy in atomic units had converged up to 6 decimals.

The adsorbed phase calculations were used to optimize the TiH bond length, as well as its height above the surface. In addition to TiH, we explicitly included the Mg and O atoms directly underneath. The other atoms in the finite surface were modeled as point charges using the lattice parameters listed above and charges $+2$ and $-2$ for the Mg and O atoms, respectively.

Initially only the MgO-TiH linear complex was included in the Molpro calculations with some Ti-H bond distance $R$ and some height $h$ above the surface. We started by choosing $h$ to be large such that TiH is essentially in the gas phase, in order to produce the correct electronic state in $C_{2v}$ symmetry. Then including MgO, we used irrep occupation 20,5,5,1 and closed orbitals 13,4,4,0 in the first CASSCF calculation, followed by one with occupation 16,5,5,1 and closed orbitals 12,4,4,0. Once the correct state was established from these CASSCF calculations, we added point charges and turned off the symmetry. Then, the height is decreased, CASSCF and MRCI are performed for each value of $h$. The energies from the MRCI calculations then provide the potential as a function of $h$, as given in Fig. 1. At the minimum of this function, we optimized the bond distance $R$. Since the optimal height $h^*$ and bond distance $R^*$ depend on each other, this process was repeated to arrive at optimal $h^*$ and $R^*$ values for each state to within a reasonable error estimate. For both electronic states, the optimal height and distance as well as the energy difference between the components are given in Table S3. These values are then used as inputs for the diabatic models.

In addition to these calculations, we performed embedded cluster calculations with TiH adsorbed on clusters as large as $Mg_9O_9$ embedded in a crystal represented by point charges. These are described in Appendix B of the main article.

### S9.5 Supplementary results

### S9.5.1 Molecular orbital diagrams

The molecular orbital diagrams of the electronic $^2\Delta$ and $^4\Phi$ states of the gas-phase TiH molecule are given in Fig. S18. The molecular orbitals are also given in a visual representation by performing *ab initio* calculations as described in S9.4.1. The visualization is done using Molden [63,64]. As can be seen, a bonding-antibonding orbital pair is formed between the Ti and H atoms. The highest occupied molecular orbitals are given by almost pure Ti d-orbitals.

To show the negligible hybridization of TiH in the $^2\Delta$ state with the surface, we provide the molecular orbitals in Fig. S19. We performed *ab initio* calculations by converging TiH towards a $Mg_9O_9$ cluster and then visualized them using Molden.

### S9.5.2 Crystal field

To give an idea of the contribution of the Coulomb interactions experienced from the surface, Fig. S20 shows the potential energy. The left picture illustrates the average energy $V_{\mathrm{s,avg}}(\theta, \phi)$ of the two real orbitals, the righthand one the difference $V_{\mathrm{s,diff}}(\theta, \phi)$ between the two energy levels. This difference is defined as the energy of the eigenstate with ${}^2\Delta_{x^2-y^2}$ as the major component minus the energy of the state with ${}^2\Delta_{xy}$ as the major component. When $\phi$ is a multiple of $45^{\circ}$, these are pure states, while for other angles the states get mixed by the rotation around the internuclear axis.

### S9.5.3 *g*-factors upon rotating TiH

In Fig. S20, one can see that the average potential is cylindrically symmetric. For $\theta = 90^{\circ}$, the orbitals get much closer to the underlying O atom and therefore have a higher energy. The difference potential clearly shows four-fold symmetry. For $\phi = 0^{\circ}$, at $\theta = 0^{\circ}$, the ${}^2\Delta_{x^2-y^2}$ state has a lower energy due to its lobes pointing towards the positive Mg point charges. Eventually, at larger $\theta$, the ground state switches to ${}^2\Delta_{xy}$, since the negatively charged O point charge below TiH increases the energy for the ${}^2\Delta_{x^2-y^2}$ state, which points towards this charge. For $\phi = 45^{\circ}$, the lowest state is always ${}^2\Delta_{xy}$.

In the main paper, *g*-factors have been given for $\theta = \phi = 0^{\circ}$, since this is the most favorable position. Fig. S21 shows the *g*-factors for different orientations of TiH. At most orientations, the determining factor for the eigenstates is the Coulomb interaction. Only at the four 'dots', where $V_{\mathrm{s,diff}}(\theta, \phi) = 0$, the spin-orbit coupling takes over.

### S9.5.4 Sensitivity of the *g*-tensor to static electric fields

As the ESR-STM experiments are extremely precise, we can consider how perturbations due to changes in local electric fields can alter the *g*-factor of the TiH molecule. The scatter of the experimental data in Fig. 3(b) could not solely be mapped to quantities like different atoms, measurement parameters or microtips. To investigate the response of the TiH molecule to electrostatic field perturbations, we employed the dynamical model and considered an additional external electric field in atomic units ($E$ = 1 a.u. = $5.14 \cdot 10^{11}$ $\mathrm{Vm^{-1}}$) interacting with the TiH dipole moment. The resulting *g*-factors for a static electric field parallel ($E_{\parallel}$) and perpendicular ($E_{\perp}$) to the MgO surface are depicted in Tables S4 and S5.

We find that the *g*-tensor of the TiH molecule is more sensitive to $E_{\parallel}$ than it is to $E_{\perp}$. Considering a single charge defect ($\pm 1e$ instead of $\pm 2e$) at one lattice site distance to the TiH, the electric field the TiH experiences is $E_{\parallel} \approx 1.5 \cdot 10^{10}$ $\mathrm{Vm^{-1}}$ = 0.03 a.u. Accordingly, $g_{\parallel}$ and $g_{\perp}$ change by ≈ 5 and 15 %, respectively. Potentially, such changes could stem from embedded defects in the crystal, vicinity to the edges of the MgO islands but also other undetected variations of the electrostatic environment. For perpendicular electric fields, the *g*-tensor remains almost unaffected, even for $E_{\perp}$ = -0.1 a.u. Therefore, even changes of the tip work function or the electric field induced by the measurement parameters assuming a simple plate capacitor model ($d$(tip-sample) ≈ 0.7 nm, $V_{\mathrm{DC}}$ = 50 mV, $I$ = 2 pA and $V_{\mathrm{RF}}$ = 8 mV, $E_{\perp} \approx$ 0.00016 a.u.) do not affect the measured *g*-tensor, as they are orders of magnitudes smaller as would be needed according to the calculations.

We conclude that we are not measuring an artificial effect from the tip on the TiH but changes in the local electrostatic environment might cause measurable variations in the *g*-tensor. We speculate, this might explain the scatter of the data we observe in Fig. 3(b), that could not be matched to any other obvious parameters.

### S9.5.5 Gas-phase and adsorbed spin-orbit coupling

The gas-phase spin orbit coupling constant as calculated from *ab initio* calculations is 110.05 $cm^{-1}$. This value has been used throughout the calculations. After convergence to the point charge surface with MgO explicitly taken into account along the *z*-axis, the spin-orbit constant is found to only change by -0.3 $cm^{-1}$ and so it only changes negligibly.

### S9.5.6 Sensitivity of charges in point charge model

We have investigated the stability of our model under changes in the charges of the surface atoms for which we had used +2 and -2 for Mg and O, respectively, for all calculations. When Bader charge analysis in DFT is considered, where we find charges of +1.7 and -1.7 for Mg and O, one might wonder how this will affect the results from the model. As it turns out, the TiH system is quite insensitive to such changes of the charges.

We carried out calculations in a similar manner as in Fig. 1 of the main text, but up to the CASSCF level. The energy splitting of the two $^2\Delta$states as a function of the height above the surface for charges $q = \pm 2$ and $q = \pm 1.7$ is given in the Fig. S22. For different charges, the optimal height above the surface changes. This is mainly due to Coulomb repulsion as higher charges lead to more repulsion and thus a larger height. However, larger charges also give rise to a larger energy splitting of the two $^2\Delta$states. These effects roughly cancel as can be seen by the colored bars along the *x*- and *y*-axes. Our model is therefore rather insensitive to the exact charges of the atoms in the ionic surface.

**FIGURES**

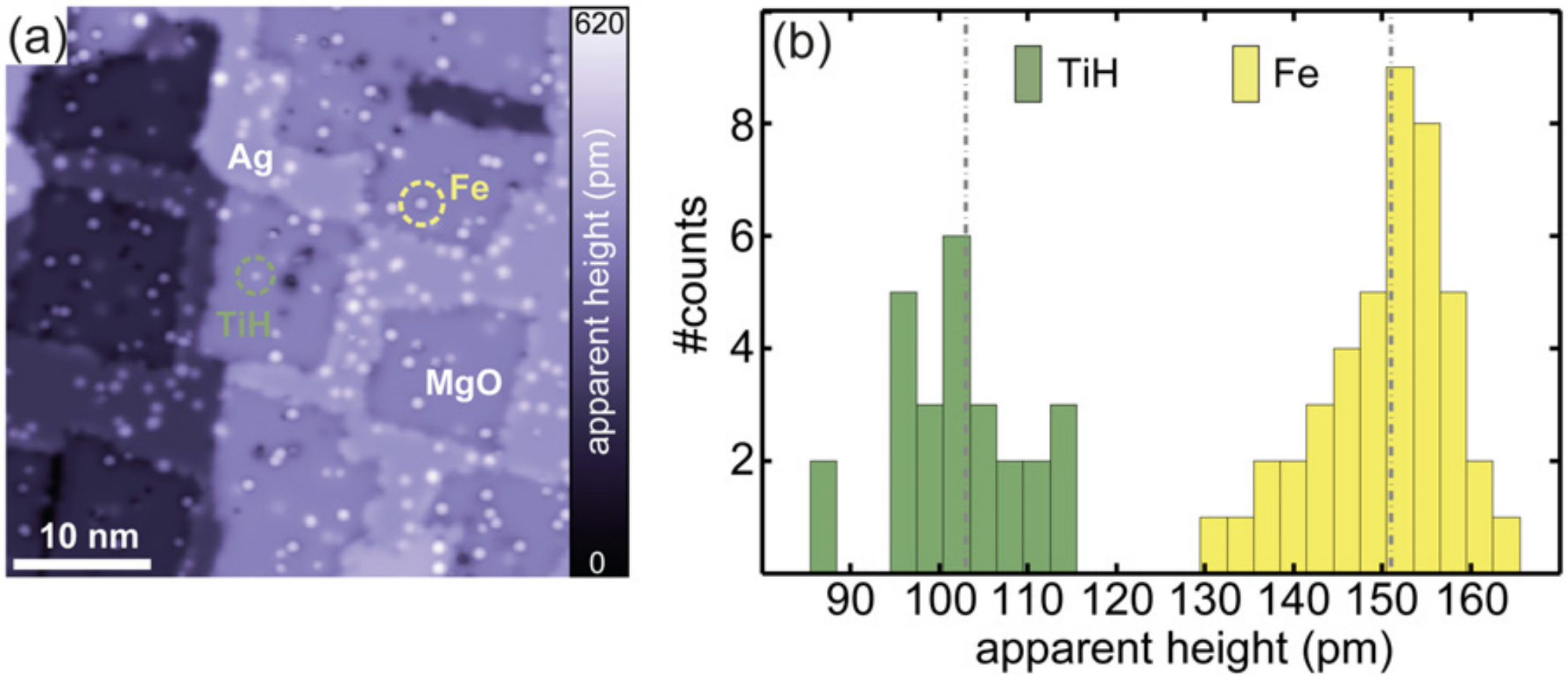

FIG. S1. STM characterization. (a) Constant-current STM image of a representative sample with MgO patches grown on Ag(100) and co-deposited Fe and TiH ($V_{DC}$ = 100 mV, $I_t$ = 10 pA). (b) Histogram representation of numerous measured apparent heights of TiH molecules (green) and Fe atoms (yellow) at $V_{DC}$ = 30 mV and $I_t$ = 10 pA. TiH molecules can easily be distinguished by their lower apparent height of 103 ± 8 pm compared to 151 ± 8 pm for Fe atoms (dashed lines represent the average value).

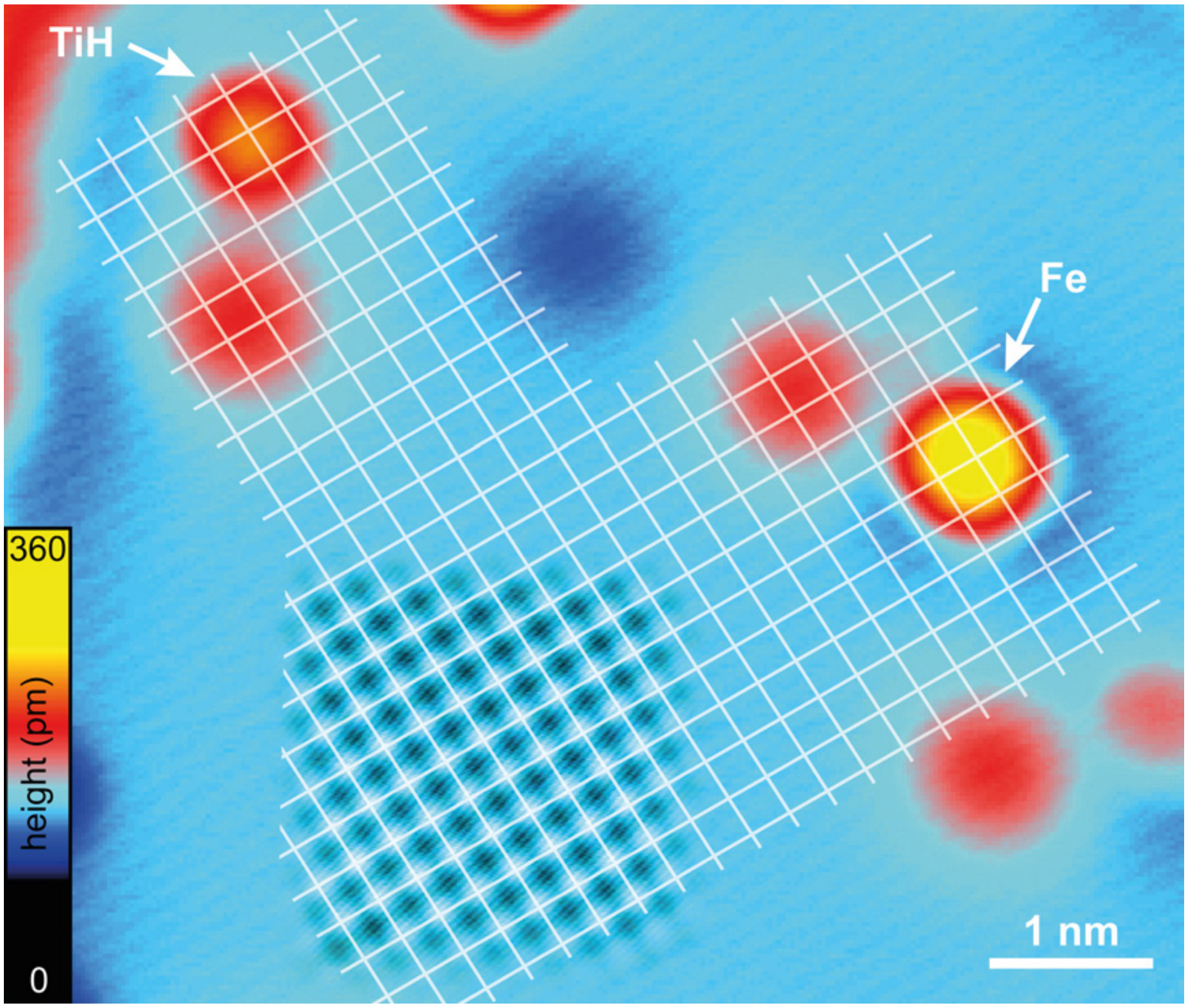

FIG. S2. Adsorption site determination. Constant-current image of a typical 2 ML MgO film with adsorbed TiH and Fe. Superimposed atomic resolution image on the indicated region of the patch was recorded at $V_{DC}$ = -5 mV and $I_t$ = 15 nA. The corresponding larger scale image was recorded at $V_{DC}$ = -5 mV and $I_t$ = 20 pA. The grid formed by the white lines reference the oxygen sites, based on the atomic lattice image.

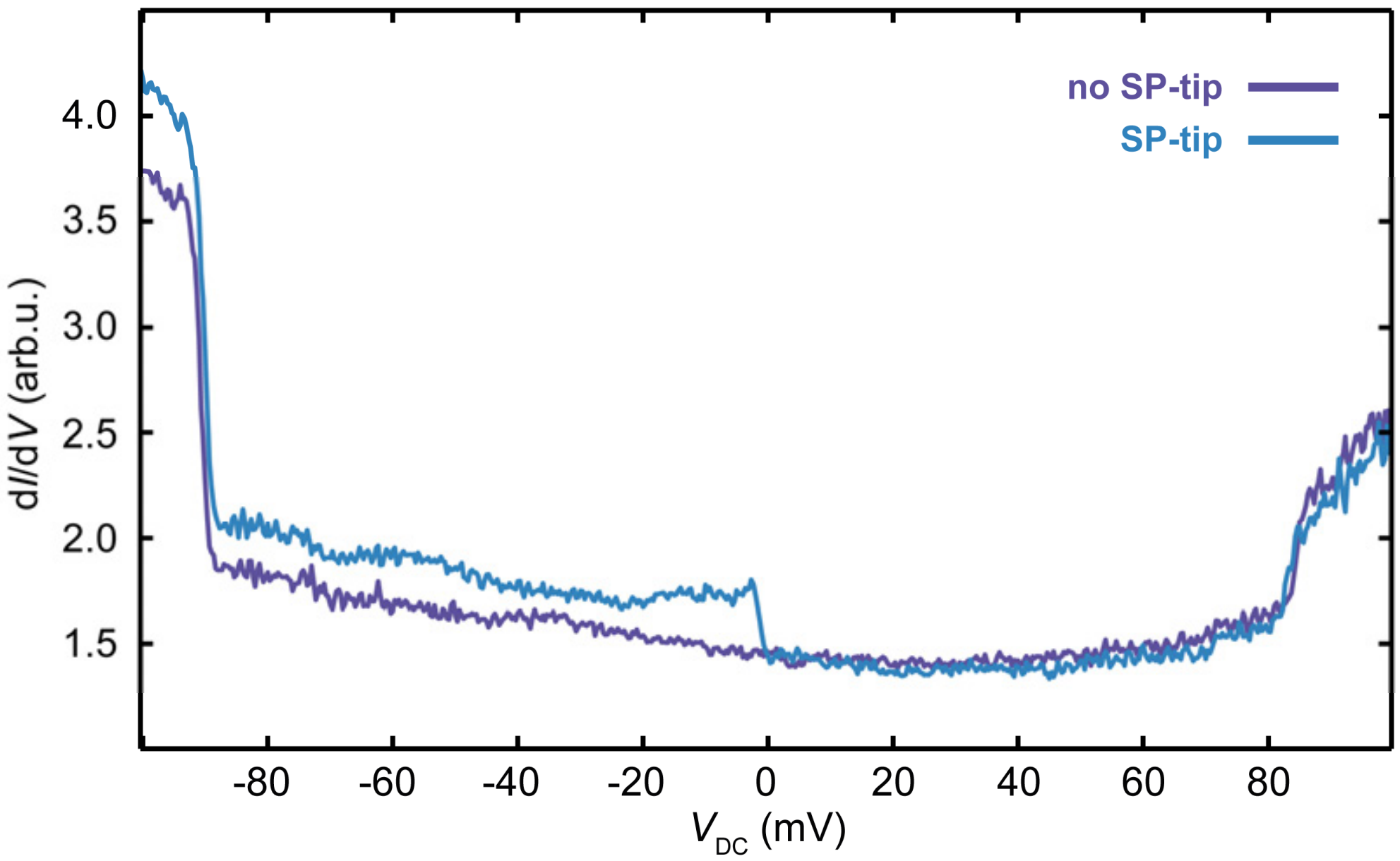

FIG. S3. Spectroscopic fingerprints of TiH. STS recorded on top of the same TiH molecule adsorbed on an oxygen site of the MgO surface, with and without a spin-polarized tip. We observe the orbital excitation at ≈ ±90 meV, as well as signatures of spin pumping for both this excitation and the low energy spin excitation (stabilized at $V_{DC}$ = 100 mV, $I_t$ = 200 pA; $B_\perp$ = 150 mT, $f_{mod}$ = 809 Hz, $V_{mod}$ = 1 mV).

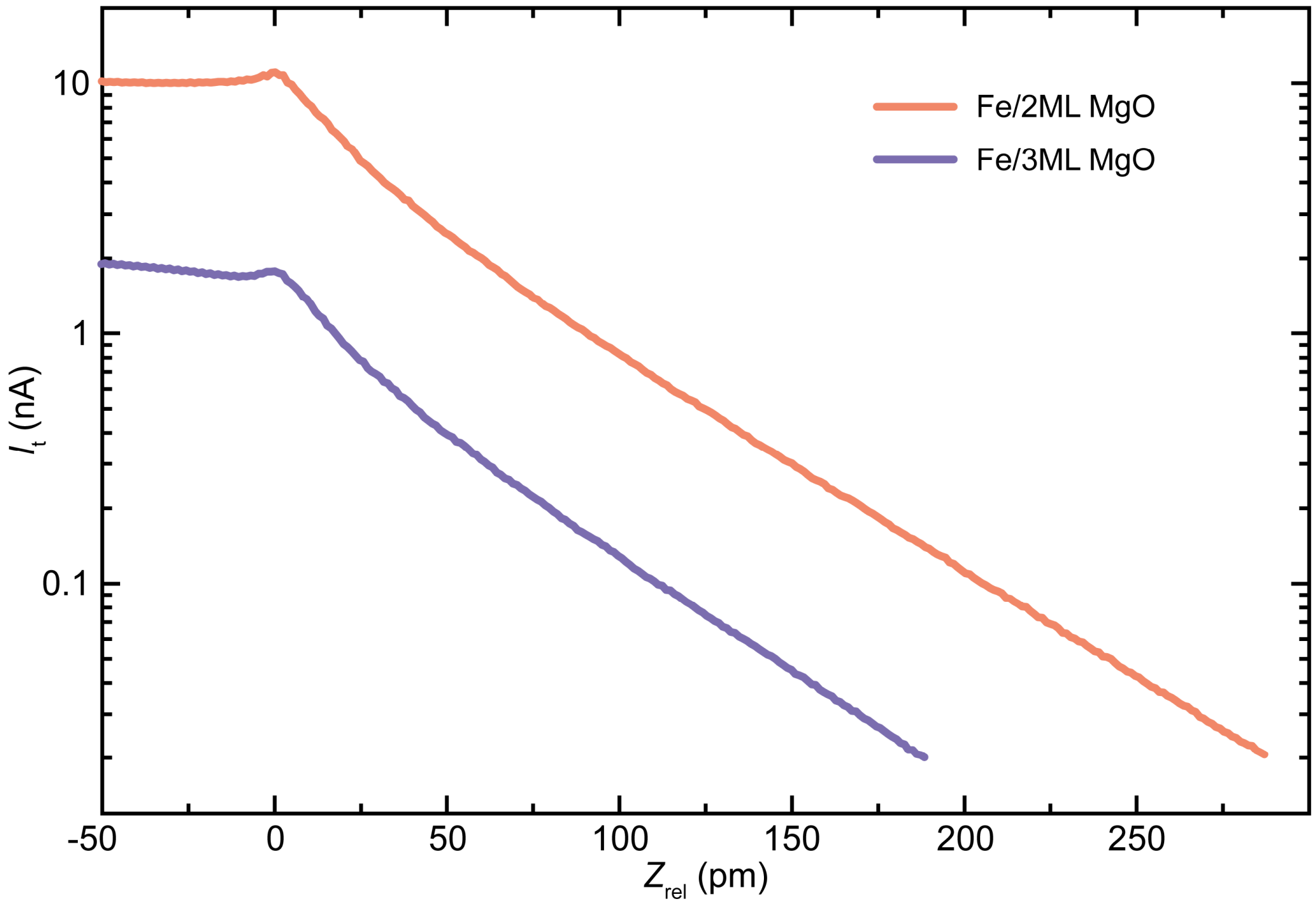

FIG. S4. MgO thickness determination. Point-contact measurements with switched off feedback loop on two different Fe atoms adsorbed on two (red) and three (blue) ML MgO (stabilized at $V_{DC}$ = 10 mV, $I_t$ = 20 pA).

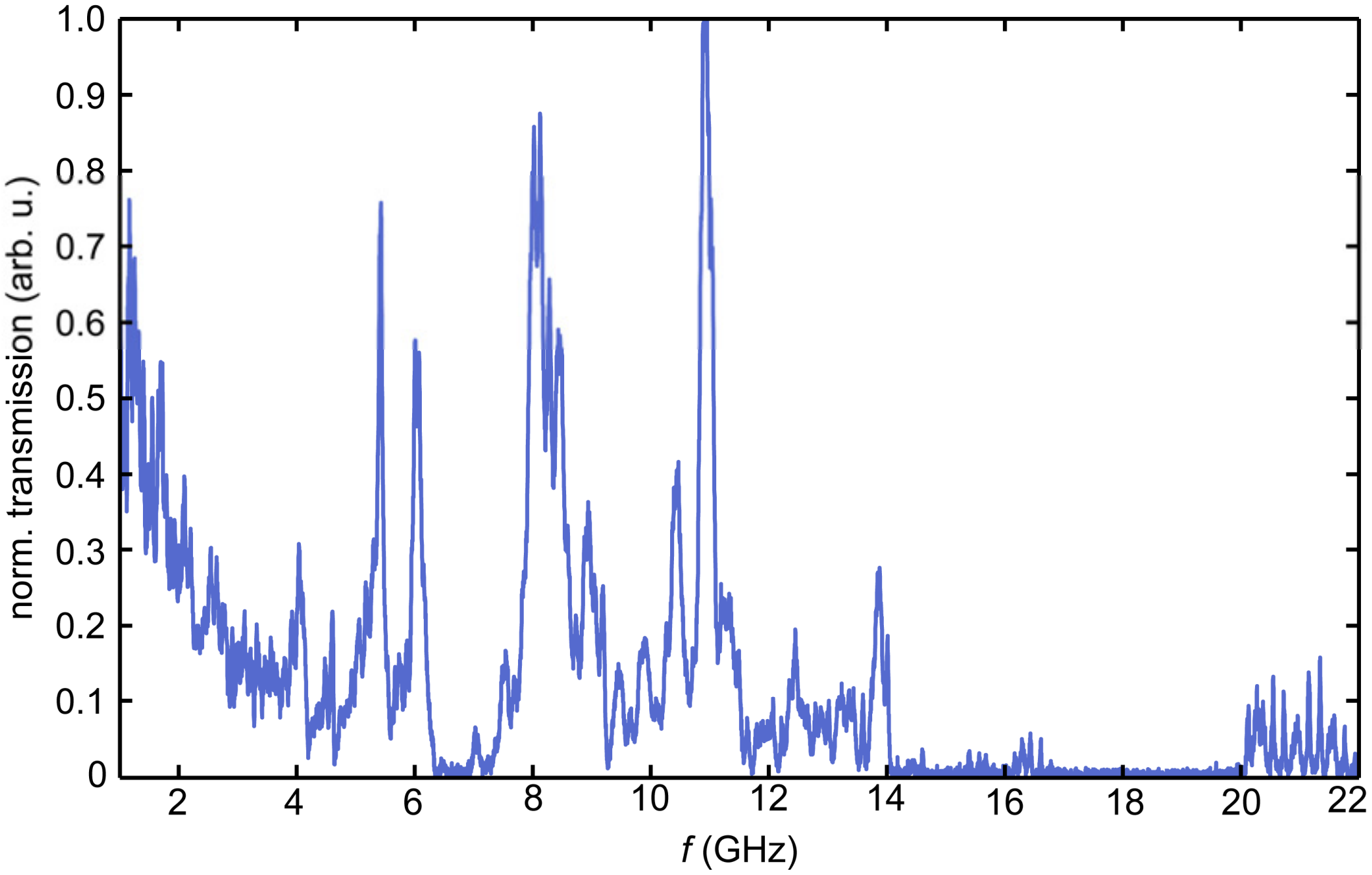

FIG. S5. Transmission of the ESR-STM. Measurement of the frequency-dependent transmission with the RF voltage applied on the tip. The recorded signal was rectified over the non-linearity in the TiH spectrum (Fig. S3) at $V_{DC}$ = -78 mV. The absolute $V_{RF}$ at the junction was not calibrated for this measurement, and therefore the signal was normalized with respect to the highest peak.

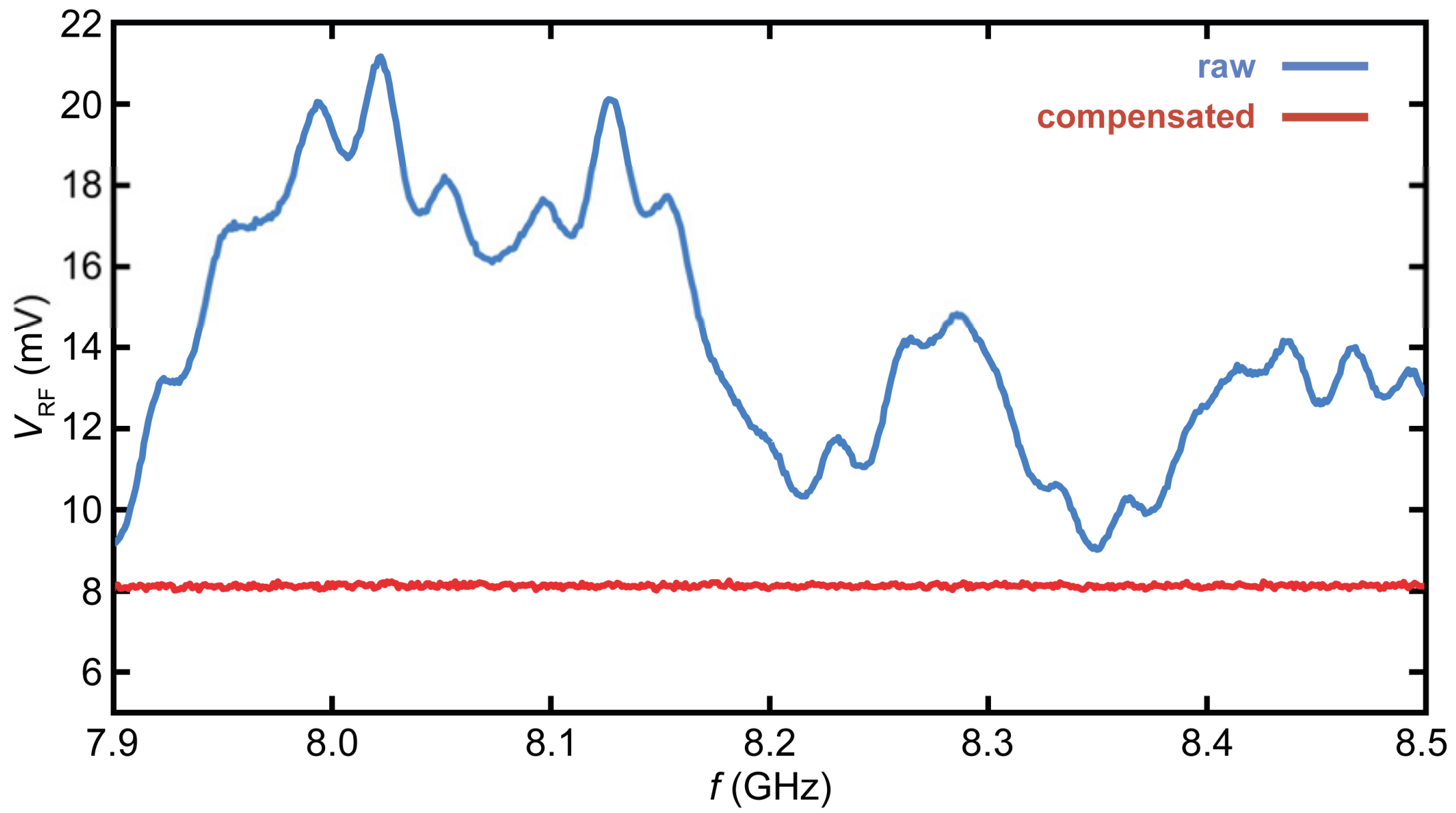

FIG. S6. Flattening of $V_{RF}$. Measured $V_{RF}$ in a region of 7.9 – 8.5 GHz before (blue) and after (red) compensating for the frequency-dependent transmission. A constant $V_{RF}$ was achieved by adjusting the output power of the generator $\tilde{P}_{RF}$ according to the frequency-dependent transmission.

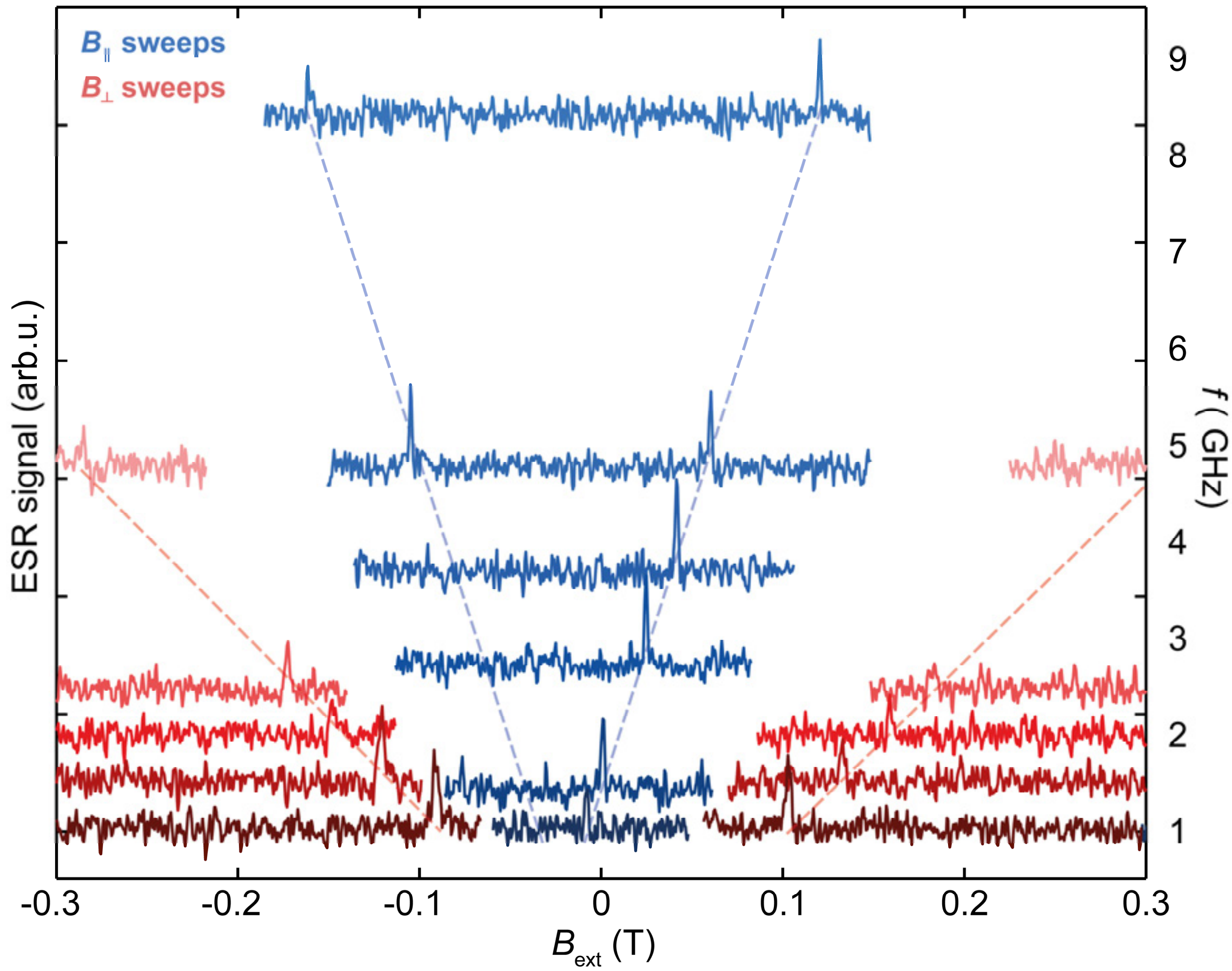

FIG. S7. Complementary ESR raw data for *B*-sweep mode. ESR *B*-sweeps on the same molecule with the same microtip complementary to Fig. 2 of the main manuscript in both field directions $B_{\parallel}$ (blue) and $B_{\perp}$ (red). The same molecule as in the $B_{\parallel}$-sweep and both *f*-sweeps of Fig. 2 was probed. After fitting each peak with a Lorentzian and linearly fitting the ESR peak positions, we obtain $g_{\parallel} = 1.80 \pm 0.01$ and $g_{\perp} = 0.49 \pm 0.01$. Note, these values were obtained prior application of the linearity criterion introduced in section S4.

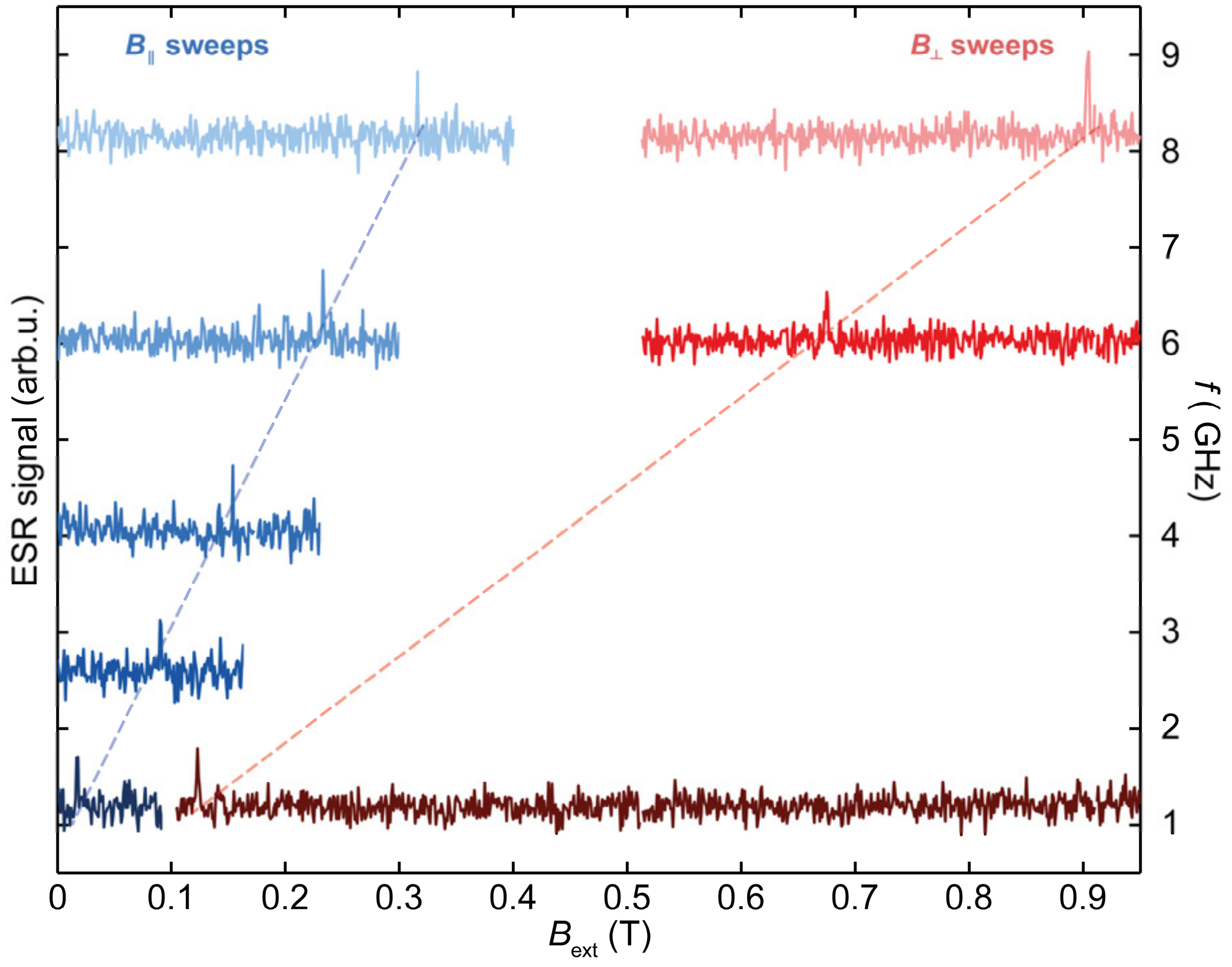

FIG. S8. Additional ESR raw data for *B*-sweep mode. Recorded in both field directions $B_\parallel$ (blue) and $B_\perp$ (red) on the same molecule and micro tip. After fitting each peak with a Lorentzian and linearly fitting the ESR peak positions, we obtain $g_\parallel = 1.82 \pm 0.02$ and $g_\perp = 0.63 \pm 0.01$.

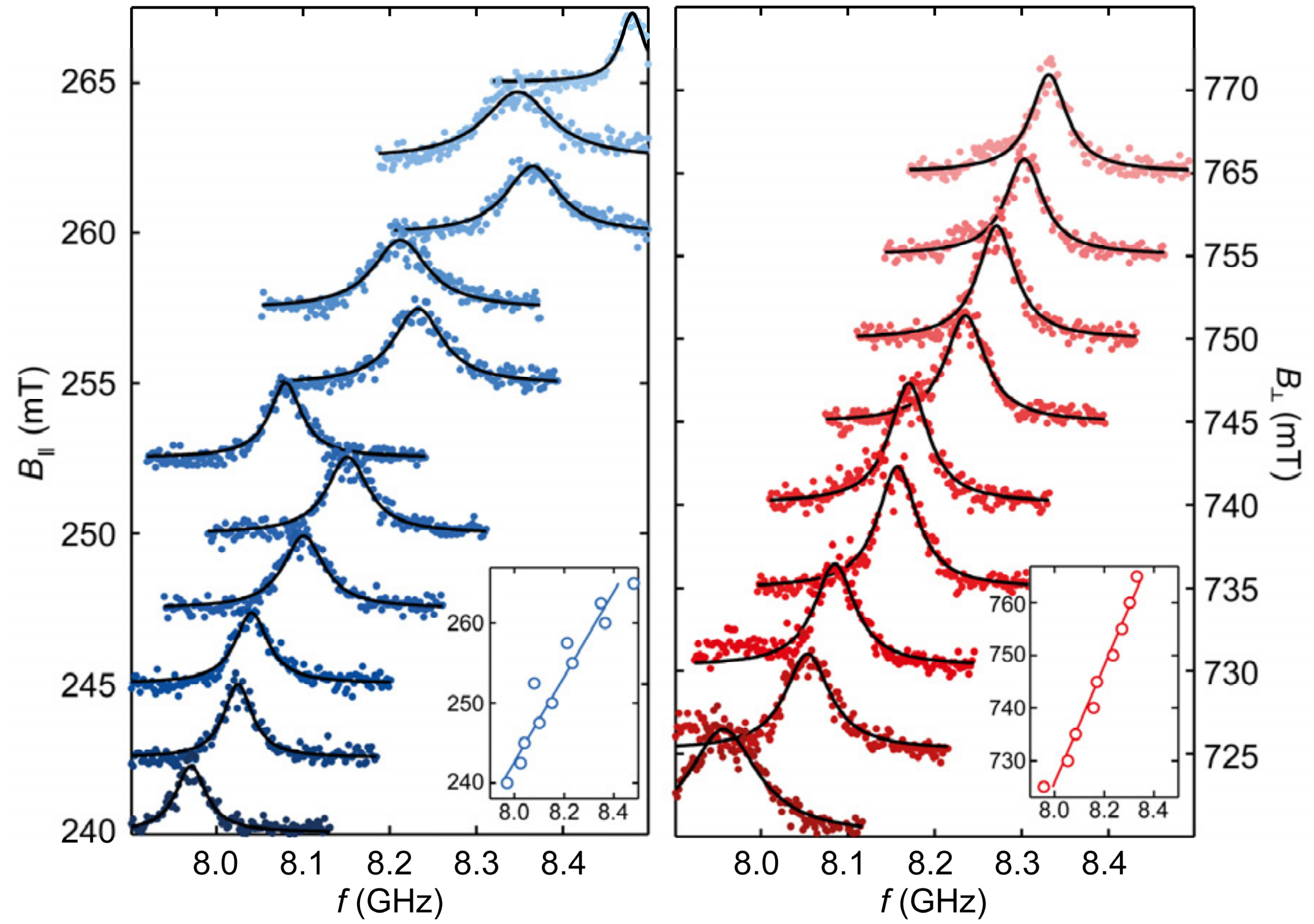

FIG. S9. Additional ESR raw data for *f*-sweep mode. Recorded in both field directions $B_\parallel$ (blue) and $B_\perp$ (red) on the same molecule and micro tip. After fitting each peak with a Lorentzian and linearly fitting the ESR peak positions (see insets), we obtain $g_\parallel = 1.33 \pm 0.14$ and $g_\perp = 0.64 \pm 0.04$.

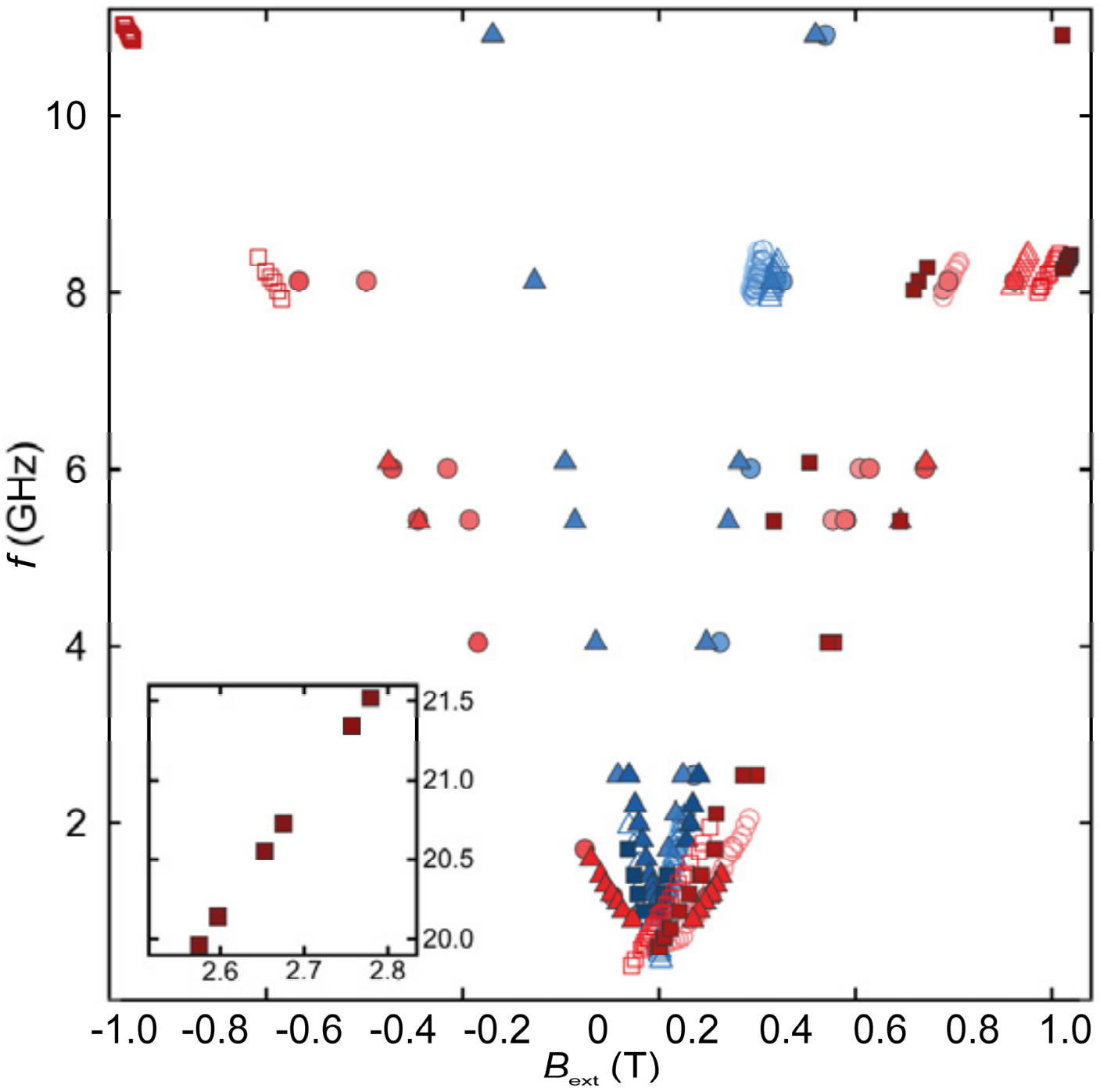

| Set | Name | $I_t$ (pA) | $V_{RF}$ (mV) |
|---|---|---|---|
| 1 | TiH A | 5 | 8 |
| 2 | TiH A | 5 | 8 |
| 3 | TiH A | 2 | 8 |
| 4 | TiH A | 2 | 8 |
| 5 | TiH A | 5 | 8 |
| 6 | TiH B | 2 | 8 |
| 7 | TiH B | 2 | 8 |
| 8 | TiH B | 2 | 7.5 |
| 9 | TiH B | 2 | 7.5 |
| 10 | TiH B | 2 | 7.9 |
| 11 | TiH B | 2 | 7.9 |
| 12 | TiH C | 2 | uncal. |
| 13 | TiH A | 5 | 8 |
| 14 | TiH A | 5 | 8 |
| 15 | TiH A | 5 | 8 |
| 16 | TiH A | 2 | 8 |
| 17 | TiH A | 5 | 8 |
| 18 | TiH A | 2 | 8 |
| 19 | TiH B | 2 | 8 |
| 20 | TiH B | 2 | 8 |
| 21 | TiH B | 2 | 7.9 |
| 22 | TiH C | 2 | 6.8 |
| 23 | TiH C | 2 | 8 |
| 24 | TiH C | 2 | 12.6 |
| 25 | TiH C | 2 | 8 |
| 26 | TiH C | 2 | 8 |
| 27 | TiH C | 2 | 10 |
| 28 | TiH C | 2 | 10 |
| 29 | TiH C | 10 | 20 |
| 30 | TiH C | 10 | uncal. |
| 31 | TiH C | 10 | 10 |

FIG. S10. ESR data sets with various stabilization parameters. Extracted ESR peak positions for the data sets used for the *g*-tensor analysis in Fig. 3B. The tip was stabilized at $V_{DC}$ = 50 mV for all measurements, and the values of $I_t$ and $V_{RF}$ are indicated in the table. For *B*-sweeps in data sets #12 and #30, $V_{RF}$ was not calibrated and $\tilde{P}_{RF}$ set to 23 dBm. In the plot, different symbols correspond to different atoms, filled or open symbols represent *B*-, or *f*-sweeps and blue or red colors represent the external magnetic field direction (∥ or ⊥ to the sample surface).

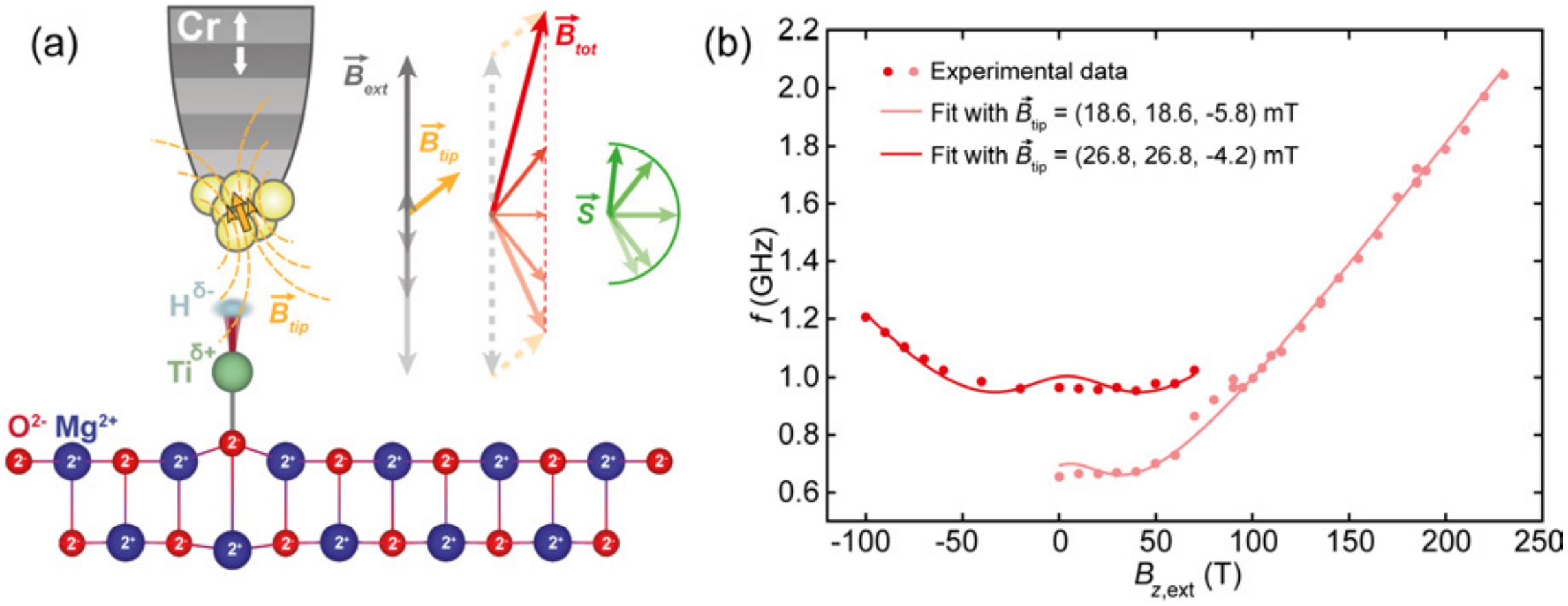

FIG. S11. Tip stray field influence on ESR-STM. (a) Illustration of the TiH molecule influenced by the stray magnetic field of the tip, $\vec{B}_{\mathrm{tip}}$, indicated by yellow dashed field lines. Vector arrows illustrate the variation of the total magnetic field $\vec{B}_{\mathrm{tot}}$ at the TiH spin and its vector decomposition in $\vec{B}_{\mathrm{ext}}$ and $\vec{B}_{\mathrm{tip}}$. The spin ($\vec{S}$) of TiH follows the direction of $\vec{B}_{\mathrm{tot}}$. (b) ESR peak positions (dots) extracted from two data sets with identical measurement parameters ($V_{DC}$ = 50 mV, $I_t$ = 2 pA, $V_{RF}$ = 8 mV), but with a different micro-tip. $\vec{B}_{\mathrm{tip}} = (B_x, B_y, B_z)$ for each data set (legend) is extracted from a global fit of all available data sets (see Fig. S12) and reveals $B_x$ and $B_y$ components that cannot be compensated with $B_{z,\mathrm{ext}}$.

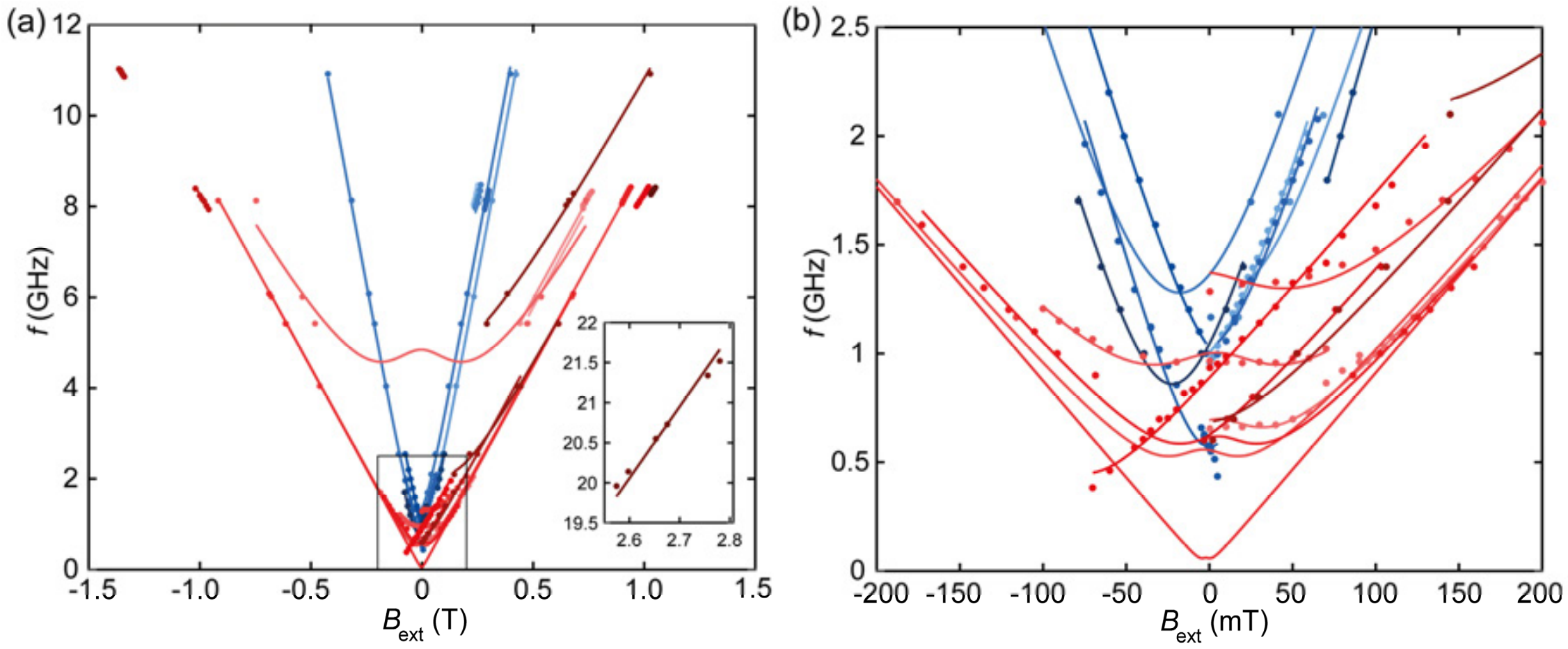

FIG. S12. Fitting measured ESR spectra considering the tip stray field. (a) Extracted ESR peaks from experimental data sets in different $B_{\text{ext}}$ directions $Z$ (red) or $Y$ (blue), fitted with an arbitrary tip stray field $\vec{B}_{\text{tip}}$. The data sets include different tips, atoms, stabilization parameters and $V_{\text{RF}}$. We obtained $g_{\parallel} = 1.896$ and $g_{\perp} = 0.638$. (b) Enlarged section of (a) (indicated with a rectangular frame) showing the non-linear trends as captured by the fit equation.

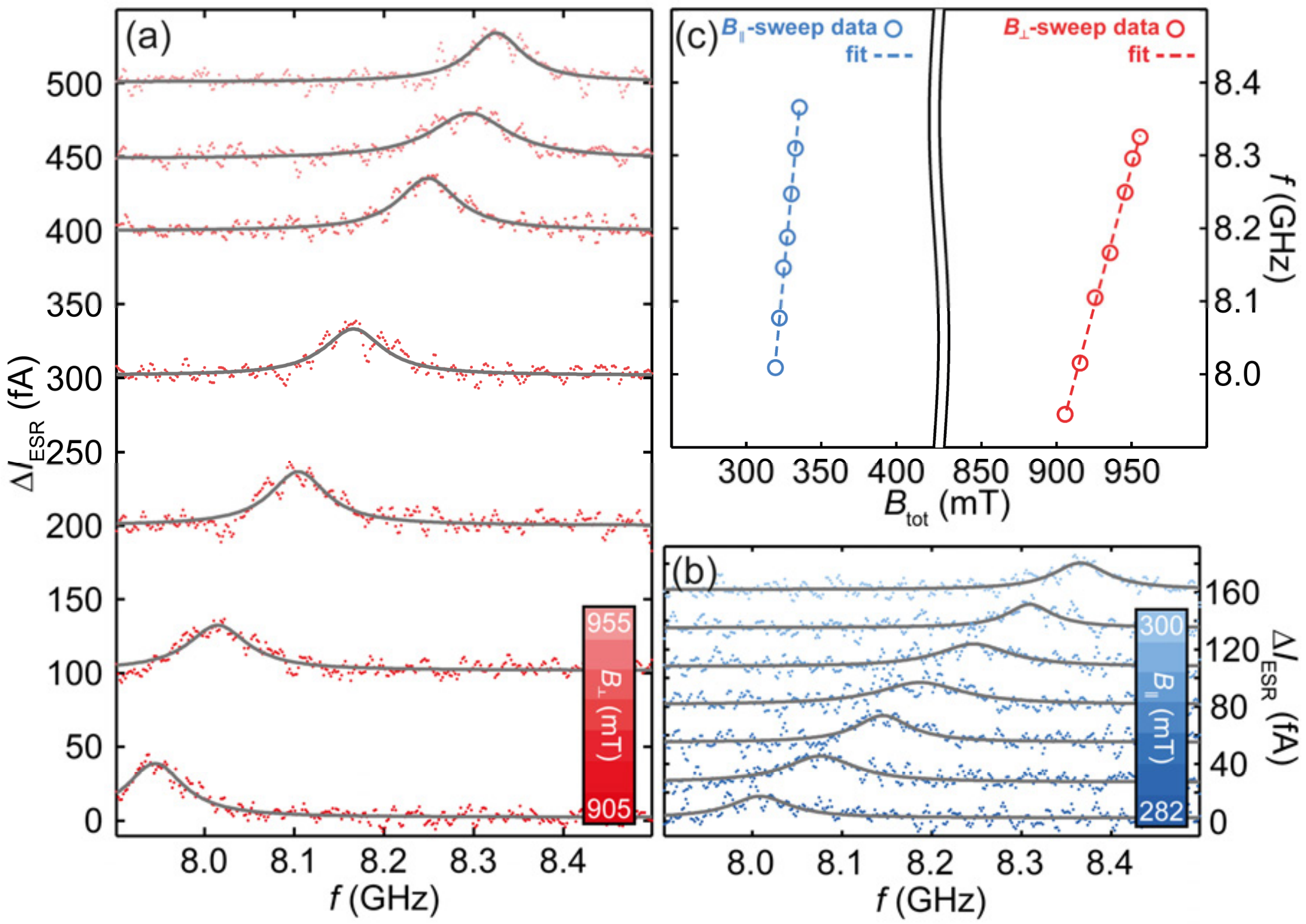

FIG. S13. ESR $f$-sweeps at 1.1 K. (a,b) ESR $f$-sweep measurements in ⊥- (A) and ∥-direction (B) on TiH at $T = 1.1$ K (stabilized at $V_{DC} = 50$ mV, $I_t = 5$ pA; $V_{RF} = 8$ mV). An additional static field of $B_\parallel = 30$ mT and $B_\perp = 150$ mT was applied in (a) and (b), respectively. Peak positions were extracted from Lorentzian fits (gray lines). (c) Extracted resonance peak positions (circles) and linear fits (dashed line) to the data. Resulting $g$-factors of $g_\parallel = 1.57 \pm 0.04$ ([21.90 ± 0.60] GHz/T) and $g_\perp = 0.55 \pm 0.05$ ([7.69 ± 0.68] GHz/T) agree well with the observations at lower temperatures.

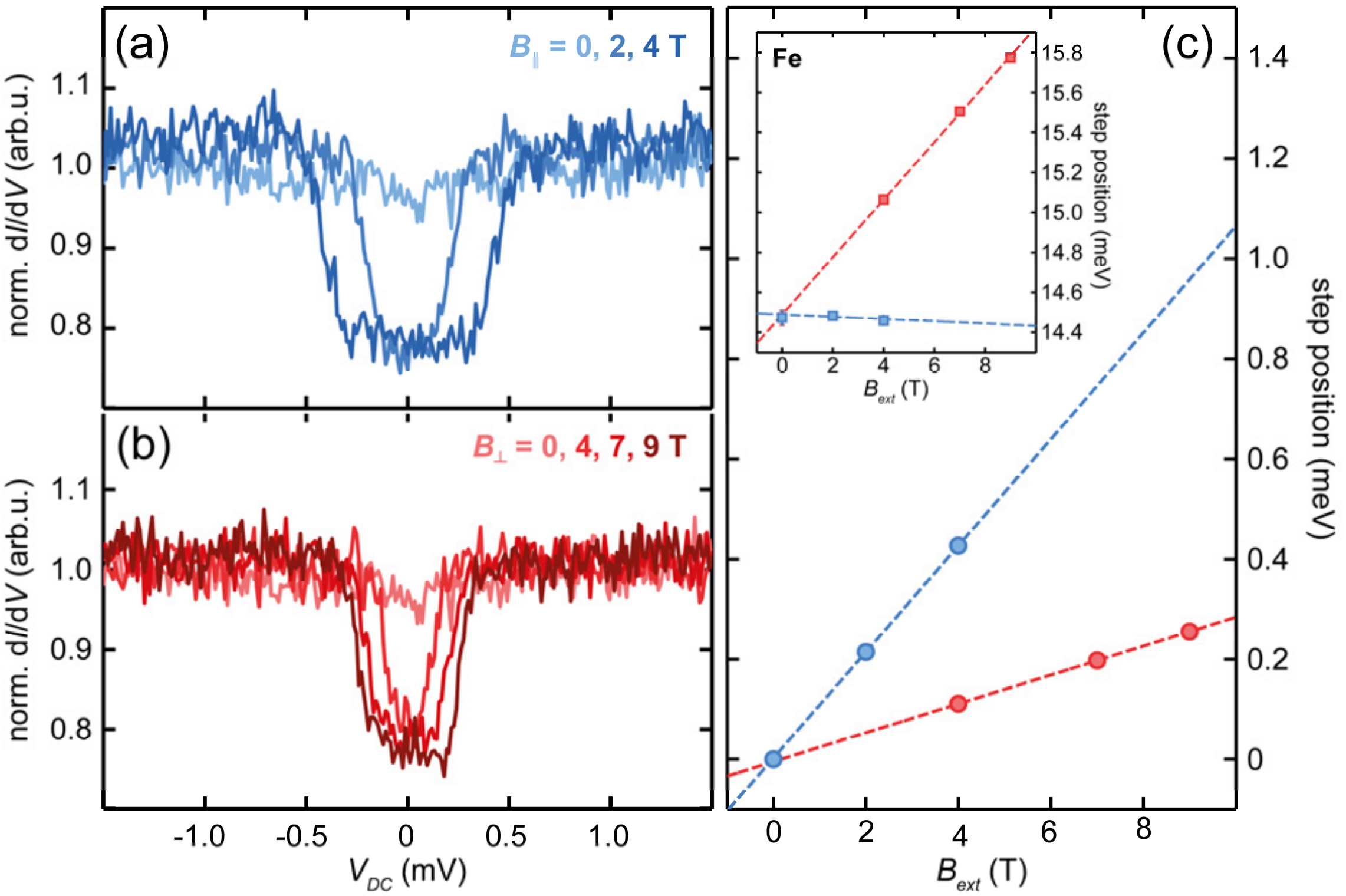

FIG. S14. ISTS of TiH and Fe measured in a vector magnetic field. ISTS of a TiH molecule in (a): in-plane field, and (b): out-of-plane field. The spin-excitation shifts linearly to higher energies for increasing fields, but with different slopes depending on the magnetic field orientation (stabilized at $V_{DC}$ = 3 mV, $I_t$ = 100 pA; $V_{mod}$ = 25 µeV, $f_{mod}$ = 809 Hz). **C** Extracted spin-excitation step position, averaged for four different TiH molecules for the out-of-plane (red) and in plane (blue) direction (error bars are smaller than the marker size). The corresponding $g$-factors are extracted from linear fits to the data and result in $g_\parallel = 1.84 \pm 0.01$ and $g_\perp = 0.50 \pm 0.01$. The inset shows the same measurements for (two averaged) Fe atoms, where $g^*_\perp = 2.48 \pm 0.18$.

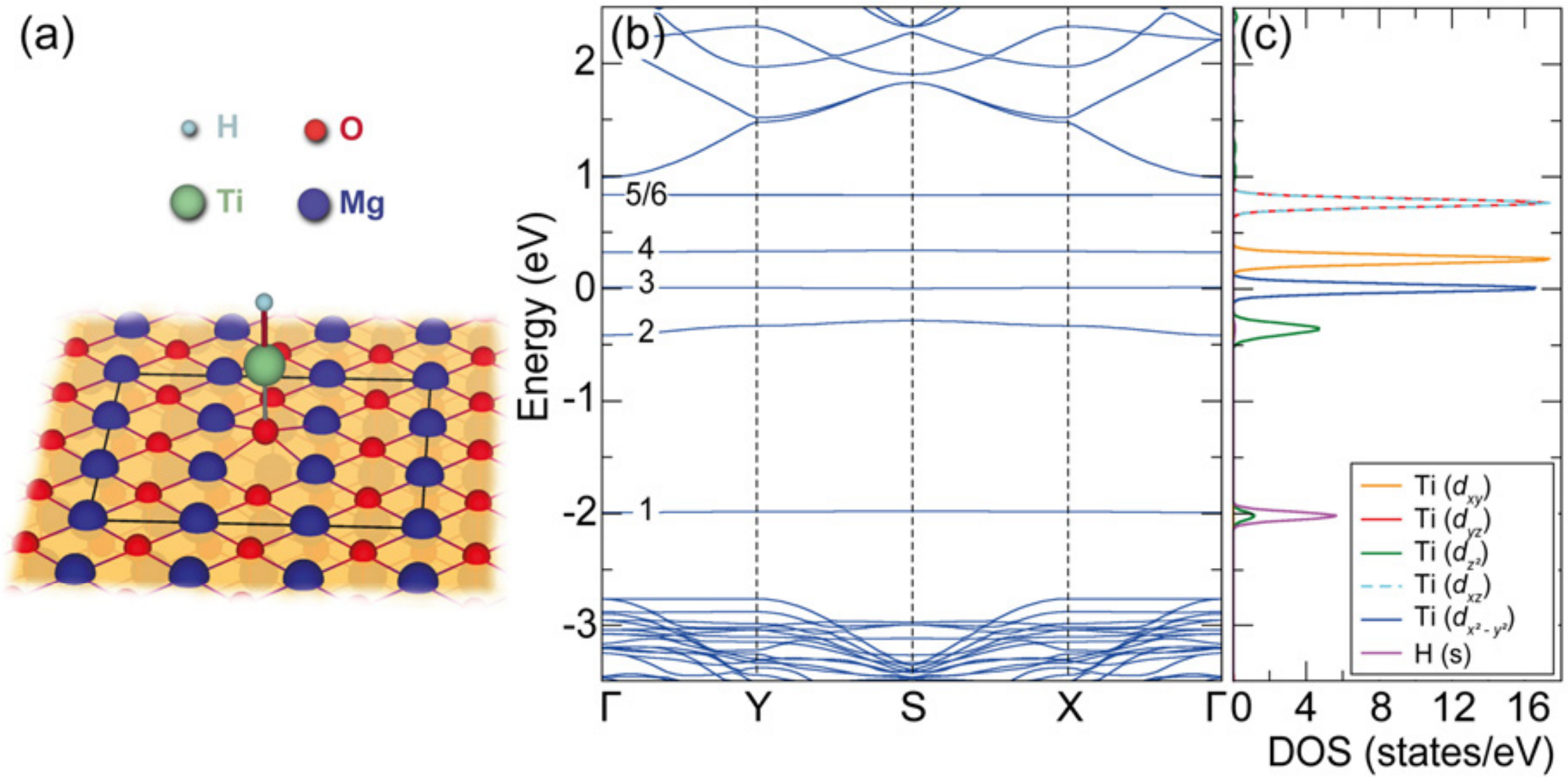

FIG. S15. Geometric and electronic structure from DFT of TiH. (a) Optimized crystal structure illustrated as ball-stick model for TiH adsorbed on the oxygen site of the MgO surface. (b) DFT ($U = 0$) band structure of the TiH/MgO system. Ti($d$) states are split in energy due to the crystal field of the $C_{4v}$-symmetric surface. Indices 1 to 6 label the Ti and H states as used in Table S1. (c) Density of states as a function of energy. Color corresponds to the various atomic orbital contributions for the energy bands of the TiH molecule to the LDOS.

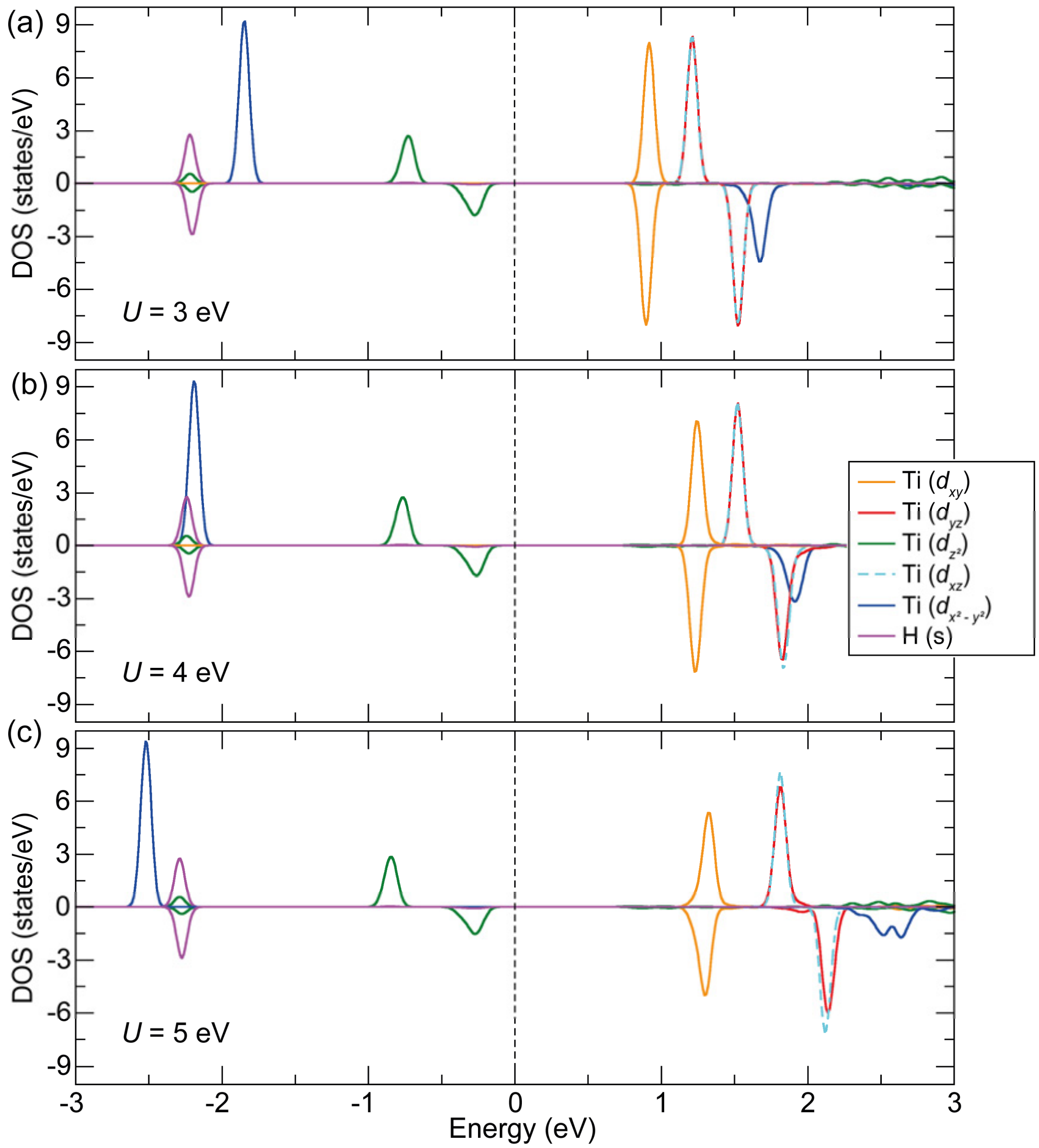

FIG. S16. DFT+$U$ density of states for various values of $U$. Evolution of projected densities of states of TiH/MgO for different values of the $U$ parameter within the DFT+$U$ calculations. For increasing $U$ (a-c), the spin up and spin down channels of the Ti($d_{x^2-y^2}$) state show a stronger splitting and all Ti($d$) states shift to higher energies. The hydrogen $s$ state remains unaffected.

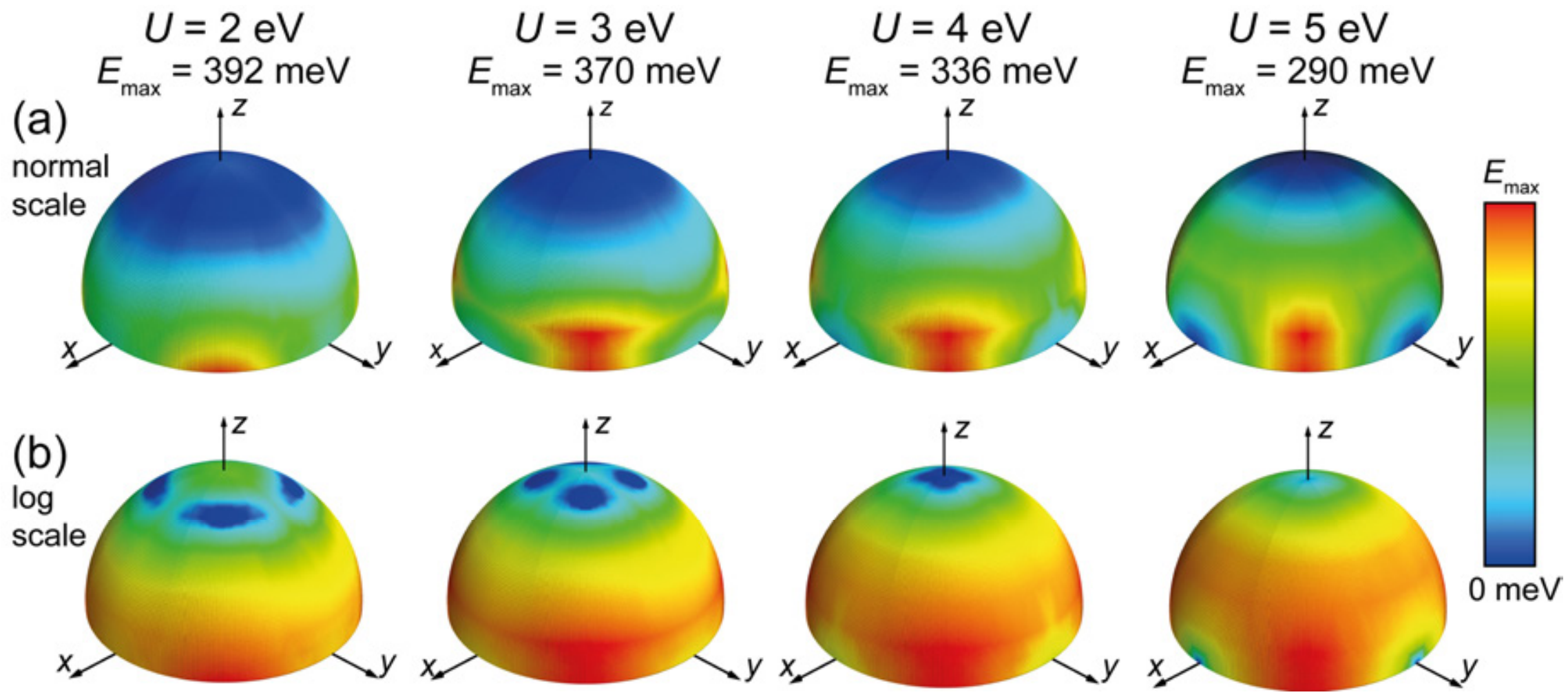

FIG. S17. Potential energy landscape for rotated H obtained from DFT+$U$. (a) 3D potential energy landscape (normal scale) for the TiH molecule adsorbed on MgO obtained by rotating the hydrogen atom around the Ti atom at a fixed Ti-H distance. Corresponding $x$- and $y$-axes are oriented towards the Mg atoms. (b) Same as in (a) but plotted on a logarithmic scale for better visualization of the energy minima.

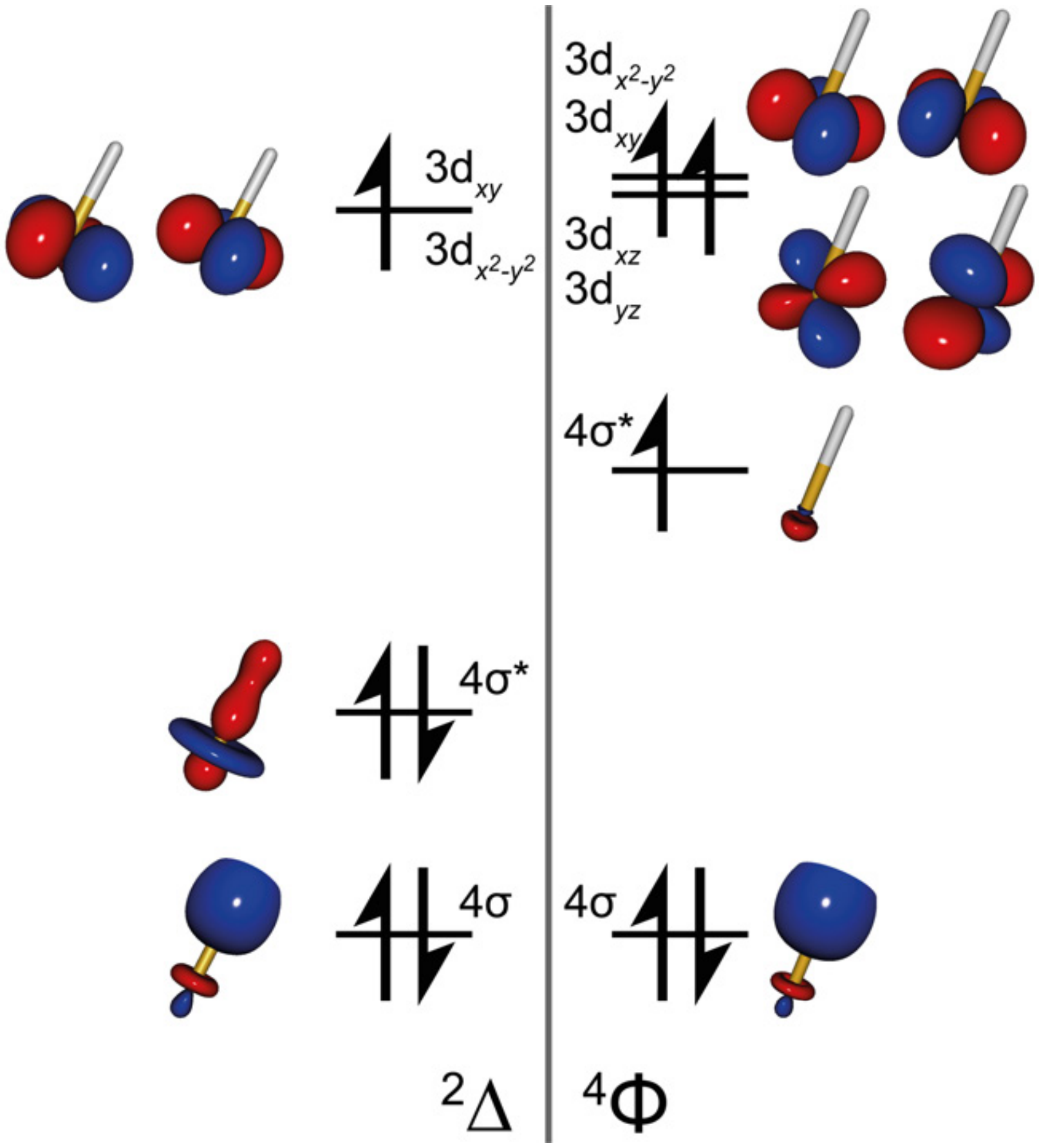

FIG. S18. Molecular orbital diagrams of the gas-phase electronic $^2\Delta$ and $^4\Phi$ states of TiH, accompanied by visual representations of the molecular orbitals.

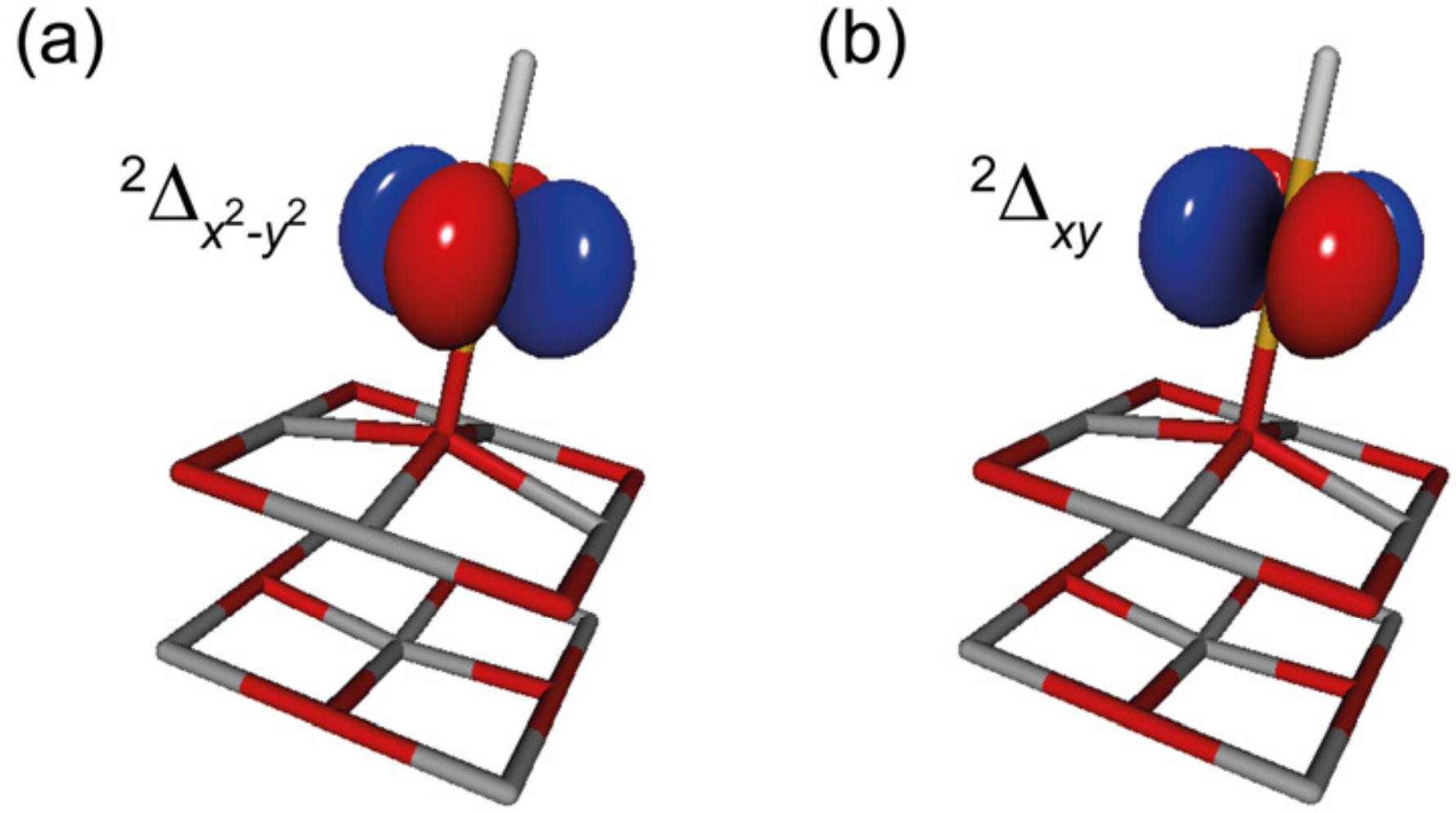

FIG. S19. Highest occupied molecular orbital of the adsorbed TiH molecule in the $^2\Delta$ state on a $Mg_9O_9$ cluster. There is negligible hybridization between the TiH and the MgO surface.

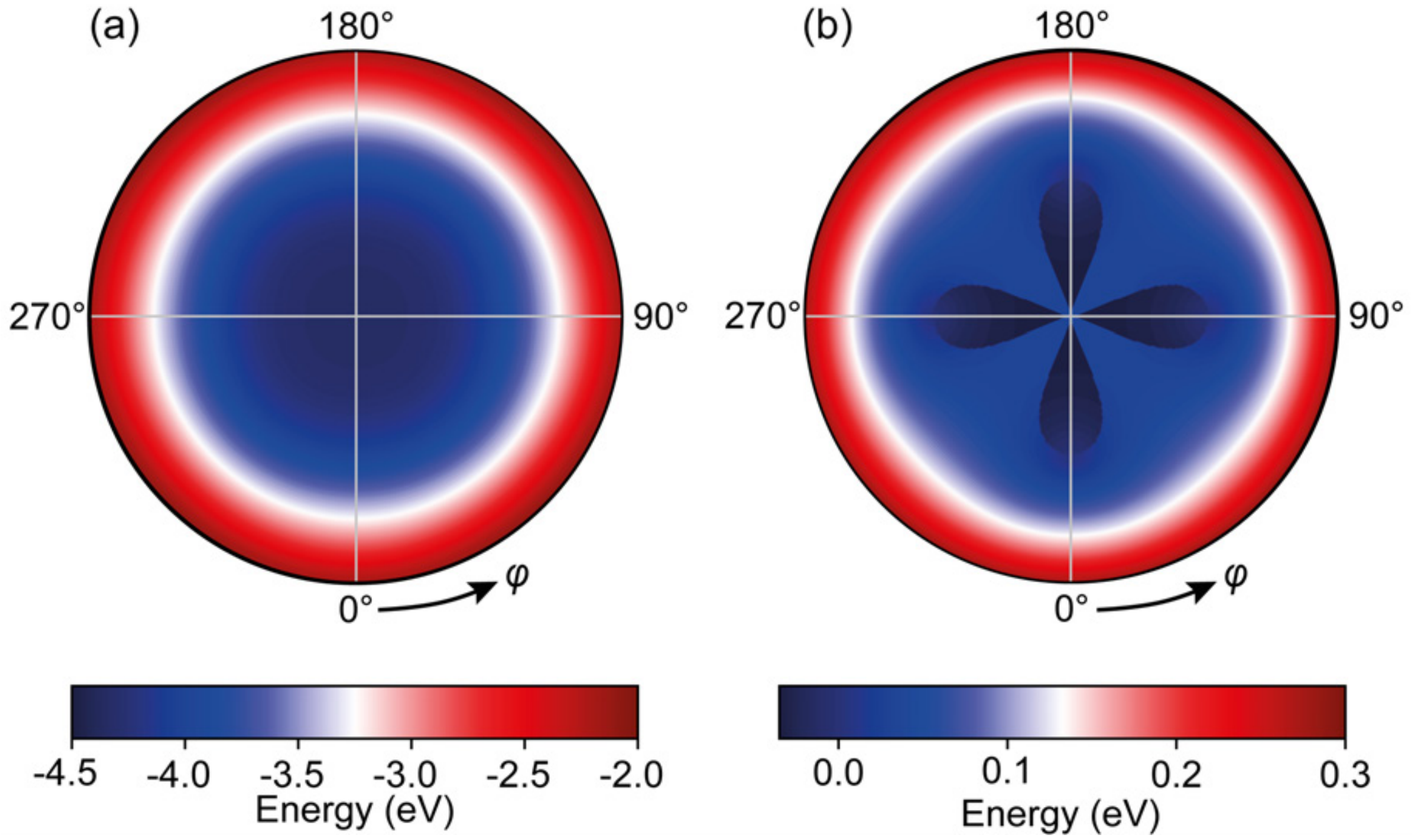

FIG. S20 Potential energy landscape for rotated H from QC. Polar plots of $V_{s,\,\mathrm{avg}}(\theta, \varphi)$ (a) and $V_{s,\,\mathrm{diff}}(\theta, \varphi)$ (b) at a height of 2.42 Å. The azimuth is represented by the angle $\varphi$, where $\varphi = 0$, 90, 180 and 270° is the rotation towards a Mg atom. The radius is given by sin $\theta$, with $\theta$ spanning the sphere in the out-of-plane direction with $\theta = 0°$ in the center and $\theta = 90°$ at the edge.

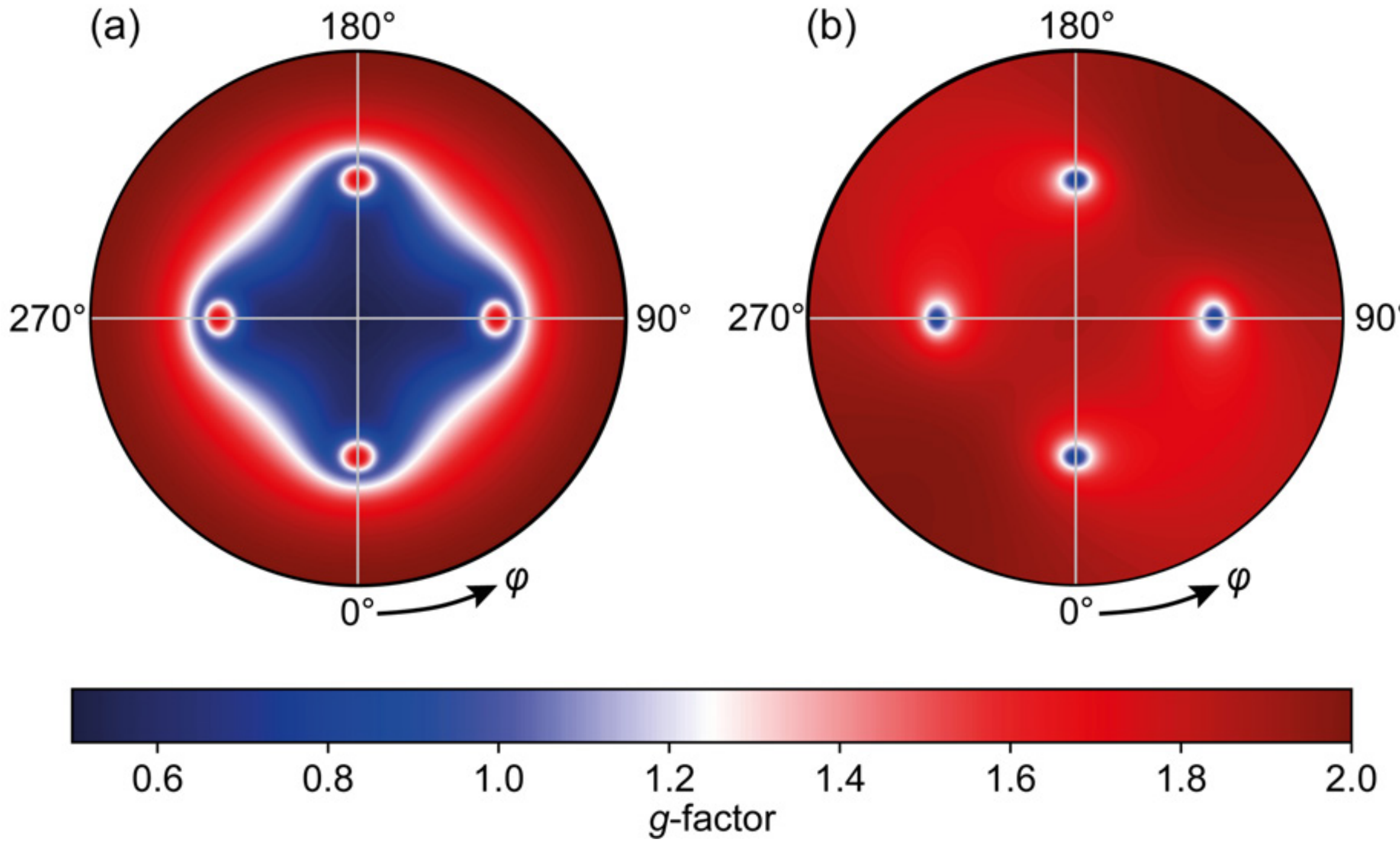

FIG. S21. Calculated $g$-factors for different orientations of TiH from QC. Polar plots of $|g_{\perp}|$ (a) and $|g_{\parallel}|$ (b) at a height of 2.42 Å. The azimuth is represented by the angle $\varphi$, where $\varphi$ = 0, 90, 180 and 270° is the rotation towards a Mg atom. The radius is given by sin $\theta$, with $\theta$ spanning the sphere in the out-of-plane direction with $\theta$ = 0° in the center and $\theta$ = 90° at the edge. For $g_{\parallel}$, the magnetic field is at an angle of $\varphi$ = 21.2°.

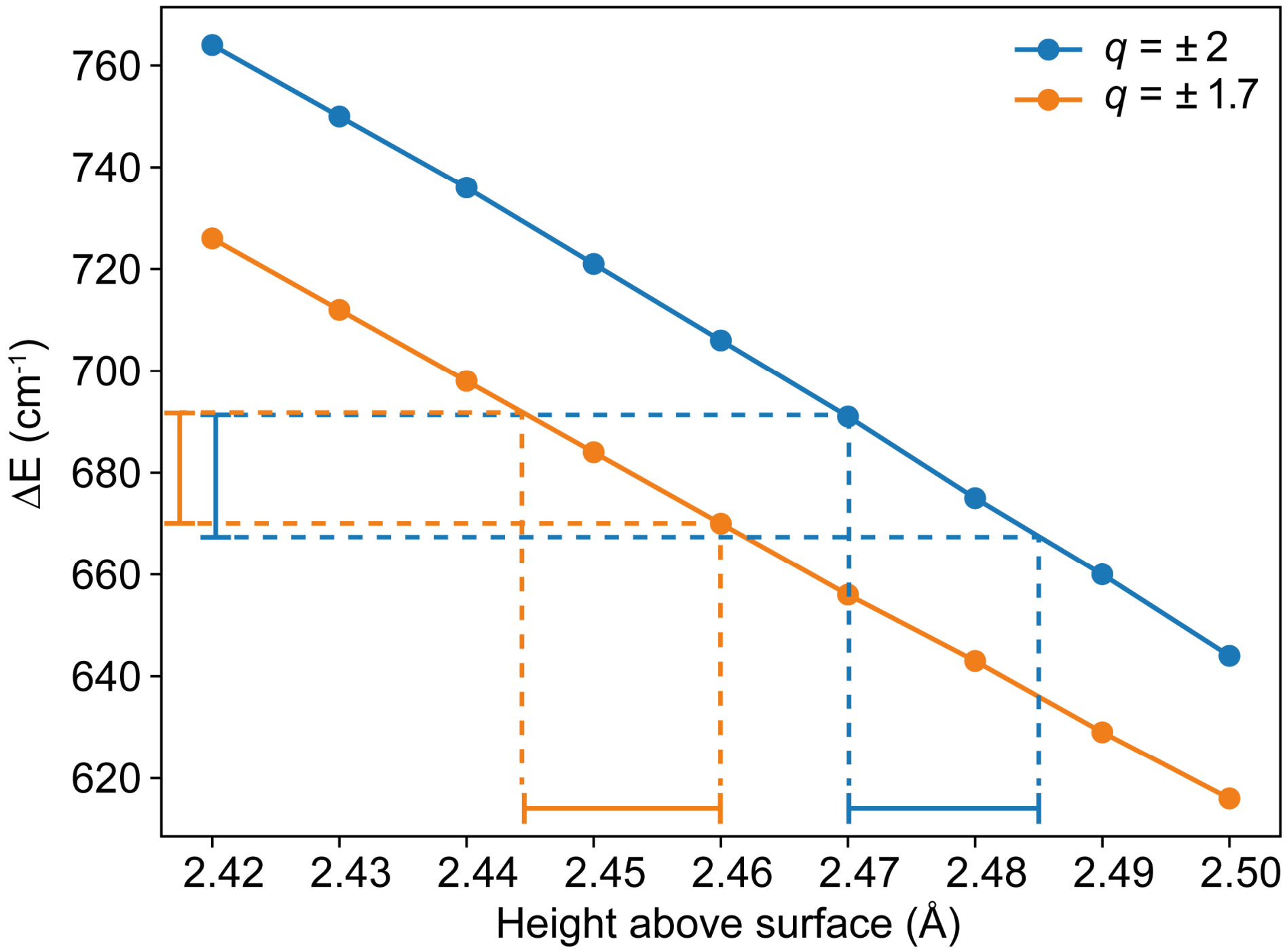

FIG. S22. Splitting of the two Δ-states for different charges. The plots show the calculated energy splitting of the two Δ-states for point charges of ±2*e* (purple) or ±1.7*e* (orange) on the Mg and O sites with respect to the TiH molecule's adsorption height above the surface. As the bars indicate, the changes resulting from a lower adsorption height for the ±1.7*e* charges, results in the same splitting of the states as for the original calculation and charges or ±2*e*.

**TABLES**

TABLE S1. Main atomic contributions to energy bands near $E_F$ from DFT. The band index (1 to 6) corresponds to the notation in Fig. S15. The decimal numbers indicate the contribution of each orbital state to the indicated energy band.

| | Hydrogen | Titanium | | | | | | | Oxygen | | |
|---|---|---|---|---|---|---|---|---|---|---|---|
| | $s$ | $s$ | $p_z$ | $d_{z^2}$ | $d_{xz}$ | $d_{yz}$ | $d_{x^2-y^2}$ | $d_{xy}$ | $p_z$ | $p_x$ | $p_y$ |
| 1 | 0.60 | 0.08 | 0.10 | 0.12 | - | - | - | - | 0.10 | - | - |
| 2 | 0.01 | 0.17 | - | 0.70 | - | - | - | - | - | - | - |
| 3 | - | - | - | - | - | - | 0.99 | - | - | - | - |
| 4 | - | - | - | - | - | - | - | 0.99 | - | - | - |
| 5 | - | - | - | - | 0.32 | 0.63 | - | - | - | 0.02 | 0.03 |
| 6 | - | - | - | - | 0.63 | 0.32 | - | - | - | 0.03 | 0.02 |

TABLE S2. Calculated spin and orbital angular momentums of TiH with DFT+$U$. The orbital moment is found to be very small and independent of the chosen $U$ parameter, which does not reflect the experimental observation.

| | moments along $Y$ ($\mu_B$) | | moments along $Z$ ($\mu_B$) | |
|---|---|---|---|---|
| $U$ (eV) | $m_S$ | $m_L$ | $m_S$ | $m_L$ |
| 3 | 0.81 | -0.04 | 0.82 | -0.04 |
| 4 | 0.84 | -0.04 | 0.83 | -0.04 |
| 5 | 0.88 | -0.04 | 0.85 | -0.04 |

TABLE S3. Optimized values for the different states in the adsorbed phase. Optimal height $h^*$, bond length $R^*$, and energy difference $\Delta E$ between different components for the $^2\Delta$ and $^4\Phi$ states.

| | $^2\Delta$ | $^4\Phi$ |
|---|---|---|
| $h^*$ (Å) | 2.42 ± 0.01 | 2.60 ± 0.01 |
| $R^*$ (Å) | 1.774 ± 0.001 | 1.834 ± 0.003 |
| ΔE (meV) | 46.26 | 0 |

TABLE S4. Calculated $g$-factors with applied parallel static electric field $E_{\parallel}$. For the calculation of $g_{\parallel}$ ($g_{\perp}$), the magnetic field was applied in [1,0,0] ([0,0,1])-direction. The [x, 0, 0]- and [0, y, 0]-directions point towards the neighboring Mg atoms. $E_{\parallel}$ = [0.03, 0, 0] compares to a single charge defect of a neighboring site to the TiH molecule and causes changes in $g_{\parallel}$ ($g_{\perp}$) on the order of ≈ 5 % (15 %).

| $E_{\parallel}$ (a.u.) | $g_{\parallel}$ | $g_{\perp}$ |
|---|---|---|
| [0, 0, 0] | 1.838 | 0.478 |
| [0.01, 0, 0] | 1.828 | 0.506 |
| [0, 0.01, 0] | 1.827 | 0.506 |
| [0.03, 0, 0] | 1.775 | 0.558 |

TABLE S5. Calculated $g$-factors with applied parallel static electric field $E_{\perp}$. For the calculation of $g_{\parallel}$ ($g_{\perp}$), the magnetic field was applied in [1,0,0] ([0,0,1])-direction. The [$x$, 0, 0]- and [0, $y$, 0]-directions point towards the neighboring Mg atoms. Compared to an applied $E_{\parallel}$, the $g$-tensor is less sensitive to changes in $E_{\perp}$.

| $E_{\perp}$ (a.u.) | $g_{\parallel}$ | $g_{\perp}$ |
|---|---|---|
| [0, 0, 0] | 1.838 | 0.478 |
| [0, 0, -0.01] | 1.839 | 0.478 |
| [0, 0, -0.1] | 1.844 | 0.491 |